\documentclass[onefignum,onetabnum]{siamonline181217}

\usepackage{bm}
\usepackage{comment}
\usepackage{setspace}
\usepackage{subcaption}
\usepackage{lipsum}
\usepackage{amsfonts}
\usepackage{graphicx}
\usepackage{epstopdf}
\usepackage{algorithmic}

\newcommand{\ti}[1]{\tilde{#1}}

\ifpdf
  \DeclareGraphicsExtensions{.eps,.pdf,.png,.jpg}
\else
  \DeclareGraphicsExtensions{.eps}
\fi

\usepackage{enumitem}
\setlist[enumerate]{leftmargin=.5in}
\setlist[itemize]{leftmargin=.5in}

\newsiamremark{remark}{Remark}
\newsiamremark{hypothesis}{Hypothesis}
\crefname{hypothesis}{Hypothesis}{Hypotheses}
\newsiamthm{claim}{Claim}

\headers{Embedded discrepancy operators in reduced models of interacting species}{R. E. Morrison}

\title{Embedded discrepancy operators in reduced models of interacting species\thanks{Submitted to the editors DATE.}}

\author{Rebecca E. Morrison\thanks{Department of Computer Science, CU Boulder, Boulder, CO
  (\email{rebeccam@colorado.edu}).}}

\usepackage{amsopn}

\ifpdf
\hypersetup{
  pdftitle={Embedded discrepancy operators in reduced models of interacting species},
  pdfauthor={R. E. Morrison}
}
\fi

\begin{document}

\maketitle

\begin{abstract} In many applications of interacting systems, we are only interested in the dynamic
    behavior of a subset of all possible active species. For example, this is true in combustion
    models (many transient chemical species are not of interest in a given reaction) and in
    epidemiological models (only certain critical populations are truly consequential). Thus it is
    common to use greatly reduced models, in which only the interactions among the species of
    interest are retained. However, reduction introduces a model error, or discrepancy, which
    typically is not well characterized. In this work, we explore the use of an embedded and
    statistically calibrated discrepancy operator to represent model error. The operator is embedded
    within the differential equations of the model, which allows the action of the operator to be
    interpretable. Moreover, it is constrained by available physical information, and calibrated
    over many scenarios. These qualities of the discrepancy model---interpretability,
    physical-consistency, and robustness to different scenarios---are intended to support reliable
    predictions under extrapolative conditions.

\end{abstract}



\section{Introduction}\label{sec:int} In the realm of computational modeling today, we---scientists,
mathematicians, and engineers---investigate, design, optimize, and make predictions and decisions
about an incredible multitude of real-world systems. In general, a computational model implements a
mathematical model; the mathematical model represents, using abstraction and simplification, the
actual system in question. There are thus two critical connections we must check to ensure the success of
computational modeling, broadly classified into the two subject areas of verification and
validation. 

\emph{Verification} is the process by which we check that any computation correctly solves the
mathematics. For example, proper verification procedures include code documentation, unit and
regression testing, and solution comparisons against a posteriori error estimates, to name a few.
While critical to the success of computational modeling, verification is not a concern of this
paper: we assume all computational implementations are correctly documented, implemented, and
executed.  For more information about verification, see e.g.,
\cite{oberkampf2010verification, prudhomme2003practical, roache2001code}.

\emph{Validation} is the process by which we check that the mathematical model faithfully represents
the system in question. At its most basic, validation compares model output to observations.
Statistical techniques that do not require knowledge of the model include, for example, goodness of
fit (computing $R^2$ values), analysis of residuals between model output and data, and $k$-fold
cross validation \cite{bruce2017practical}. However, a validation process may require a more nuanced
procedure, depending on what one plans to do with the model. In \cite{oliver2015validating}, Oliver
et al.\@ describe a sophisticated approach to model validation for predictions of unobservable
quantites. Their framework relies on knowledge of the model and system under study, and takes into
account the behavior of the model over different scenarios. There are  other approaches that go
beyond purely statistical tests. For example, in \cite{bayarri2007framework}, Bayarri et al.\@ describe
a comprehensive framework for the validation of computer experiments. This work describes detailed
processes such as determining appropriate domains of model inputs, guarding against overfitting, and
accounting for bias in the simulation output. As another example, in \cite{farrell2017adaptive},
Farrell-Maupin and Oden describe an adaptive method for model calibration and validation using
increasingly complex models. Additional richness is only introduced to the model after the simpler
version is shown to be invalid.

Note that all validation procedures rely on access to observations, which should (hopefully)
include a description of the associated measurement error. If there is some mismatch between the
model and the observations, the source of the discrepancy could either be the model, or the
observations, or both. Reliable experimental practices and proper data reduction techniques ensure
that all observations are reported correctly, with quantified measurement uncertainty. In this paper
we assume that any discrepancy between the model and observations is not caused by faulty
experimental procedures or reporting. In this way, we may focus on what to do when
the model itself causes the discrepancy.

Once a discrepancy that cannot be reasonably explained by measurement error has been observed, then
we would like to improve the model directly. Indeed, much of basic science aims to do exactly that:
When one hypothesis is shown invalid, a scientist or team of scientists proposes a new model
based on updated evidence. This evidence could be theoretical, experimental, or even computational. Then
of course this new model would again be subject to validation processes.

However, a direct improvement of the model may not be possible, due to time, financial, or
computational limitations. Additionally, one may not know why the model is invalid or how to improve
it. Despite the model error, a needed prediction or decision may require the model's use, before it
can be directly improved. In this case, we must try to model the discrepancy itself.

Consider that the discrepancy is revealed by comparing some set of model output to the corresponding
set of observations. If a bias is perceived between the two, a natural first step would be to attach
a discrepancy function or stochastic process to the model output, which can then be calibrated to
correct the model. In fact, this type of discrepancy model, which we term \emph{response
discrepancy model}, has been duly investigated, starting with the fundamental work of Kennedy and O'Hagan
in 2001 \cite{kennedy2001bayesian}. Since then, the response discrepancy model has been adapted into
fields as diverse as climate modeling \cite{stainforth2007confidence}, hydrology
\cite{renard2010understanding}, and cardiology \cite{mirams2016uncertainty}, among many others. A
response discrepancy model can be useful and relatively quick to develop when one only need
interpolate between data points.

Recall, however, our goals for computational modeling: to investigate, design, optimize, and make
predictions and decisions. To achieve these goals, we must be able to trust the model output beyond a
specific calibration scenario---otherwise we could just rely on observations without need for a
model! To this end, we aim to represent the model discrepancy with a
discrepancy operator embedded within the model itself, i.e., an \emph{embedded discrepancy operator}.

There are several advantages of an embedded discrepancy operator. First, the operator
can be constrained by physical information such as conservation laws, symmetries, fractional
concentrations, nonnegativity constraints, and so on. Second, as a function of state
variables or other existing model variables, then the action of the discrepancy operator is physically
interpretable. Third, the operator can be calibrated over many different scenarios, such as initial
conditions, boundary conditions, or simulation geometries. Because of these
qualities---physical-consistency, interpretability, and robustness to different scenarios---an
embedded discrepancy model could be valid for extrapolative predictions.

Embedded, or intrusive, approaches have been previously investigated. In
\cite{sargsyan2015statistical}, Sargsyan et al.\@ allow for model error by endowing model parameters
with random variable expansions. As an approach to model discrepancy, this does not break physical
constraints, and the random parameters can be calibrated over many scenarios. However, not all model
error can by captured in this way. With complex computational models, it is certainly possible that some
discrepancy is caused by missing physics or misspecified physics---problems that cannot be
accounted for with these types of parameter expansions.

When the discrepancy model becomes part of the model itself---yielding an augmented or
\emph{enriched model}---then the specific form of the discrepancy model depends on the modeling
context. In \cite{morrison2018representing}, the authors investigate this type of inadequacy
operator in the context of chemical kinetics for combustion. Portone et al.\@ developed an embedded operator in
\cite{portone2017stochastic} for porous media flow models of contaminant transport. In this work, we
propose and analyze a class of embedded discrepancy operators in the context of the generalized
Lotka-Volterra (GLV) equations.

The GLV equations describe the interactive behavior of any number $S$ of different species. The
concentration of each species is represented by a variable $x_i, i=1,\dots S$; there is one
differential equation for each $x_i$ whose right-hand side (RHS) includes a linear growth rate term
and nonlinear interaction terms. This framework is used to describe many types of physical systems.
For example, reaction models in chemical kinetics \cite{steinfeld1989chemical}, ecological models
\cite{barabas2016effect}, and epidemiological models \cite{dantas2018calibration} often take this
form. In these applied fields, it is common to use a reduced model that only includes $s < S$
species. For example, there are over 50 chemical species believed to be involved in methane
combustion \cite{grimech}; in practice, often merely five to ten species are in fact included in a
reduced model \cite{bilger1990reduced}. In epidemiology models, it is common to include humans and
the disease carriers, e.g., mosquitoes (with subpopulations of susceptible, exposed,
infected, and recovered), while omitting many others such as cattle or non-human
primates \cite{childs2019mosquito}. As reduced models, built about $s <S$ species, are quite common, we are thus
interested in the model discrepancy caused by the use of this type of reduced model.

One immediate question that arises at this point is the following: Given a system of $S$ species,
suppose only $s$ are truly of interest. What is the \emph{best reduced model}, using only
information about those (given) $s$ species? Although a natural first step, the question of model
reduction is beyond the scope of this work. For now, note that common model reduction techniques may
be undesirable or unused: some techniques will output quantities that do not directly correspond to
individual species, such as eigenvalue methods. Also, many field scientists or modelers may already
be working with commonly used reduced models, because the detailed models are either too
computationally expensive or are in fact unknown. (This is indeed the case in the combustion and
disease modeling examples above.) Therefore, the objective of this work is to instead answer the
following question: \emph{Given a reduced model with $s$ species, meant to act in place of a detailed one
of $S$ species, where $s < S$, how can we represent the resulting model error?} 

Previous work shows how a set of $S$ coupled Lotka-Volterra equations can be converted to a set of
$s$ equations, $s < S$, using algebraic substitutions and/or integration \cite{morrison2019exact}.
For example, two coupled differential equations for state variables $x$ and $y$ can be converted to
a single differential equation for $x$, but the resulting equation will either need depend on higher
derivatives of $x$, or on its complete time history. There are now two important points to consider:
1) We are interested in the setting where a reduced model over $s$ species replaces a detailed model
of $S$ species, and 2) A system of $S$ ODEs can be converted to a system of $s$ coupled differential
equations, but these remaining equations include either higher derivatives or time history of the
remaining $s$ variables. Connecting these two concepts together motivates the particular discrepancy
models in this work.

The paper is organized as follows. A brief review of the GLV equations along with a description of
the detailed and reduced models is given in Section~\ref{sec:glv}. Section~\ref{sec:mco} reviews the
types of model conversion that motivate the proposed discrepancy models. Section~\ref{sec:edo}
proposes a class of embedded discrepancy operators and how to enforce physics-based constraints.
The details of calibration and validation for the enriched models are given in
Section~\ref{sec:calval}. Numerical results are in Section~\ref{sec:num}, and Section~\ref{sec:con}
provides a concluding discussion.

\section{Generalized Lotka-Volterra equations}
\label{sec:glv}
The generalized Lotka-Volterra equations are sets of coupled ordinary differential equations, used to model the
time dynamics of any number of interacting quantities. In particular, the Lotka-Volterra framework
allows for linear (growth rate) and quadratic (interaction) terms.  Let the $S$-vector $x$ represent
species concentrations. Here, the units of a particular $x_i$ refer to the number of specimens per
unit area, but specific units are omitted in this paper. The GLV equations are written succintly as:
\begin{equation}
    \frac{dx}{dt} = \text{diag}(x)(r + Ax),
\end{equation}
where the vector $r \in \mathbb{R}^S$ represents the intrinsic growth rates, and the matrix $A \in
\mathbb{R}^{S\times S}$ collects the interaction rates. That is, the $ij$-th entry of $A$, $a_{ij}$,
indicates how species $j$ affects the concentration of species $i$. The term \emph{intraspecific}
refers to interactive behavior within a particular species (the $a_{ii}$ are intraspecific terms),
while the term \emph{interspecific} refers to the behavior between two different species (the
$a_{ij}, i \neq j$, are interspecific terms).

\subsection{Detailed and reduced models}\label{ssec:dnr}
Let us now introduce some notation and terminology. The objects in the detailed models will be
denoted with a $\hat \,$ symbol. The detailed model is referred to as $\mathcal{D}$, as in
\begin{equation}
    \frac{dx}{dt} = \mathcal{D}(x) = \text{diag}(x)(\hat r + \hat A x).
\end{equation}
Since this model is completely determined by the vector $\hat r$ and $\hat A$, we also say $D = \{\hat A, \hat
r \}$. The species included in the detailed model are called the \emph{detailed set}.

Given a detailed model, there are various ways one could arrive at a reduced model. In this paper,
the reduced model is comprised of all terms involving the reduced species set, i.e., by subsampling
the detailed one. For example, suppose $S=3$, $s=2$.  Then the detailed model, written out, is:
\begin{subequations}
    \begin{align}
        \dot x_1 &= r_1 x_1 + (a_{11} x_1 + a_{12}x_2 + a_{13}x_3) x_1 \label{eq:31}\\
        \dot x_2 &= r_2 x_2 + (a_{21} x_1 + a_{22}x_2 + a_{23}x_3) x_2 \label{eq:32}\\
        \dot x_3 &= r_3 x_3 + (a_{31} x_1 + a_{32}x_2 + a_{33}x_3) x_3 \label{eq:33}
    \end{align}
\end{subequations}
and the reduced model is:
\begin{subequations}
    \begin{align}
        \dot x_1 &= r_1 x_1 + (a_{11} x_1 + a_{12}x_2 ) x_1 \label{eq:r31}\\
        \dot x_2 &= r_2 x_2 + (a_{21} x_1 + a_{22}x_2 ) x_2 \label{eq:r32}.
    \end{align}
\end{subequations}
Likewise, the reduced model is referred to as $\mathcal{R}$, so that
\begin{equation}
    \frac{dx}{dt} = \mathcal{R}(x) = \text{diag}(x)( r +  A x),
\end{equation}
and $R = \{A, r\}$.

In this example, the growth rate vectors and interaction matrices are:
\begin{equation}
    \hat A = \begin{bmatrix} a_{11} & a_{12} & a_{13}\\a_{21} & a_{22} & a_{23}\\ a_{31} & a_{32} &
    a_{33}\end{bmatrix},\quad
    \hat r = \begin{bmatrix} r_1 \\r_2\\r_3\end{bmatrix},\quad
    A = \begin{bmatrix} a_{11} & a_{12} \\a_{21} & a_{22} \end{bmatrix},\quad
    r = \begin{bmatrix} r_1 \\r_2\end{bmatrix}.
\end{equation}

The species included in the reduced model are called the \emph{reduced
set} and sometimes also the \emph{remaining species}, that is, remaining after the reduction process.
Note that notation in the vector $x$ is overloaded---sometimes it refers to the detailed model,
sometimes to the reduced, and later to the enriched model. The meaning should be clear from context.

While this may seem a naive reduced model, there are a few reasons why this is reasonable here, at
least in this investigation of model discrepancy.  First, if a growth rate for one species or an
interaction rate between two is known, then this value might reasonably be used by modelers no
matter what set of species are present in the model. Second, the species and their derivatives in
the reduced model (and how their derivatives are subsequently modified by the discrepancy model) can
exactly correspond to those in the detailed model; i.e., the reduction does not pervert what the
variables in $x$ and $\dot x$  represent. Third, as a study in model discrepancy, this type of
reduction allows a clear objective: Represent the effects of the missing growth rate and interaction terms using
only information about the remaining species.

\subsection{Defining the scope}\label{ssec:scope}
As the objective of this paper is to understand model discrepancy in the context of reduced GLV
models, we must define the scope of this context. There are a few considerations to keep in mind.
First note that, as explained above, the reduced model considered here follows immediately from the
detailed model. Thus, when determining the scope of models under investigation, it suffices to
determine the detailed model(s). Then, given a detailed model, we investigate all possible reduced
models from $s=1$ to $s=S-1$.

Second, the GLV equations of course encompass an infinite number of specific models, or model
realizations, as $S$ can be any integer $\geq 2$, and the entries of $r$ and $A$ can be, in theory,
any real numbers. Moreover, any two GLV models, determined by a specific $A$ and $r$, may behave
very differently from one another.  At one extreme---the most specific---all model parameters are
fixed, yielding a single fixed pair of detailed and reduced models, and we could then investigate
the model discrepancy therein. At the other extreme---the most general---many models are supplied
via highly unconstrained realizations of the model parameters, and we could hope to thus discover highly
general results about the model discrepancy. In this paper, by aiming somewhere in between these
two extremes, we examine a moderately general random class of LV models. This class is determined by
specifying appropriate distributions for the entries of $r$ and $A$. 

Third, since this is an initial exploration into representing model discrepancy in the GLV context,
let us narrow the scope in order to examine well-behaved models.  Stability is most easily achieved
by using a symmetric interaction matrix $A$ with non-positive entries. This constraint says that all
interactions between and within species are competetive, not cooperative. The matrix $A$ can always
be stabilized by making its diagonal entries larger in magnitude than the sum of off-diagonal
entries in the same row (or column), ensuring diagonal dominance. So, here we consider models whose
interaction matrices are diagonally dominant and have non-positive entries. The distributional form
characterizing entries of $r$ and $A$ will be given in the following subsection. Then in
Section~\ref{sec:num}, specific values are given and analyzed through numerical examples.

\subsection{Creating the GLV detailed and reduced models}\label{sec:algs} This subsection summarizes
and refines the information above algorithmically.  Algorithm~\ref{alg:det} generates a realization
of a detailed model, and Algorithm~\ref{alg:red} provides the corresponding reduced model.  Recall,
to differentiate between the two, we use $\hat \,$ to denote a quantity of the detailed model.

\begin{algorithm}[H]
\caption{Generating a realization of the detailed model}
\label{alg:det}
    \begin{algorithmic}[1]
      \STATE {Initialize $S$}
        \STATE{ Sample $ B_{ij} \sim \log\mathcal{N}(0,\sigma^2_B), 1 \leq i < j \leq S$}
        \STATE{ Set $ B_{ji} = B_{ij}$}
        \STATE{ Sample $ C_{ii} \sim \log\mathcal{N}(0,\sigma^2_C) + \sum_{k \neq i} B_{ki}, 1 \leq i
        \leq S$}
        \STATE{ Set interaction matrix $\hat A = -( B +  C)$}
        \STATE{ Set growth rate vector $\hat r = \max\{C\} \bm{1}_S$}
        \RETURN $D = \{\hat A, \hat r\}$
\end{algorithmic}
\end{algorithm}

\begin{algorithm}[H]
    \caption{Subsampling the reduced model}
    \label{alg:red}
    \begin{algorithmic}[1]
        \STATE {Initialize $s<S$, $D$}
        \STATE {Fix $s < S$}
        \STATE {Set $A$ as submatrix: $A = \hat A_{1:s,1:s}$}
        \STATE {Set $r$ as subvector: $r =  \hat r_{1:s}$}
        \RETURN $R = \{A, r\}$
\end{algorithmic}
\end{algorithm}
Note that stability of the reduced model follows directly from stability of the detailed model.


\section{Model conversion}\label{sec:mco}
A system of $S$ coupled ordinary differential equations can sometimes be converted (decoupled) to a system of
$s$ differential equations, where $s < S$, without loss of information. Possible structures of the
resultant set, comprising $s$ equations, motivates the functional form of the proposed model
discrepancy here. This section will briefly review two methods of model conversion, and what the
application of each method yields in the GLV context. For more information about these types of
model conversion, or exact model reduction, see \cite{morrison2019exact, hernandez2019algebraic}.

\subsection{Algebraic method}
In \cite{harrington2017reduction}, Harrington and van Gorder present a method to algebraically
convert systems of coupled differential equations from one form to another. As an example from that
paper, consider the Lorenz system of three ODEs:
\begin{subequations}
\begin{align}
    \dot{x} &= a (y - x)\label{eq:lx}\\
    \dot{y} &= x(b - z) - y\label{eq:ly}\\
    \dot{z} &= xy - cz.\label{eq:lz}
\end{align}
\end{subequations}
Through algebraic substitutions, this can be converted to a single third-order nonlinear differential equation in
only the variable $x$ and its derivatives. After substituting expressions for $z$ and $y$ in terms of
$x$ and its derivatives, we have:
\begin{equation}
    \left(\frac{d}{dt} + c\right)\left(b - \frac{1}{ax}(\ddot x + (1 + a)\dot x + ax)\right) -
    x\left(\frac{\dot{x}}{a} + x\right) = 0.
\end{equation}
In this example, variables $y$ and $z$ have been ``exchanged'' for derivatives of $x$.

In the GLV setting, we can perform a similar exchange. For example, consider the following system
for $x$ and $y$:
\begin{subequations}
\begin{align}
    \dot x &=  r_1  x + ( a_{11} x +  a_{12} y) x \label{eq:dx}\\
    \dot y &=  r_2  y + ( a_{21} x +  a_{22} y) y \label{eq:dy}.
\end{align}
\end{subequations}
This is in fact equivalent to the following single differential equation for $x$:
\begin{align} \left(\frac{d}{dt}\right)\left( \frac{1}{a_{12}} \left( \frac{1}{x} \left( \dot x - r_1 x\right) - a_{11}x
    \right)\right)\label{eq:alg1} &=\\  \frac{r_2}{a_{12}} \left( \frac{1}{x} \left( \dot x - r_1 x\right) - a_{11}x
        \right) &+ \left(a_{21}x + \frac{a_{22}}{a_{12}} \left( \frac{1}{x} \left( \dot x - r_1 x\right) - a_{11}x
        \right)\right)\nonumber.
\end{align}
Equation~\ref{eq:alg1} can also be written more compactly as
\begin{equation}
    \dot z =  r_2  z + ( a_{21} x +  a_{22} z) z,\label{eq:alg2} \end{equation}
where 
\begin{equation}z = \frac{1}{a_{12}} \left( \frac{1}{x} \left( \dot x - b_1 x\right) - a_{11}x
\right).\end{equation}
While this single differential equation is quite messy, it is
now written entirely in terms of $x$ and its derivatives.

\subsection{Memory method}
Similarly, the Mori-Zwanzig approach to model reduction makes an exchange, but here variables may be
exchanged for time history, or memory, of the remaining variables. Again, a simple example starts with a
two-variable system:
    \begin{align} \frac{dx}{dt} &= f(x,y) + \alpha (x,y) \frac{dU}{dt}\\
        \frac{dy}{dt} &= g(x,y) + \beta (x,y) \frac{dV}{dt},
    \end{align}
    where $U, V$ are noise processes.
This system of two equations can be converted to one by introducing the memory kernel $K$:
    \begin{equation}
        \frac{dx(t)}{dt} = \bar f(x(t)) + \int_0^t K\left(x(t-s),s\right)ds +
        n\left(x(0),y(0),t\right)
    \end{equation}
    where  $\bar f(x(t))$ represents a Markovian term that depends only on the current state of $x$,
    the integral $\int_0^t K\left(x(t-s),s\right)ds$ depends on the entire history of $x$ between
    $0$ and $t$, and the final term $n\left(x(0),y(0),t\right)$ satisfies an auxiliary equation.
    Further details about this process are beyond the scope of this paper, but
    \cite{givon2004extracting} provides an excellent review.

Let us return to the 2-variable system introduced in lines~\cref{eq:dx,eq:dy}. The analogue of a
Mori-Zwanzig type process in the GLV setting yields:
\begin{equation}
    \dot x =  r_1  x + ( a_{11} x +  a_{12} \chi) x \\
\end{equation}
where
\begin{equation}
    \chi =  \int_0^t b_2  z + ( a_{21} x +  a_{22} z) z
\end{equation}
and $z$ is defined as in the previous subsection.  Note that, like the algebraic reduction, we now
have a single differential equation in terms of $x$. In this case, the variable $y$ is ``exchanged''
for the memory of $x$.


In each GLV example above, we have
\begin{equation}
    \dot x = \mathcal{F}(x, \dot x, \ddot x,  K(x)),
\end{equation}
that is, the derivative of $x$ can be written in terms of itself and extra information about
it---such as higher derivatives or its memory.  This motivates an approximation of $\mathcal{F}$
with the available reduced model $\mathcal{R}$ and a discrepancy model $\Delta$ that is a function of
either the derivatives or memories of the remaining variables. That is, we seek a model for
the reduced set of variables as:
\begin{equation}
    \frac{dx}{dt} \approx \mathcal{R}(x) + \Delta(x, \dot x, \ddot x,  K(x)).
\end{equation}
The particular form of $\Delta$ will be specified in the next section.

\section{Embedded Discrepancy Operator}\label{sec:edo}
Motivated by the examples in the previous section, we are now ready to explore possible formulations
for an embedded discrepancy operator.

\subsection{Linear EDO, $S=2, s=1$ example}
First let us examine the $S=2, s=1$ case again. The detailed model is:
\begin{subequations}
\begin{align}
    \dot x_1 &= r_1 x_1 + (a_{11} x_1 + a_{12} x_2) x_1\\
    \dot x_2 &= r_2 x_2 + (a_{21} x_1 + a_{22} x_2) x_2.
\end{align}
\end{subequations}
As described in \S~\ref{ssec:dnr}, the reduced model is given by retaining only the submatrix and
subvector associated with the remaining variable $x_1$:
\begin{equation}    \dot x_1 = b_1x_1 + a_{11}x_1^2. \end{equation}
In this case, the exact discrepancy is $a_{12} x_2 x_1$. We aim to approximate the effect
of this term, using an expression in terms of $x_1$ and extra information about $x_1$:
    \begin{equation*}
        \dot x_1 = r_1 x_1 + a_{11}x_1^2 + \Delta(x_1,\dot x_1, \ddot x_1,
        \mathcal{K},\dots).
    \end{equation*}
As a very simple example, let us try a linear polynomial in $(x_1, \dot x_1)$:
    \begin{equation}
        \Delta(x_1,\dot x_1) = \delta_{10} x_1 + \delta_{11}\dot x_1.
    \end{equation}
    The subscripts on each $\delta_{ij}$ are chosen so that $i$ indicates that this coefficient
    appears in the RHS of the variable $x_i$, and $j$ indicates that this coefficient is multiplying
    the $j$th derivative of $x_i$. This notation will be more useful in the next subsection, when
    we generalize to arbitrary $S$ and $s$.

A major advantage of an embedded operator, as opposed to a response discrepancy model, is that the
operator can (and should) be constrained by any available information about the physical system. In
this simple example, we do have some information about the system that implies constraints on the
introduced discrepancy parameters $\delta_{10}, \delta_{11}$.  First, we make the modeling ansatz
that these discrepancy parameters should not depend explicitly on time.  A result of this ansatz is
then that the parameters  be constrained independently. Next, the species concentrations are
non-negative and reach a stable equilibrium. We also assume knowledge of the fact that all
interspecific interactions are competetive. In particular, note that $a_{12}x_2x_1 < 0$ because
$a_{12} < 0$ and $x_1, x_2 \geq 0$.  Thus, we enforce that $\Delta(x_1, \dot x_1) \leq 0$. Thus,
specific information about the high-fidelity physical system implies constraints as described below.
 \begin{itemize}
    \item The inequality $\Delta \leq 0$ must be true as $x_1$ and $\dot x_1 \rightarrow 0$. 
    \item We know $x_1 \geq 0$ which imples $\delta_0 \leq 0$.
    \item The constraint on $\delta_{11}$ is slightly less clear since the sign of $\dot x_1$ could be
        positive or negative. Thus, we could set $\delta_{11} = \ti{\delta}_{11} \text{sgn}(\dot
        x_1)$, where $\ti{\delta}_1 \leq 0$.  Equivalently, we can write the discrepancy as
        \begin{equation} \Delta(x_1, \dot x_1) = \delta_{10} x_1 + \delta_{11} \left|\dot
        x_1\right|.\end{equation} Then set $\delta_{11} \leq 0$ and the constraint is satisfied.
\end{itemize}
Because of this final constraint, the discrepancy operator is no longer linear in $\dot x$, but
rather in $|\dot x |$. We still refer to such a formulation as linear.

Finally, the introduced discrepancy parameters $\delta_{10}, \delta_{11}$ are calibrated, using
observations of species concentrations generated by the detailed model. Indeed, the strength of the
embedded operator approach stems from  two properties: 1) the ability to constrain the formulation by
available physical information, and 2) the ability to leverage information from the detailed system
by calibrating the model discrepancy parameters. Moreover, we calibrate over a range of
initial conditions, denoted $\phi_i, i = 1,\dots, n_{\phi_c}$. Note that we also validate over
a range of initial conditions $\phi_i, i = n_{\phi_c}+1,\dots,n_{\phi_v}$, where $n_{\phi_c} +
n_{\phi_v} = n_\phi$. Each $\phi_i$ specifies the species initial
concentrations:
\begin{equation}
    \phi_i = (x_1(0), x_2(0),\dots,x_s(0)), \quad i = 1,\dots, n_\phi.
\end{equation}
By calibrating with observations from all $n_{\phi_c}$ scenarios, the goal is to build a more robust discrepancy model that is
valid over several scenarios instead of only calibrated to a very specific dataset. This property of
the model discrepancy construction further allows for the possibility, at least, that such an enriched
model could be used in extrapolative conditions, such as a prediction in time, or in scenarios given
by different initial conditions.

Note that the actual observations used to calibrate the parameters are specified in
Section~\ref{sec:calval}, along with the particulars of the calibration itself. We have tried to separate
what is essential to the formation of the discrepancy operator from the calibration details, which
could reasonably change based on the example at hand.

At this point, it is worthwhile to point out a subtle difference between this and previous work by
the author and co-authors. In \cite{morrison2018representing}, there is a single fixed detailed and reduced model
(of hydrogen combustion), and the embedded operator is referred to as an ``inadequacy operator.''
In that work, the embedded inadequacy operator is our most honest attempt to account for the uncertain error of the reduced
model as compared to the detailed model. There, the exact form of the inadequacy is unknown
and so the operator is truly stochastic---the entries of the operator are random variables of
unknown distributions. During the calibration process, it is the hyperparameters of these
distributions that must be calibrated using a hierarchical Bayesian scheme.  This is an added layer
to the inadequacy operator that is not explored in the current work. One could still treat the
discrepancy here as deserving of a stochastic operator, and calibrate hyperparameters, but the focus
of this paper is slightly different: Here we want to understand the ability of these simple linear
embedded operators to account for the discrepancy between reduced and detailed models, in a very
general and deterministic sense. For this reason, we use the term ``discrepancy operator'' and calibrate
its parameters directly. If we were to apply these techniques to a specific
reduced model again, likely we would need return to the stochastic framework.

\subsection{Linear EDO}\label{ssec:edo-prop}
Let us now generalize the above example to arbitrary $S, s$. This is the main propsed
form of model discrepancy in this work, and the numerical examples in Section~\ref{sec:num} are based
on it. However, other possible formulations will be presented in the next
subsection~\ref{ssec:edo-forms}.

Recall that the detailed model is:
\begin{equation}
    \frac{dx}{dt} = \mathcal{D}(x) = \text{diag}(x)(\hat r + \hat A x).
\end{equation}
and the reduced model:
\begin{equation}
    \frac{dx}{dt} = \mathcal{R}(x) = \text{diag}(x)( r +  A x).
\end{equation}

Now, to build the discrepancy operator, 
let $\delta_0 = (\delta_{10}, \delta_{20},\dots, \delta_{s0})^T$ be the vector of coefficients which
multiply elements of $x$ (the zero-th derivative of $x$), and
$\delta_1 = (\delta_{11}, \delta_{21},\dots, \delta_{s1})^T$ be the vector of coefficients which
multiply elements of $|\dot x|$.  Then the enriched model is
\begin{align}
    \frac{dx}{dt} &= \mathcal{E}(x, \dot x)\\
    &= \mathcal{R}(x) +  \text{diag}(x)\delta_0 + \text{diag}(|\dot x|) \delta_1 \\
    &= \mathcal{R}(x) + \Delta(x, |\dot x|).
\end{align}
The constraints follow immediately from the $S=2$ example: All $\delta_{ij} \leq 0, i=1,\dots,s;
j=0,1$.

\subsection{Possible formulations}\label{ssec:edo-forms}
There are a number of related possible formulations of the model discrepancy, and here we present a
few in terms of a single $x_i$; that is, $\Delta_i$ denotes the $i$th component of the discrepancy
operator $\Delta$. Some options are the following:
\begin{enumerate}
    \item An affine expression up to the $N$th derivative:
     \begin{equation}
         \Delta_i =  \lambda_i + \sum_{j=0}^N
         \delta_{ij}\frac{d^j}{dt^j}(x_i)
     \end{equation}
    where $\Delta_i$ refers to the $i$th component of the discrepancy operator.
 \item A quadratic expression up to $N$th derivative.  Let \[\bm{q} =
        \left\{\frac{d^0}{dt^0}(x_1),\dots,\frac{d^N}{dt}(x_1),\dots,\frac{d^0}{dt^0}(x_s),\dots,\frac{d^N}{dt^N}(x_s)\right\}.\] Then
     \begin{equation}
         \Delta_i =  \lambda_i + \sum_{i,j}^{s(N+1)}
        \delta_{ij}(q_i q_j).
     \end{equation}
 \item A memory expression, such as:
     \begin{equation}
         \Delta_i(t) = \lambda_i + \beta_i \int_{s=0}^{t} x_i(s)ds
     \end{equation}
     for some $\beta_i \in \mathbb{R}$.
        \footnote{There is an interesting similarity between this formulation and the linear
        approximations to closure models in the Kuramoto-Sivashinshy equation developed by Lu et al.\@ in \cite{lu2017data}.}
\end{enumerate}
Each of the above formulations includes a constant off-set term, $\lambda_i$. Whether or not such a
constant term would be advantageous when all the missing dynamics terms are state-dependent is not
immediately clear.

Of course one could also propose some combination of the above formulations as an embedded
discrepancy operator. Investigating the numerical advantages and limitations of many such
discrepancy operators is beyond the scope of the current paper, but will be addressed in future
work. For now, numerical results are presented in Section~\ref{sec:num} about the proposed linear
embedded discrepancy operator (as described in Subsection~\ref{ssec:edo-prop}).

\section{Calibration and validation}\label{sec:calval}
This section contains all relevant details about the calibration and validation processes. First,
for both of these, it is necessary to know what observations are available.

\subsection{The observations}\label{ssec:obs}
The data sets used to calibrate and validate the discrepancy model includes observations from the
detailed model trajectories of the $s$ species included in the reduced model. From each trajectory,
$T$ observations are taken, and there is a new trajectory for each initial condition $\phi$, so that
the observations can be summarized as
\begin{equation}
    \mathcal{O} = \{ y_{ijk} \}, i= 1,\dots,s; j= 1,\dots,T; k = 1,\dots,n_\phi
\end{equation}
where $y_{ijk}$ is the observation of $x_i(t_j)$ given the initial condition $\phi_k$. 
This observed value $y^*$ is given by the true value $y^t$ with additive measurement error $\epsilon$:
\begin{equation}
    y^* = y^t + \epsilon, 
\end{equation}
where the distribution of measurement error is normal: $p_\epsilon =  \mathcal{N}(0,
\sigma^2_{\epsilon})$.

Finally, this set of observations is partitioned into two sets, one for calibration and the other
for validation. Let us partition as so:
\begin{align}
    \text{Calibration data:}\quad &\mathcal{O}_c = \{ y_{ijk} \}, i= 1,\dots,s; j= 1,\dots,T; k =
    1,\dots,n_{\phi_c}\\
    \text{Validation data:}\quad &\mathcal{O}_v = \{ y_{ijk} \}, i= 1,\dots,s; j= 1,\dots,T; k =
    n_{\phi_c}+1,\dots,n_\phi.
\end{align}
That is, $n_{\phi_c}$ initial conditions are used for calibration, and the remaining $n_{\phi_v}$
are designated for validation, where $n_{\phi_c} + n_{\phi_v} = n_\phi$.

\subsection{Calibration details}\label{ssec:cal}
The calibration is done using a Bayesian approach, and the details of the calibration problem are
as follows.
\begin{itemize}
    \item \textbf{Prior:} We set a uniform prior distributions on the discrepancy parameters
        $\theta$\footnote{One might expect a negative lognormal distribution for these priors, and
        this was in fact the first choice. However, the uniform priors performed much better during
        the sampling process, and all of the parameter chains in MCMC were well-contained by the
        uniform bounds. Why the lognormal priors led to poor mixing will be investigated further
        in future work.}:
\begin{align}
    p(\theta) &= \prod_{\stackrel{i=1,\dots s}{j=0,1}} p(\delta_{ij}),
\end{align}
where
\begin{align}
    p(\delta_{ij}) &= \mathcal{U}(-100,0)\quad i=1,\dots,s; j= 0,1.
\end{align}
    \item \textbf{Likelihood:} The likelihood is determined by the measurement error:
        \begin{equation}
            p(\mathcal{O}_c | \theta ) = \prod_{l = 1,\dots, |\mathcal{O}_c|} p_\epsilon( y_l -
            y_{l,\mathcal{E}})
        \end{equation}
        where the observations have been reindexed from 1 to $|\mathcal{O}_c|$ (to avoid triple
        subscripts here) and $y_{l,\mathcal{E}}$ is the corresponding model output from the enriched
        model $\mathcal{E}$.
    \item \textbf{Posterior:} Given the prior and likelihood distributions above, the posterior
        distribution follows as:
        \begin{equation}
            p(\theta | \mathcal{O}_c ) \propto p(\mathcal{O}_c | \theta )p(\theta).
        \end{equation}
\end{itemize}
Specifically the calibration is performed according to the DRAM method, introduced in
\cite{haario2006dram} and implemented in the statistical library \textsc{QUESO}
\cite{prudencio2012parallel}.

\subsection{Validation metric}\label{ssec:val}
Next we must define an appropriate quantitative validation metric.

First, we quantify the agreement between the enriched model output and the corresponding
observation: We compute how probable the observation is as a realization of the model output. The
probability of observing some $y^*$, given the data $\mathcal{O}_c$, is
\begin{equation}
    p(y^*|\mathcal{O}_c) = \int_{y^t} p_\epsilon(y^t - y^*) \left( \int_\theta p(y^t | \theta)
    p(\theta|\mathcal{O}_c) d\theta \right) d y^t. \label{eq:p1}
\end{equation}

Now we can compare this probability to the rest of possible model outputs. In particular, we are
interested in how much of the distribution corresponds to model outputs less likely than the one
above in~(\ref{eq:p1}). This amount is exactly given by the so-called $\gamma-$value
\cite{oliver2015validating}:
\begin{equation}
    \gamma_{y^*} = \int_{y \in \mathcal{S}} p(y|\mathcal{O}_c) dy
\end{equation}
where $\mathcal{S} = \{ y: p(y|\mathcal{O}_c) \leq p(y^*|\mathcal{O}_c) \}$.

\begin{figure}[htb]
  \centering
  \includegraphics[width=.5\textwidth]{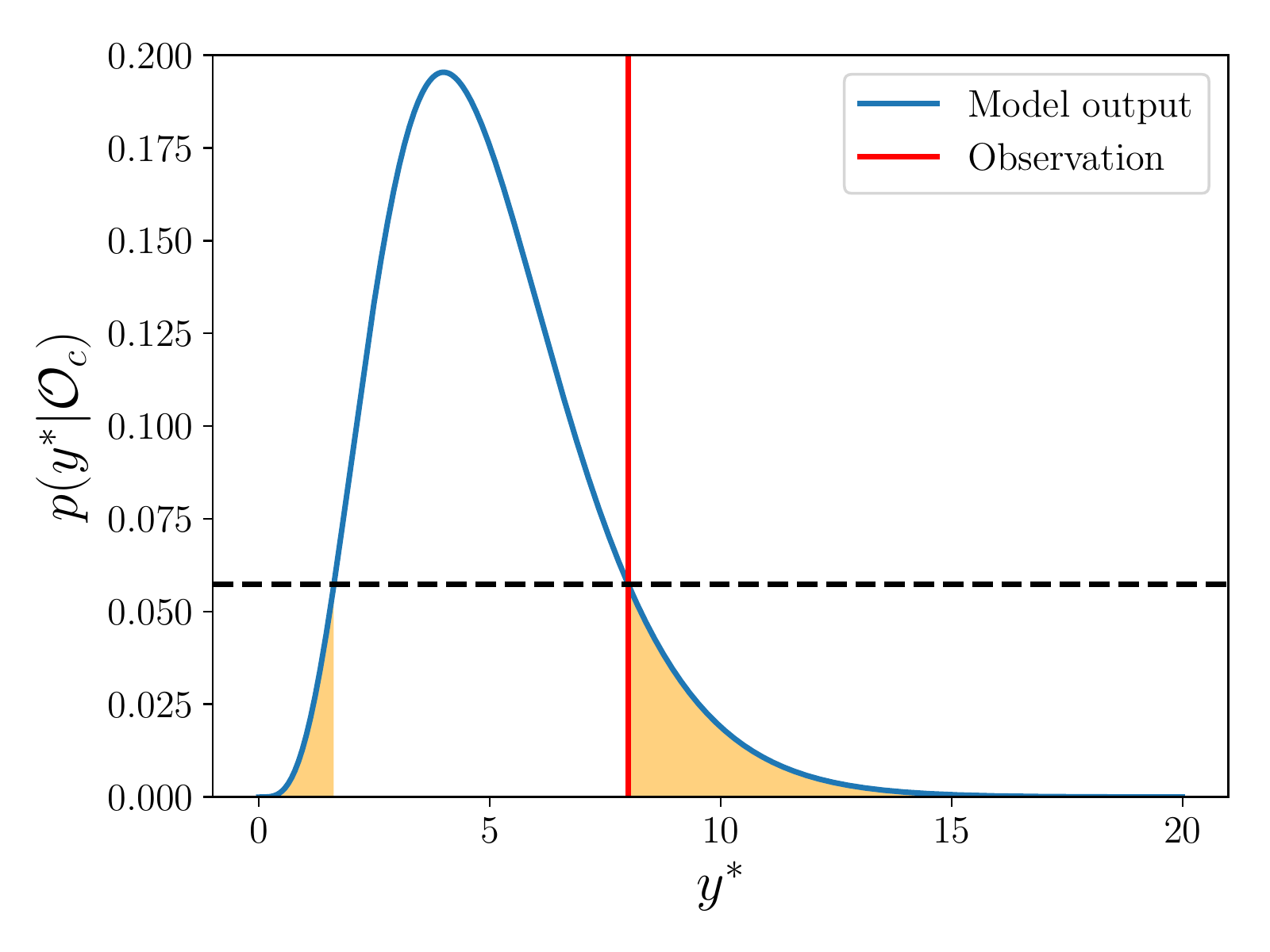}
    \caption{The $\gamma$-value corresponds to the shaded area.\label{fig:gamma}}
 \end{figure}

An example of the area corresponding to this integral is given in Figure~\ref{fig:gamma}.
For a more thorough discussion about $\gamma-$values, see \cite{oliver2015validating}, and for
another example of this used in practice as a validation metric, see
\cite{morrison2018representing}. Note that a very low $\gamma$-value implies that the observation is
less probably an outcome of this model than most possible outcomes. In contrast, values that are not
very low demonstrate consistency between the model and observation. In this work, we compute the
fraction of $\gamma$-values below a given threshold $\tau$.

\section{Numerical results}\label{sec:num}
Let us now investigate the numerical performance of the proposed linear EDO presented in
Section~\ref{ssec:edo-prop}.
\begin{figure}[htb]
  \centering
    \begin{subfigure}{.24\textwidth}
  \centering
  \includegraphics[width=\textwidth]{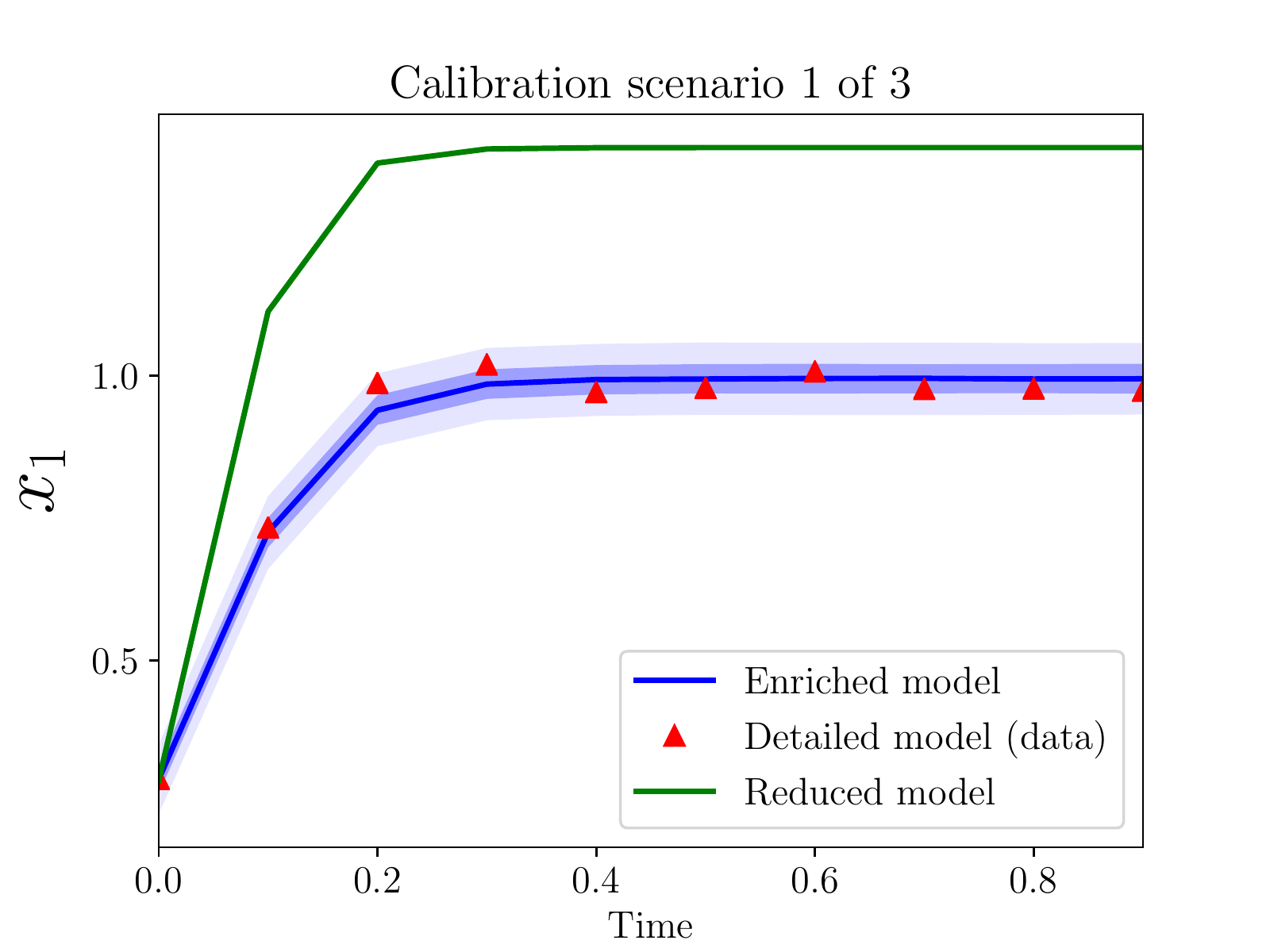}
    \end{subfigure}
    \begin{subfigure}{.24\textwidth}
  \centering
  \includegraphics[width=\textwidth]{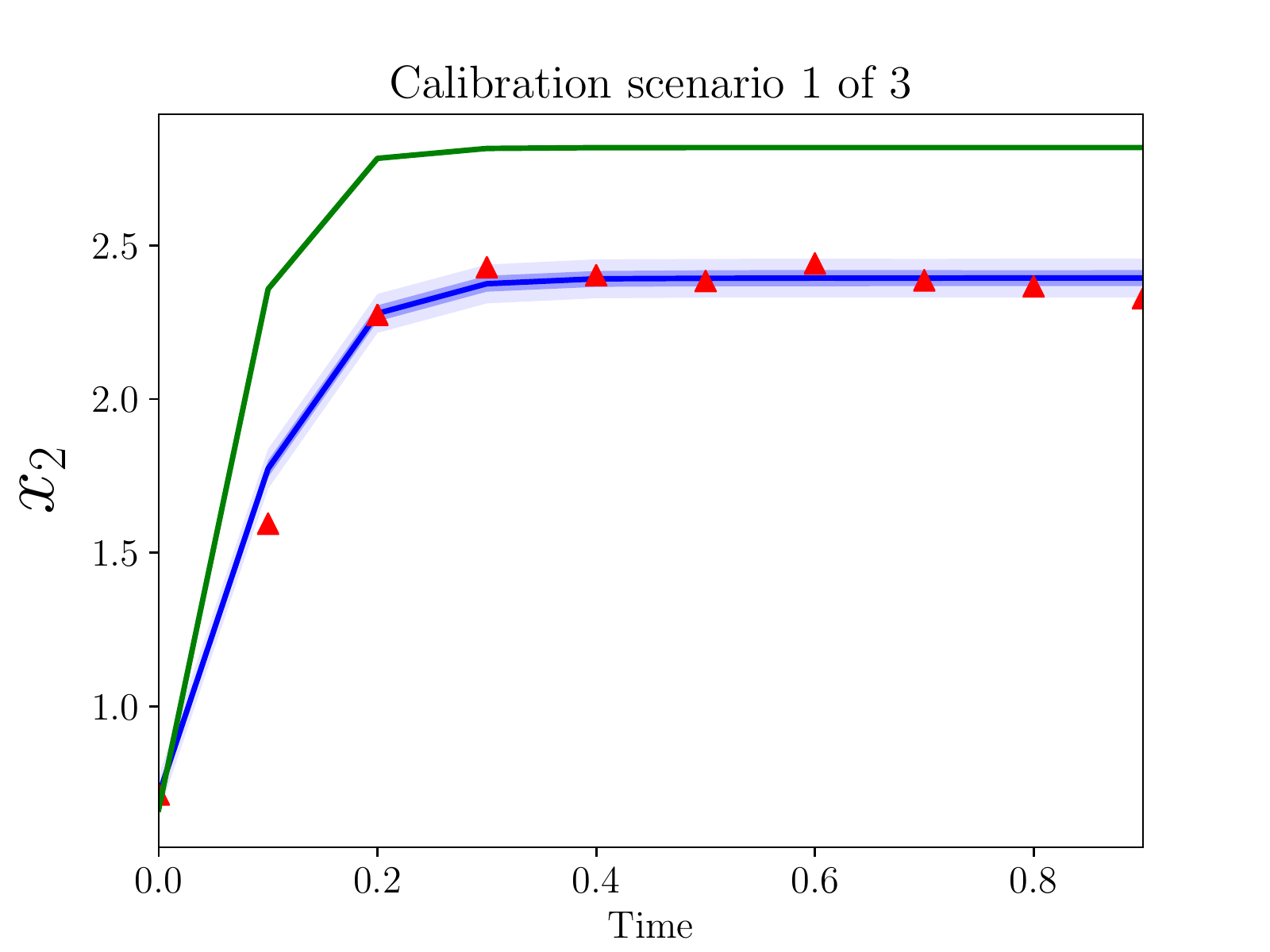}
    \end{subfigure}
    \begin{subfigure}{.24\textwidth}
  \centering
  \includegraphics[width=\textwidth]{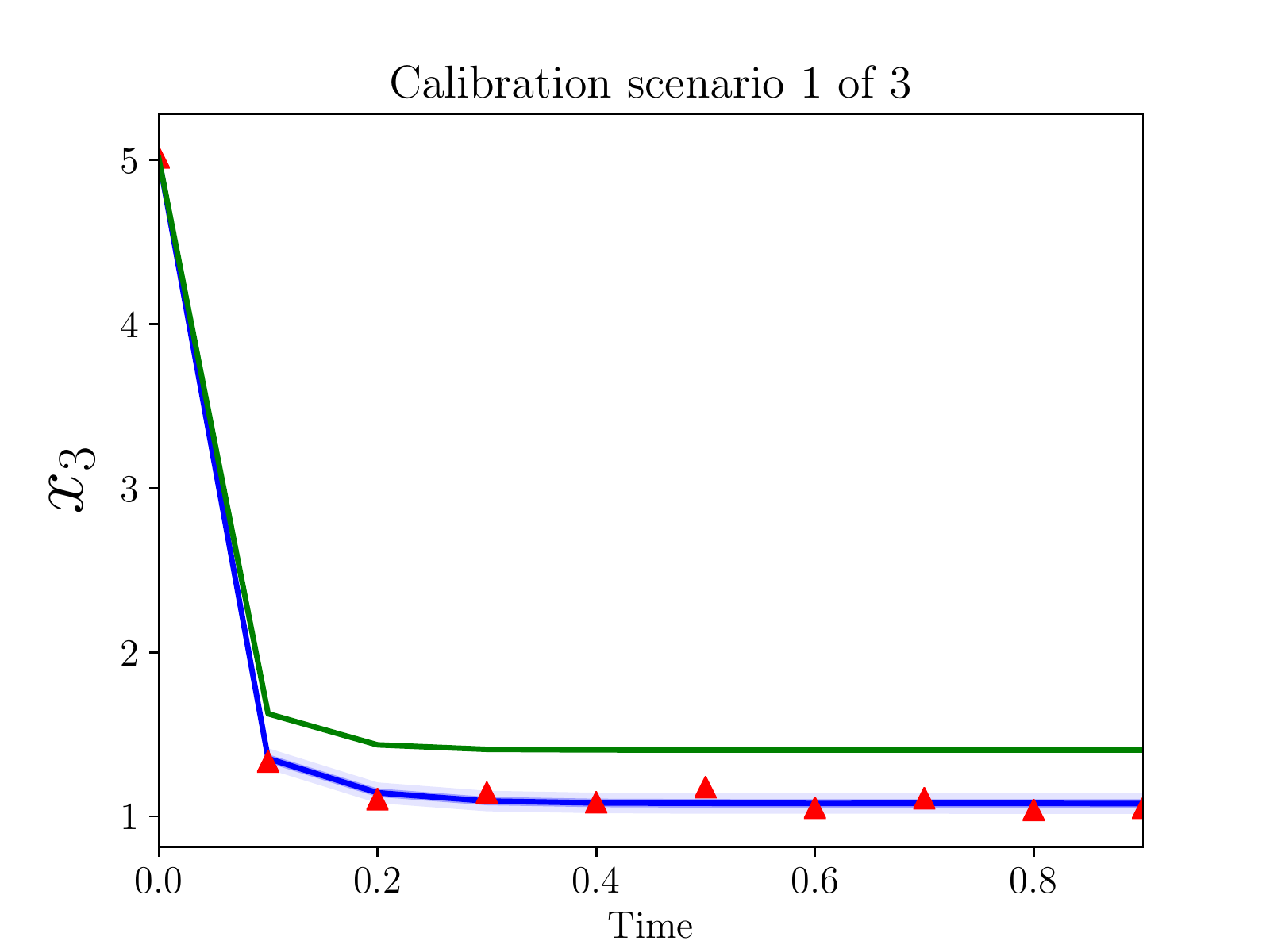}
    \end{subfigure}
    \begin{subfigure}{.24\textwidth}
  \centering
  \includegraphics[width=\textwidth]{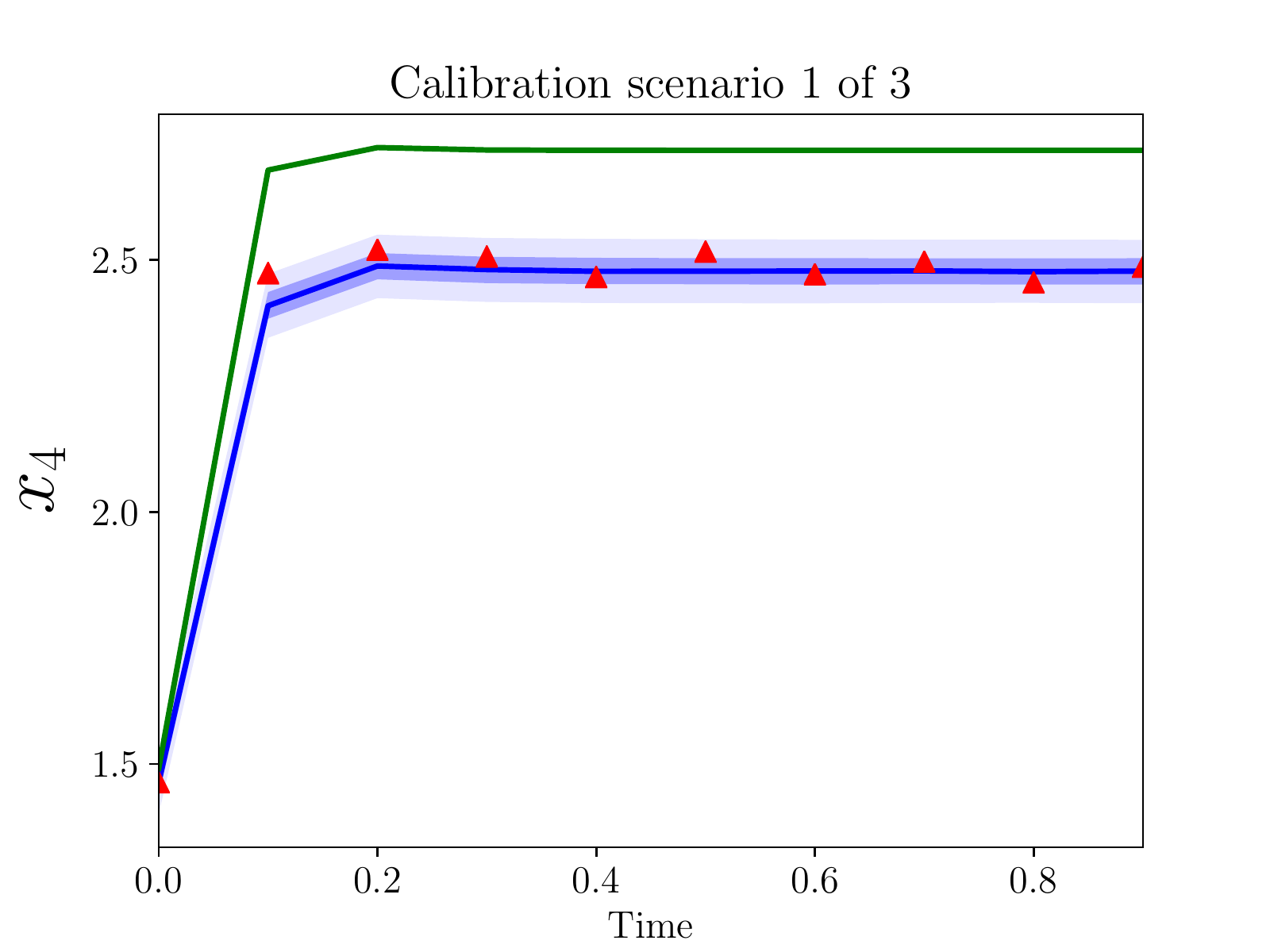}
    \end{subfigure}
    \begin{subfigure}{.24\textwidth}
  \centering
  \includegraphics[width=\textwidth]{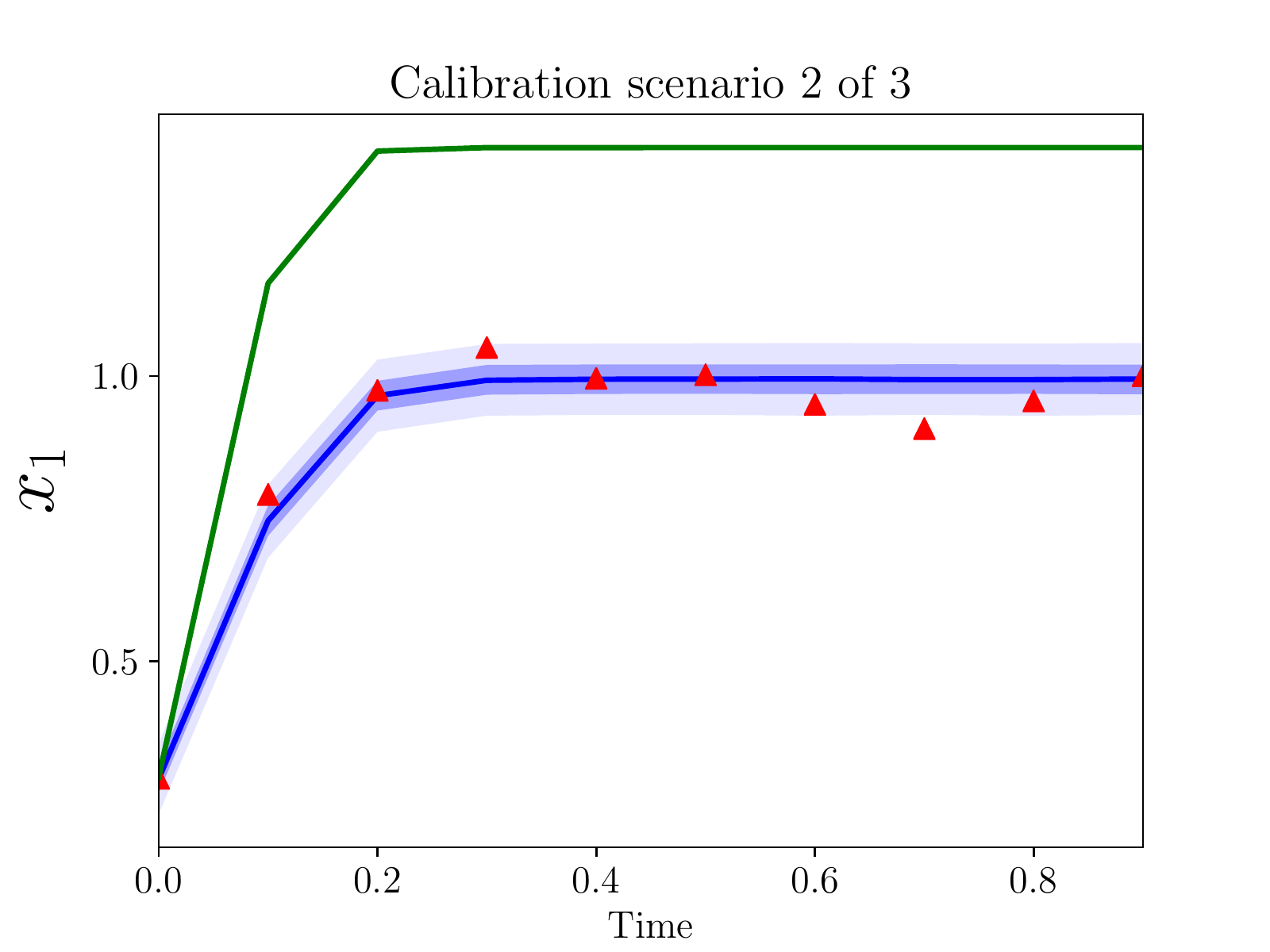}
    \end{subfigure}
    \begin{subfigure}{.24\textwidth}
  \centering
  \includegraphics[width=\textwidth]{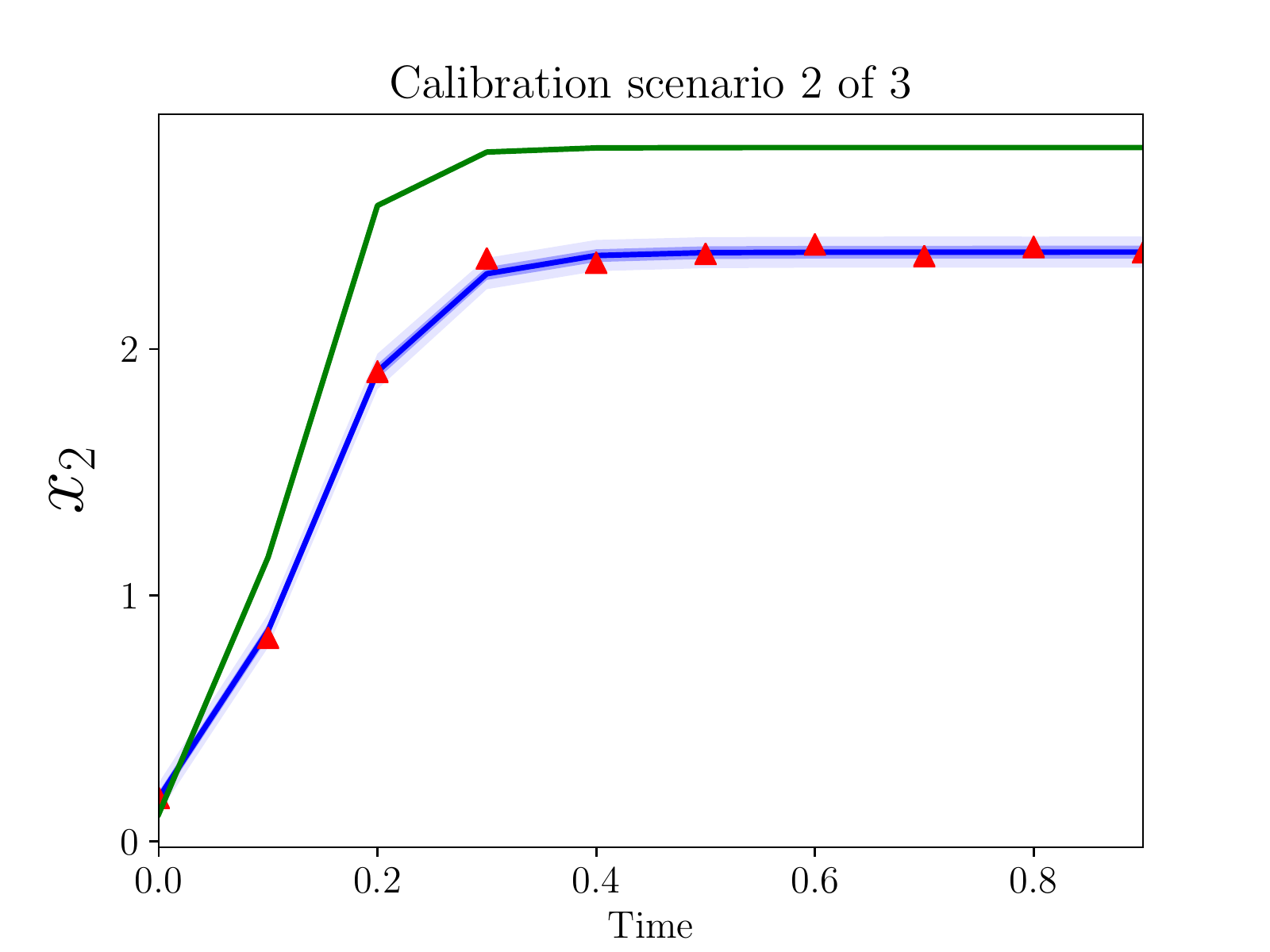}
    \end{subfigure}
    \begin{subfigure}{.24\textwidth}
  \centering
  \includegraphics[width=\textwidth]{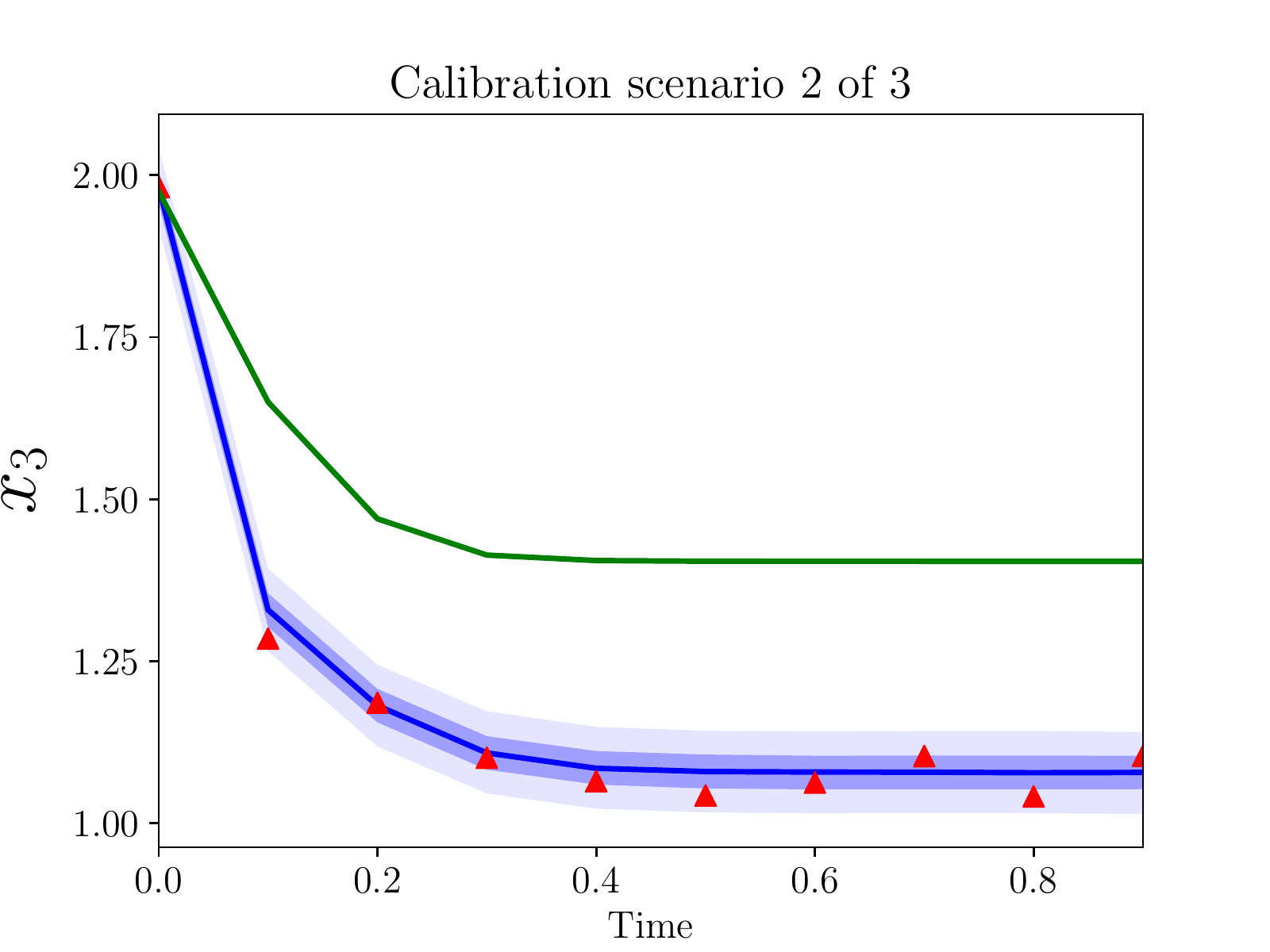}
    \end{subfigure}
    \begin{subfigure}{.24\textwidth}
  \centering
  \includegraphics[width=\textwidth]{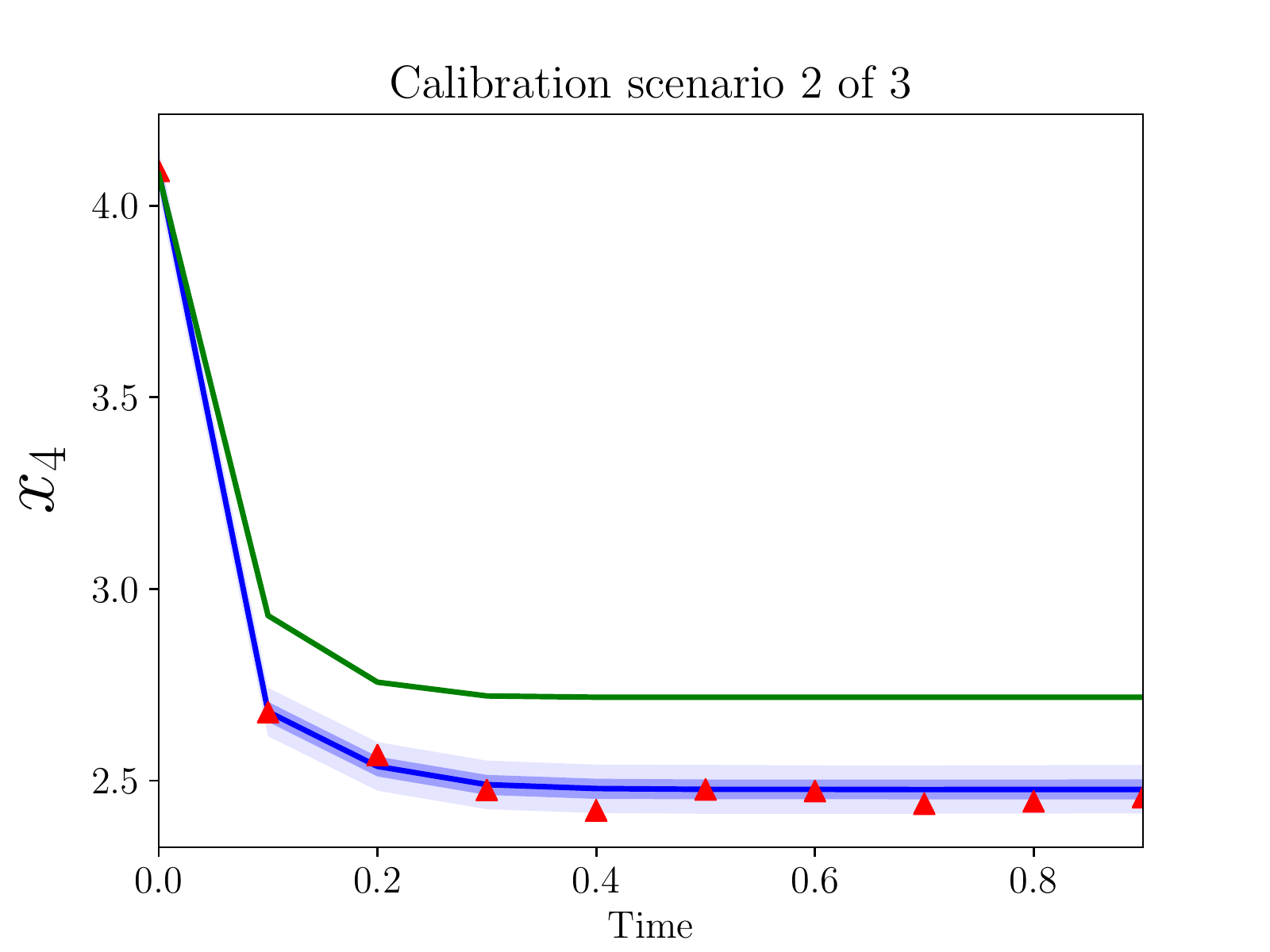}
    \end{subfigure}
    \begin{subfigure}{.24\textwidth}
  \centering
  \includegraphics[width=\textwidth]{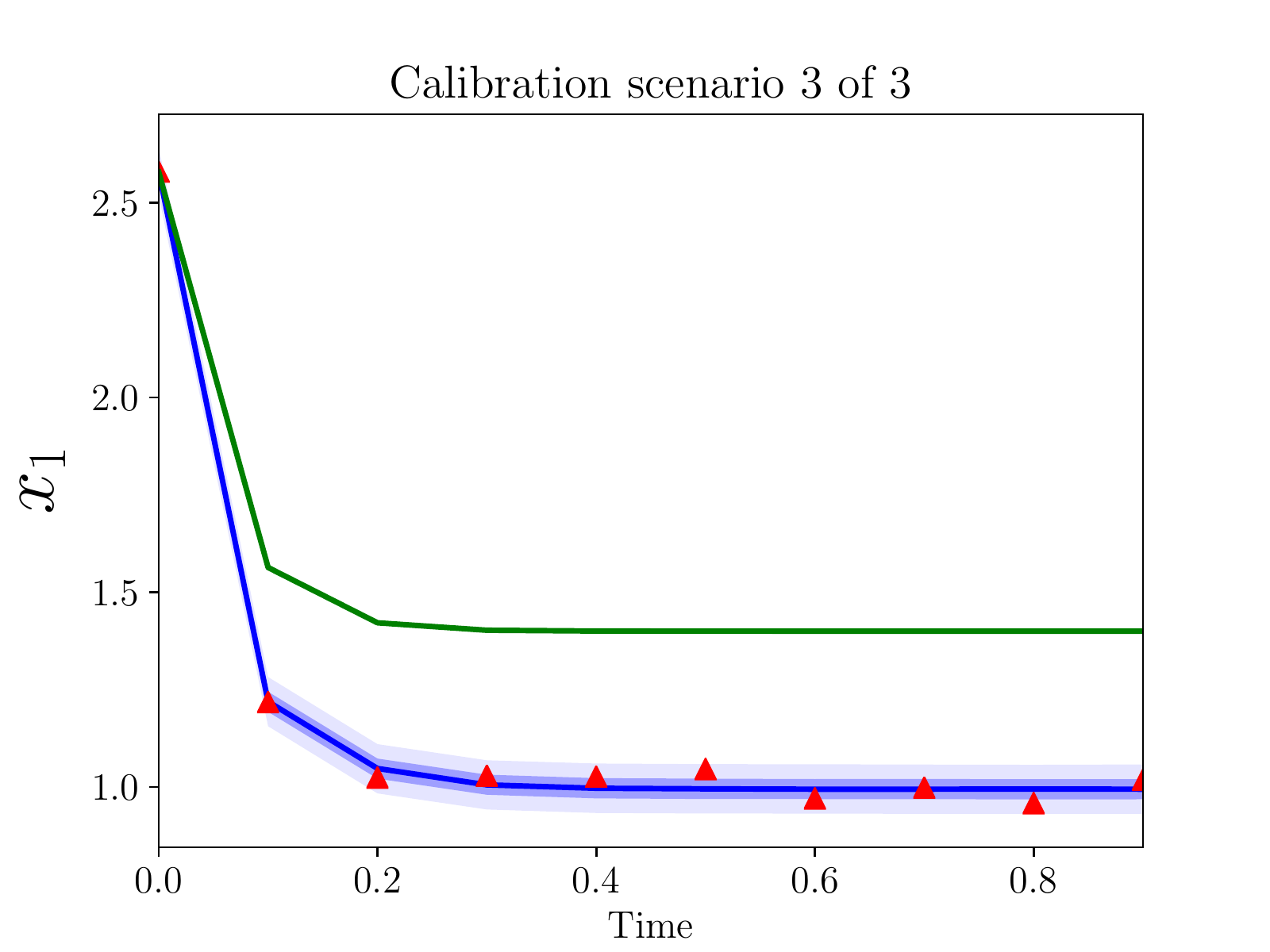}
    \end{subfigure}
    \begin{subfigure}{.24\textwidth}
  \centering
  \includegraphics[width=\textwidth]{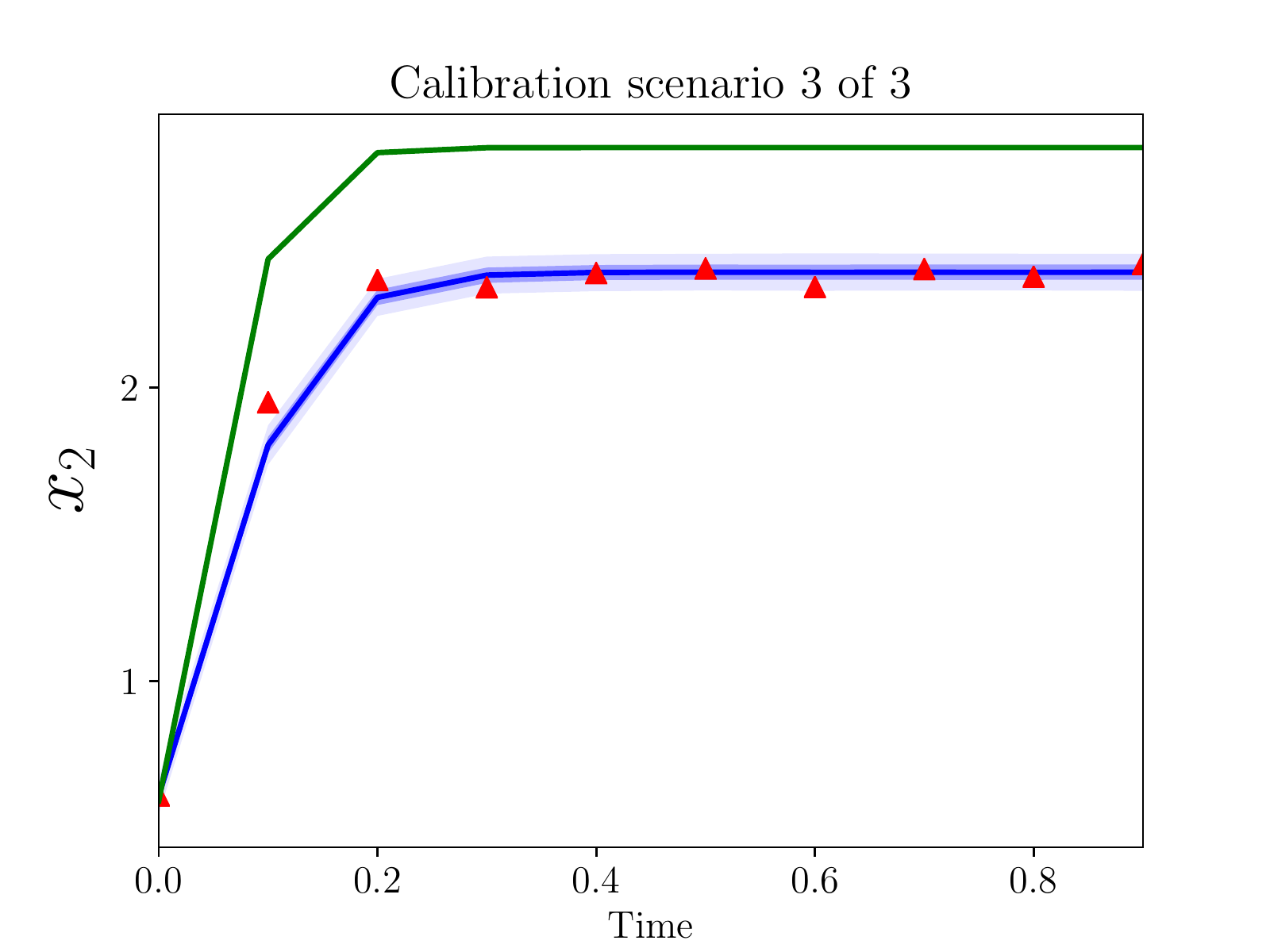}
    \end{subfigure}
    \begin{subfigure}{.24\textwidth}
  \centering
  \includegraphics[width=\textwidth]{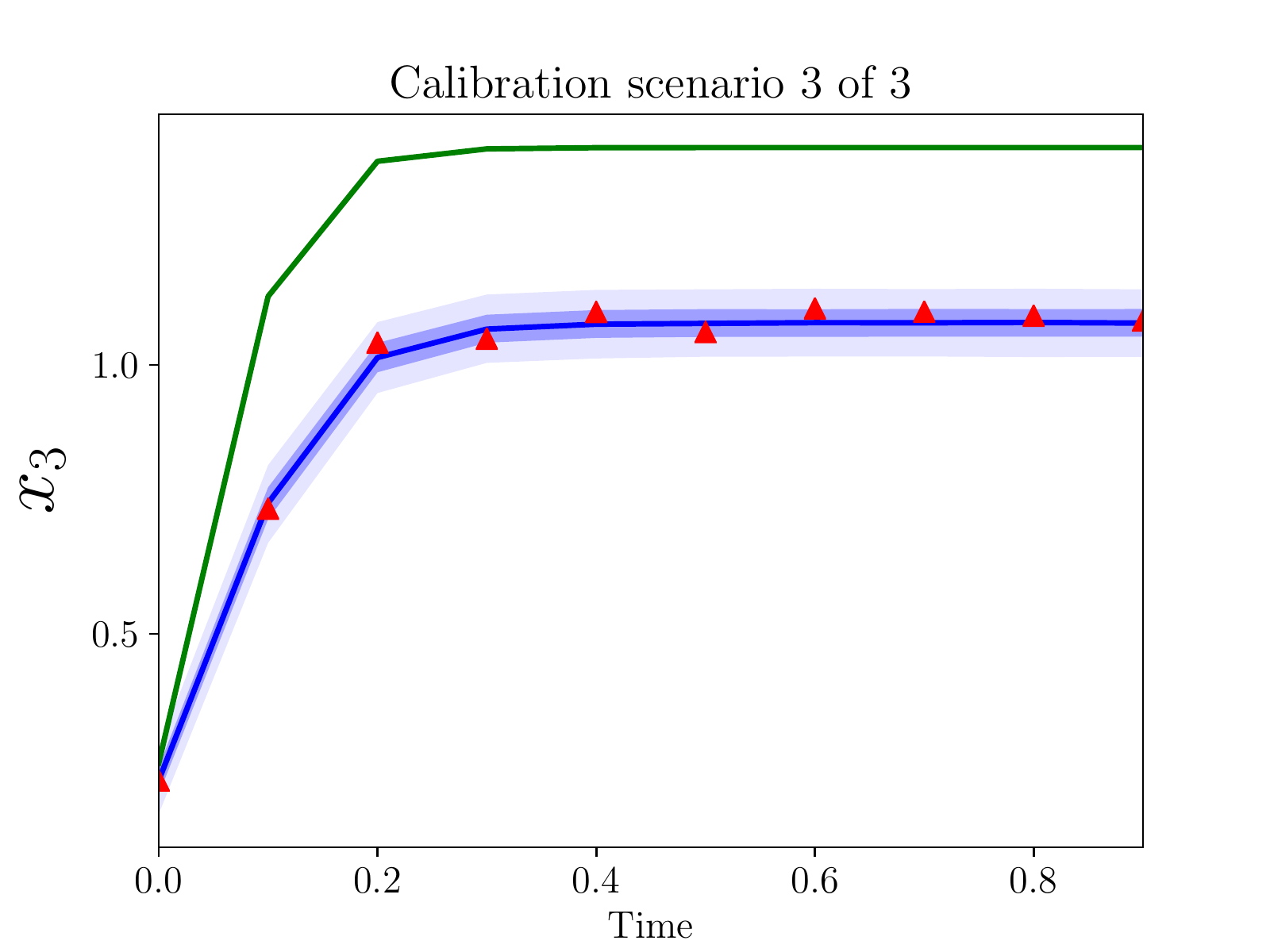}
    \end{subfigure}
    \begin{subfigure}{.24\textwidth}
  \centering
  \includegraphics[width=\textwidth]{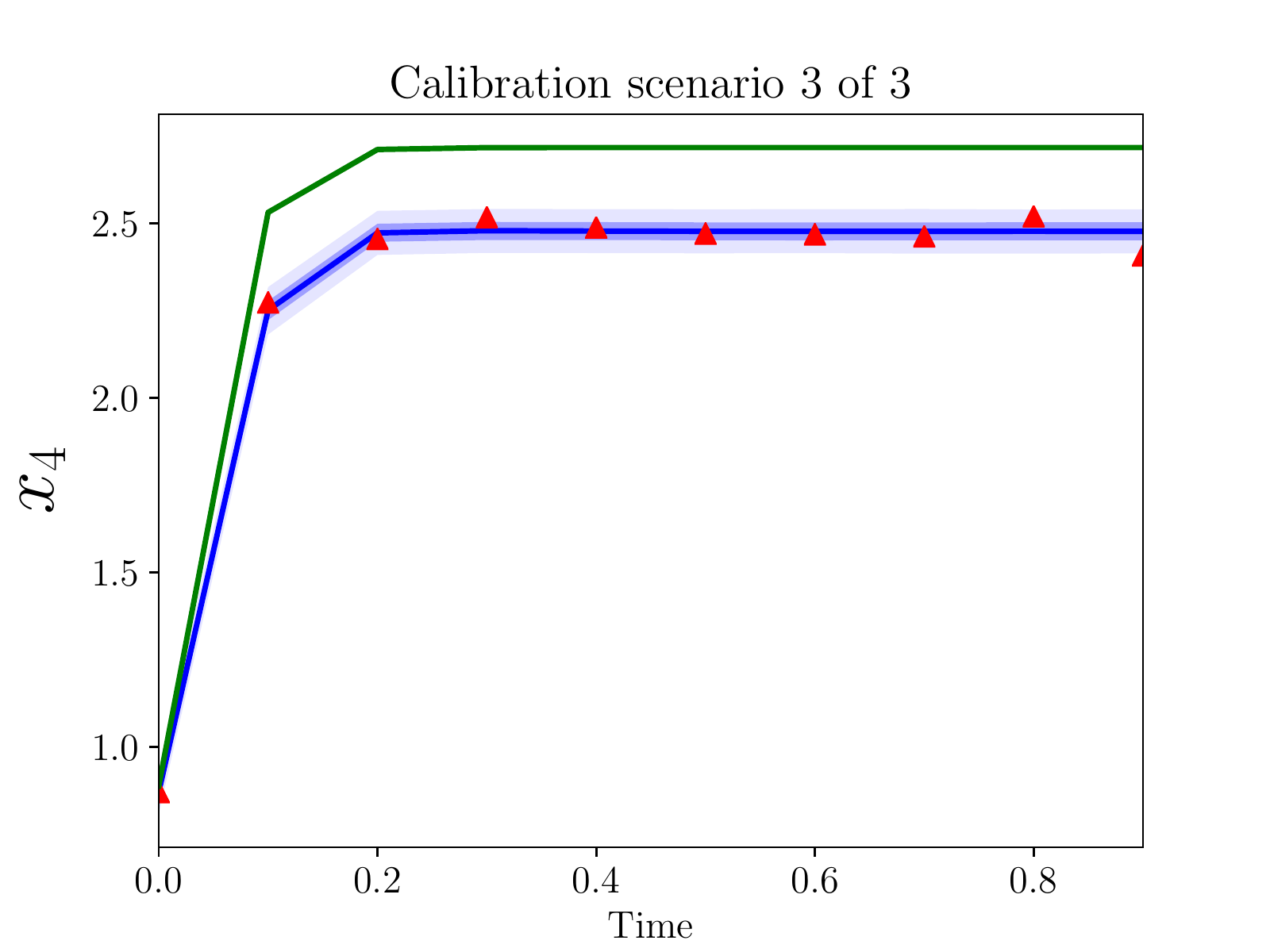}
    \end{subfigure}
    \caption{Reduced and enriched models, compared to observations, over three calibration
    scenarios. $S=10, s=4$. \label{fig:S10s4cal3}}
\end{figure}

\begin{figure}[htb]
  \centering
    \begin{subfigure}{.24\textwidth}
  \centering
  \includegraphics[width=\textwidth]{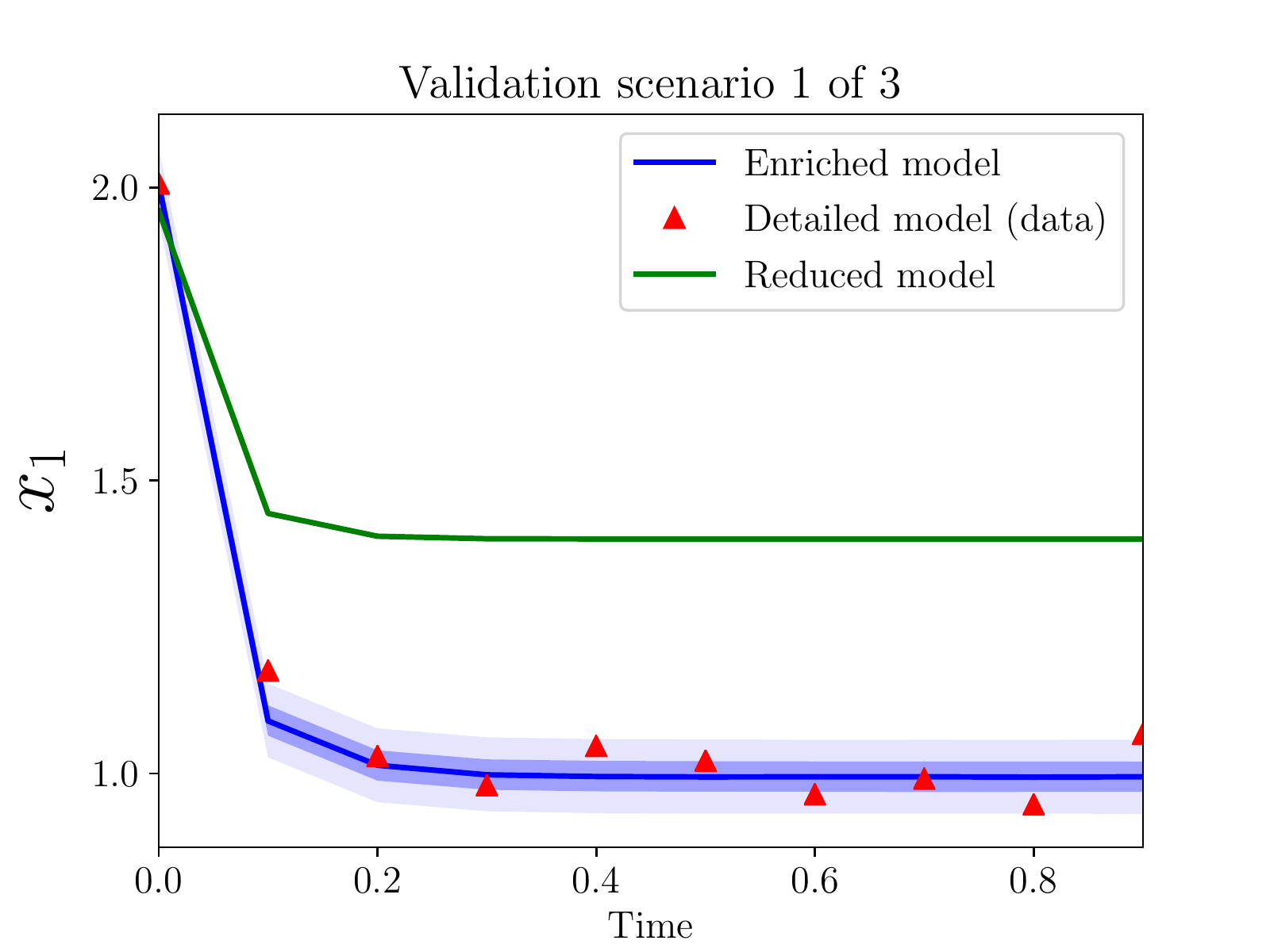}
    \end{subfigure}
    \begin{subfigure}{.24\textwidth}
  \centering
  \includegraphics[width=\textwidth]{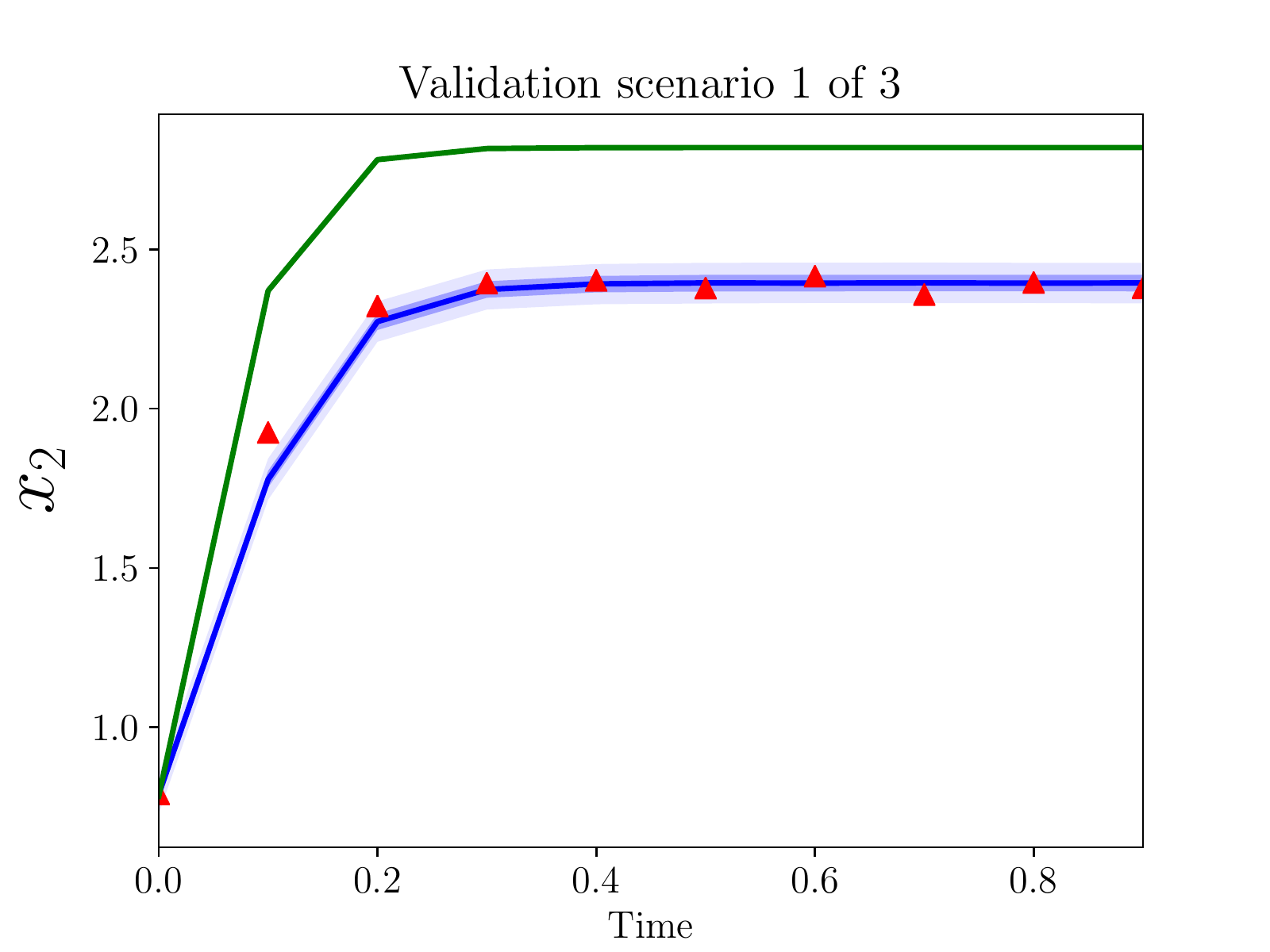}
    \end{subfigure}
    \begin{subfigure}{.24\textwidth}
  \centering
  \includegraphics[width=\textwidth]{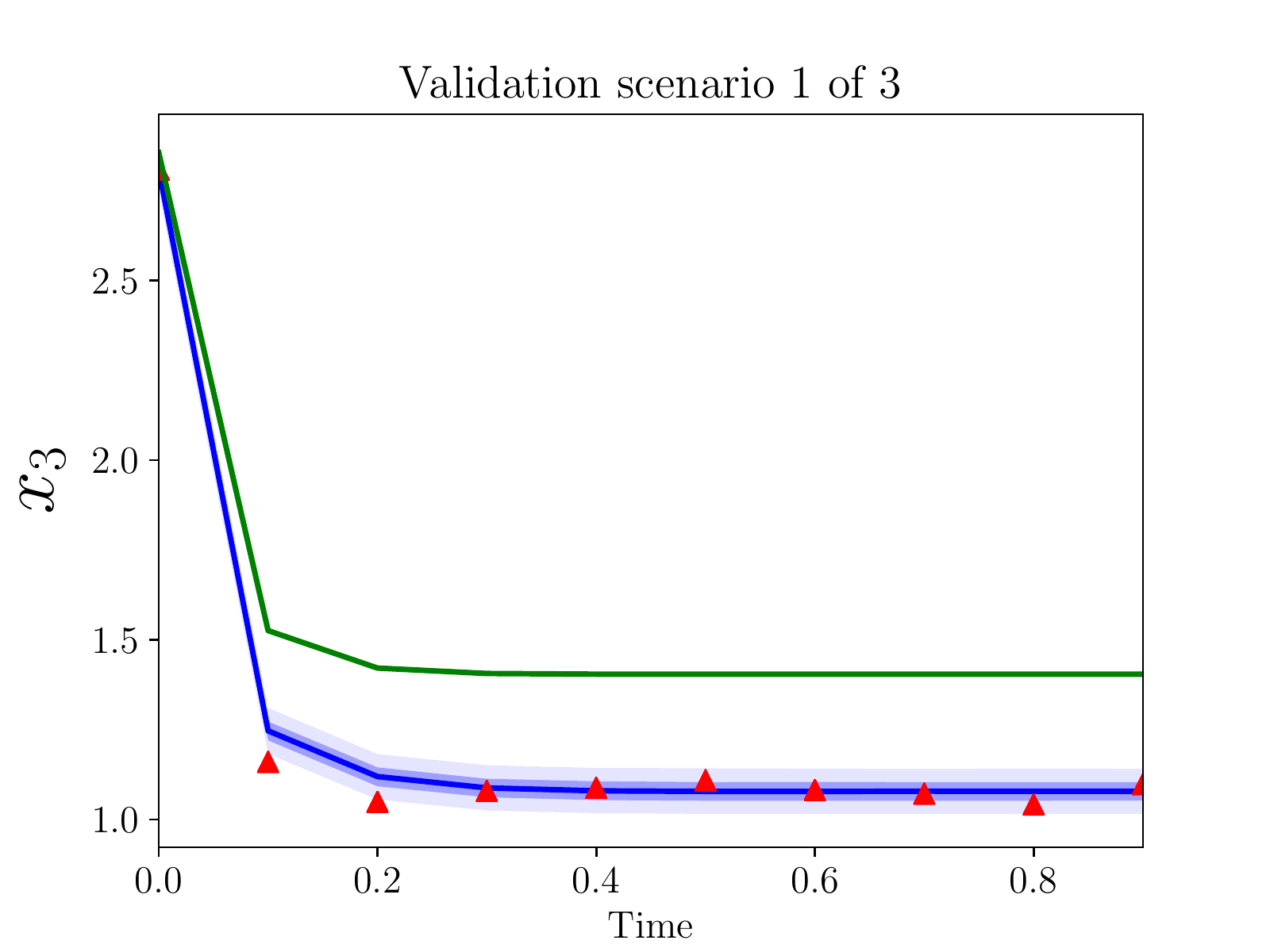}
    \end{subfigure}
    \begin{subfigure}{.24\textwidth}
  \centering
  \includegraphics[width=\textwidth]{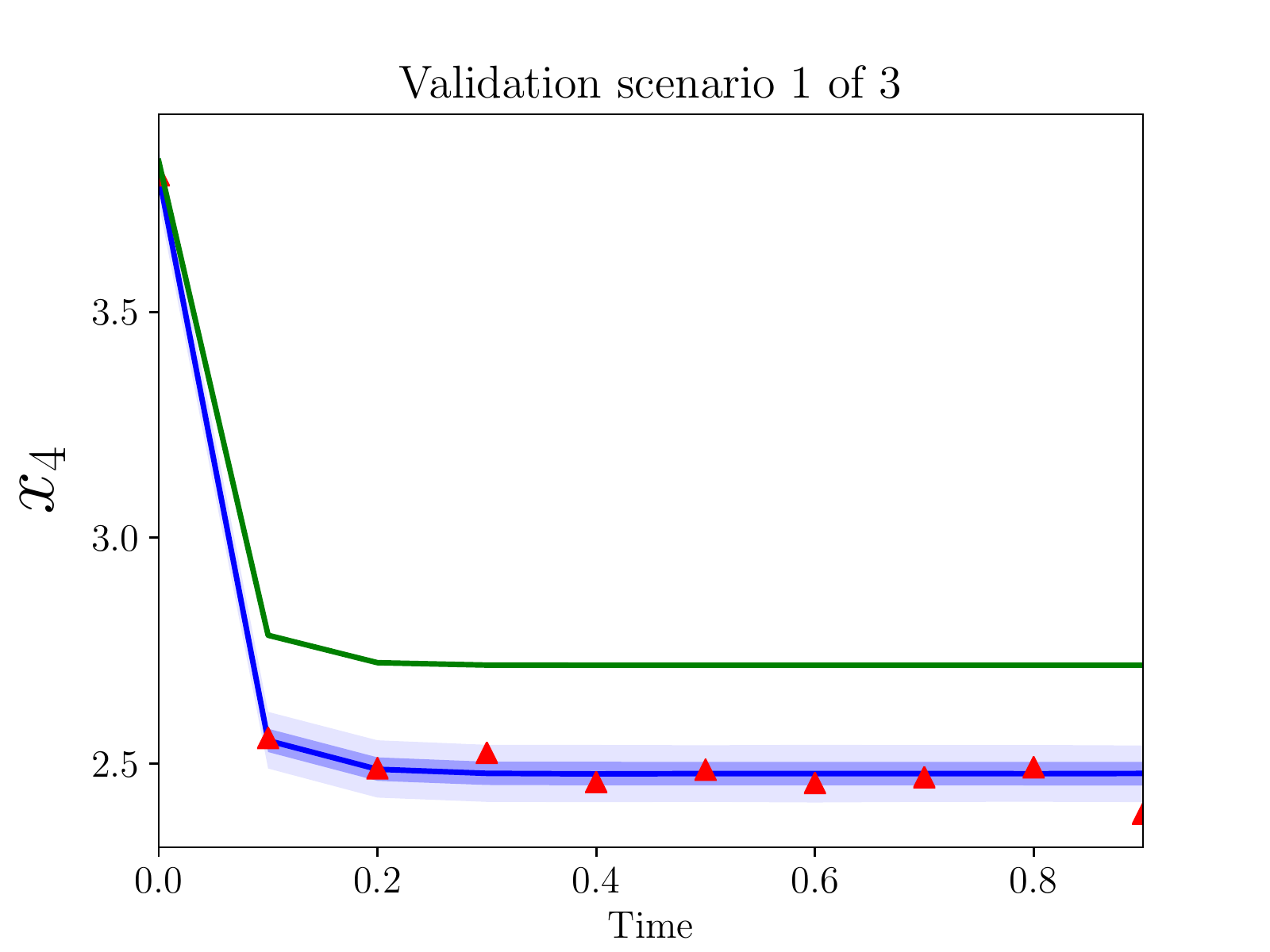}
    \end{subfigure}
    \begin{subfigure}{.24\textwidth}
  \centering
  \includegraphics[width=\textwidth]{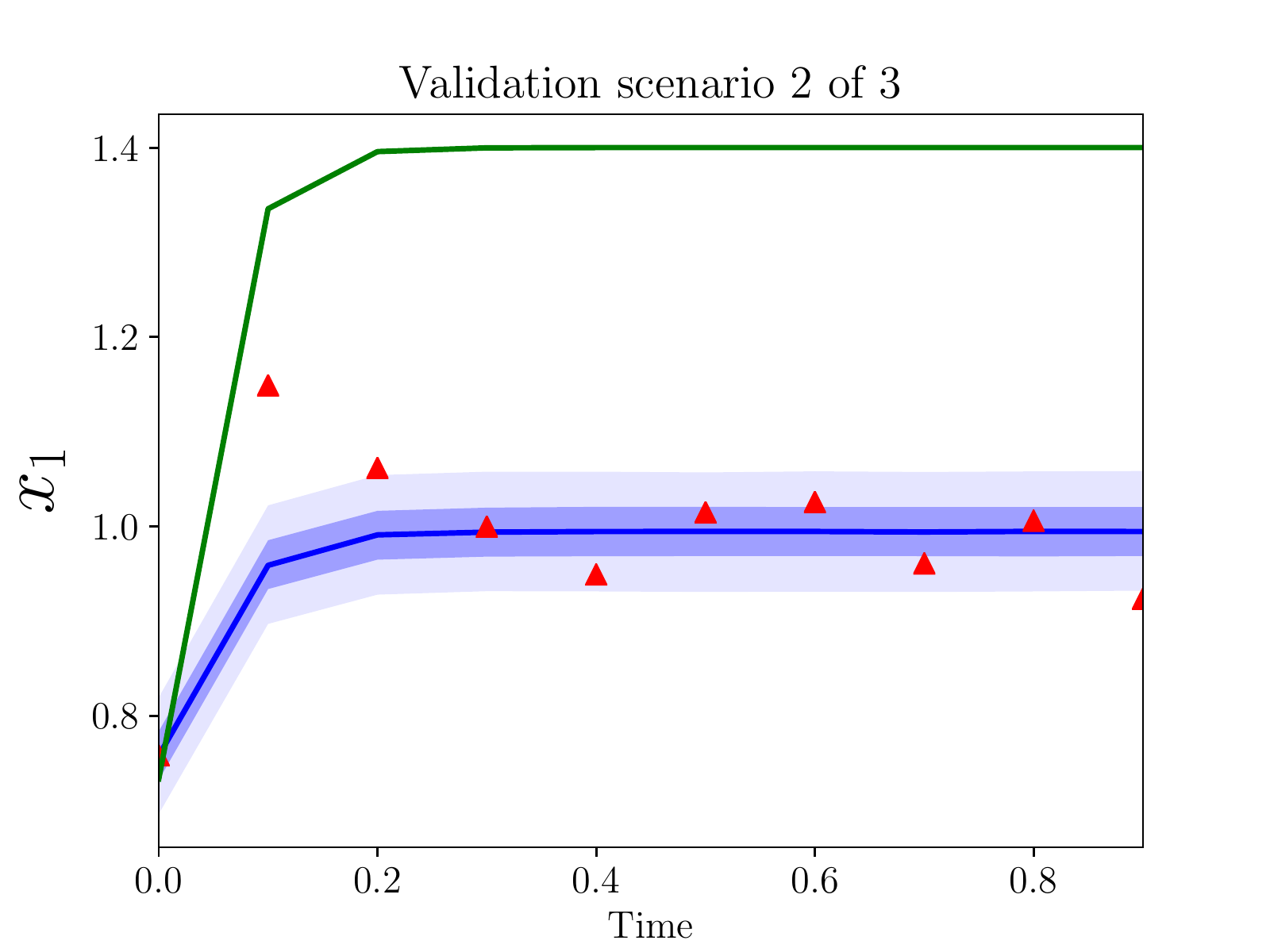}
    \end{subfigure}
    \begin{subfigure}{.24\textwidth}
  \centering
  \includegraphics[width=\textwidth]{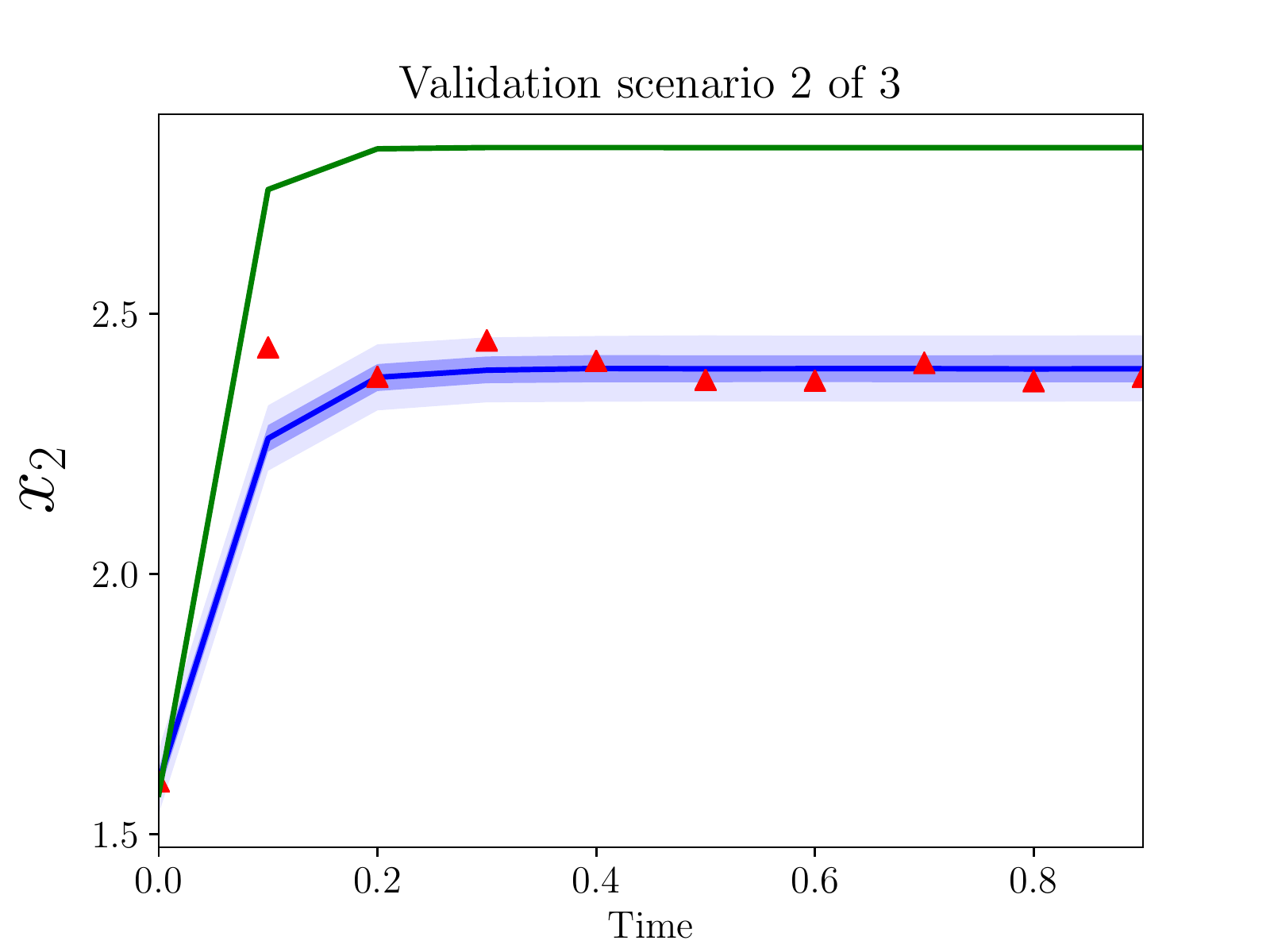}
    \end{subfigure}
    \begin{subfigure}{.24\textwidth}
  \centering
  \includegraphics[width=\textwidth]{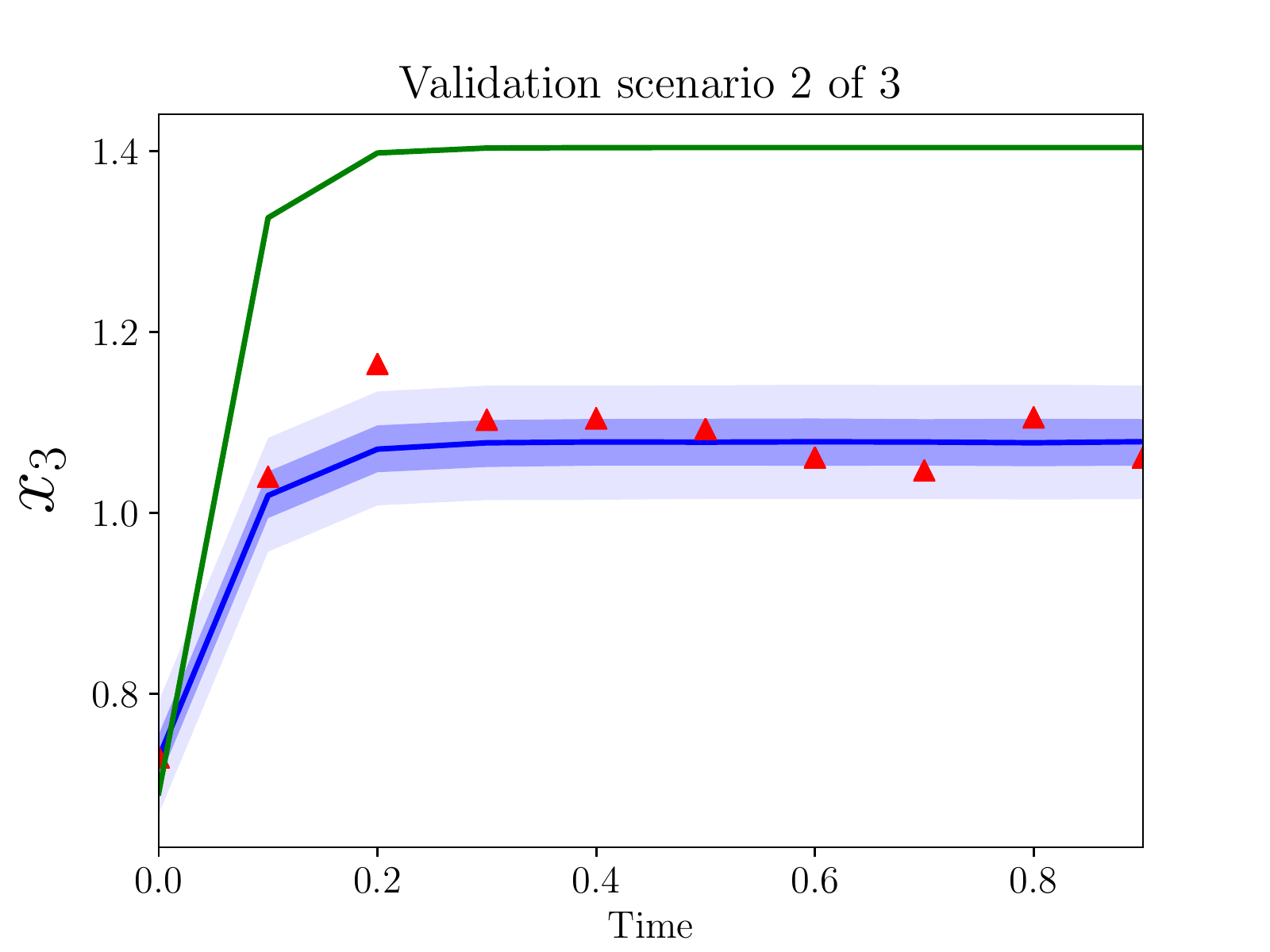}
    \end{subfigure}
    \begin{subfigure}{.24\textwidth}
  \centering
  \includegraphics[width=\textwidth]{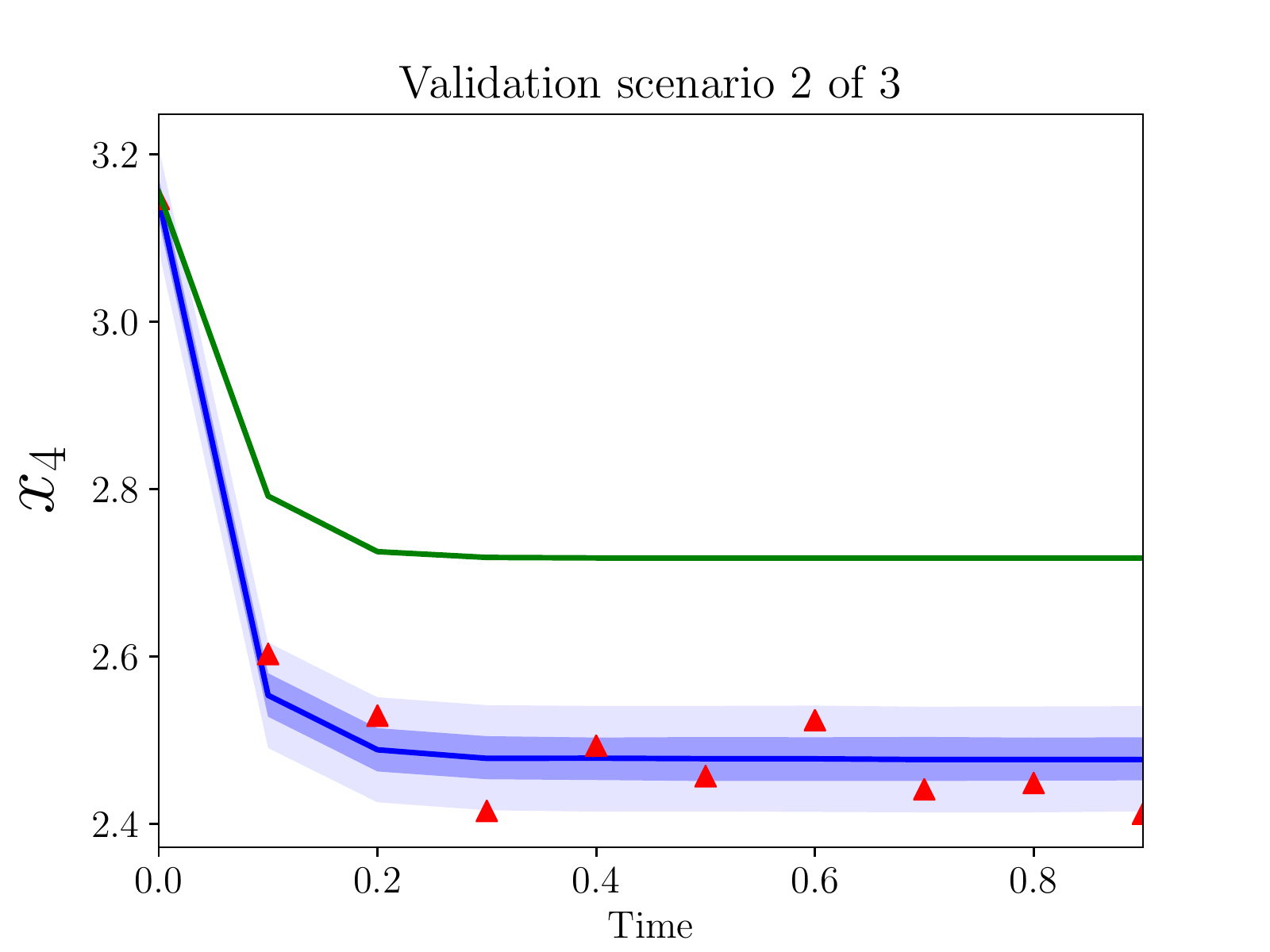}
    \end{subfigure}
    \begin{subfigure}{.24\textwidth}
  \centering
  \includegraphics[width=\textwidth]{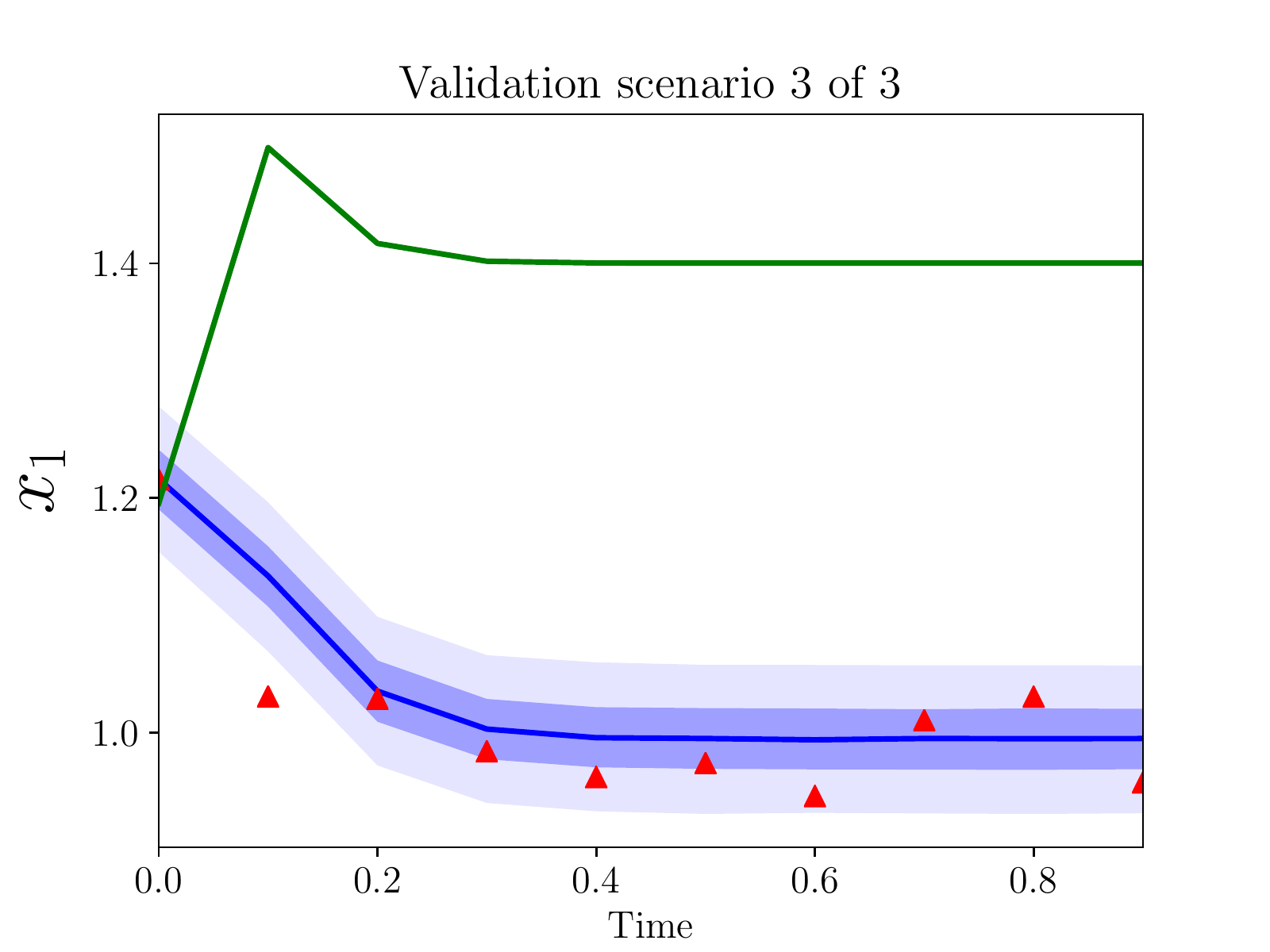}
    \end{subfigure}
    \begin{subfigure}{.24\textwidth}
  \centering
  \includegraphics[width=\textwidth]{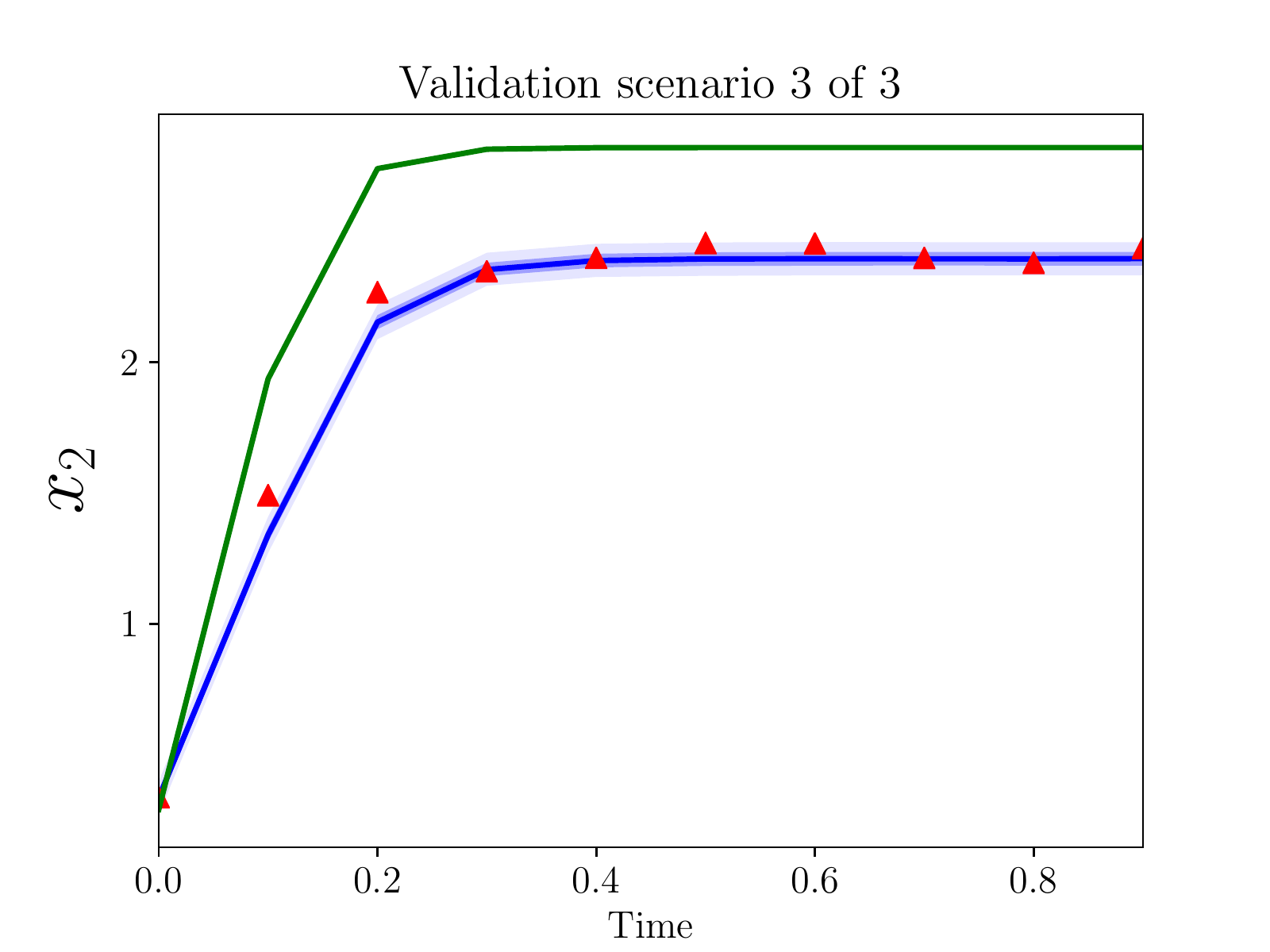}
    \end{subfigure}
    \begin{subfigure}{.24\textwidth}
  \centering
  \includegraphics[width=\textwidth]{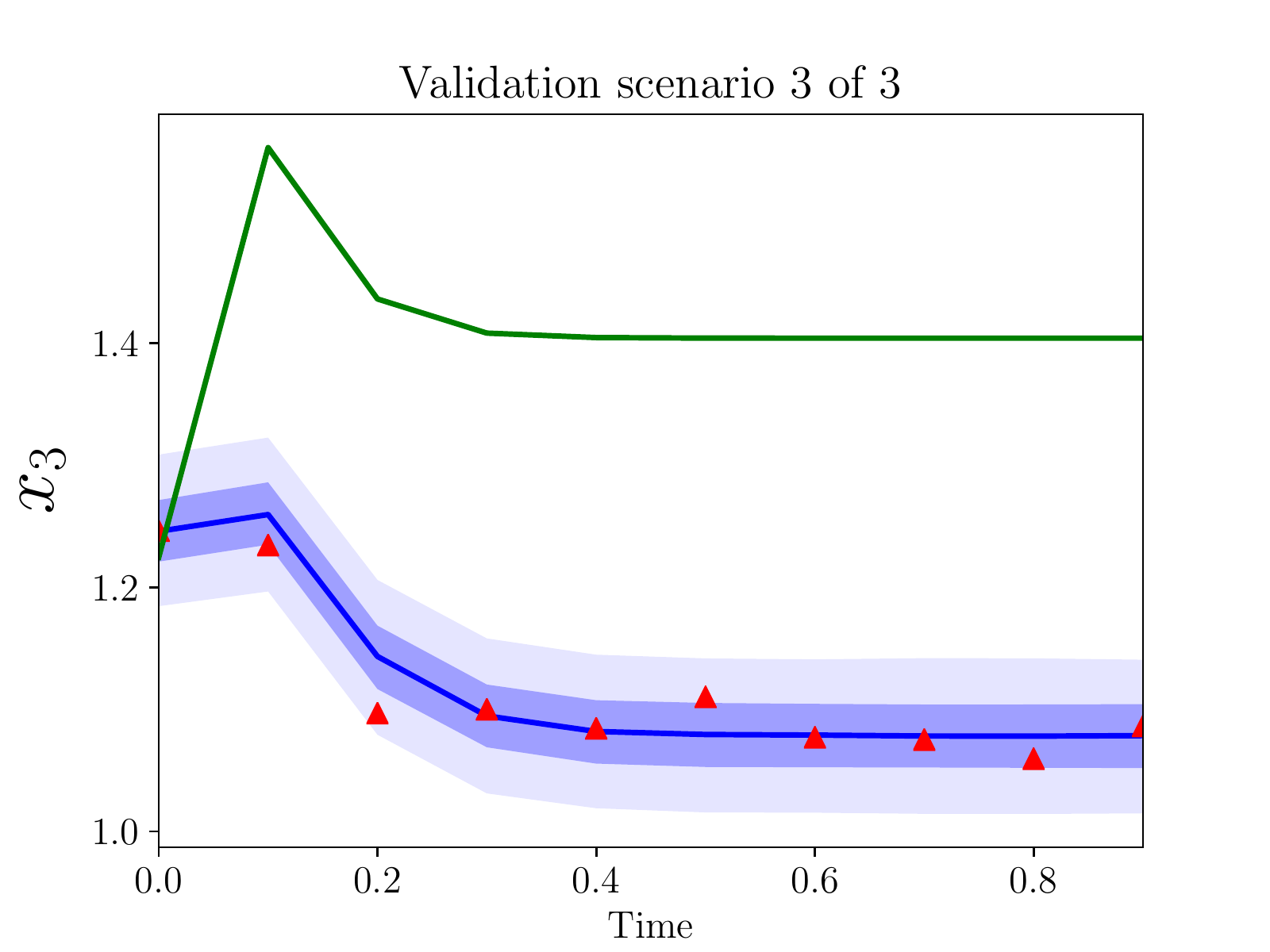}
    \end{subfigure}
    \begin{subfigure}{.24\textwidth}
  \centering
  \includegraphics[width=\textwidth]{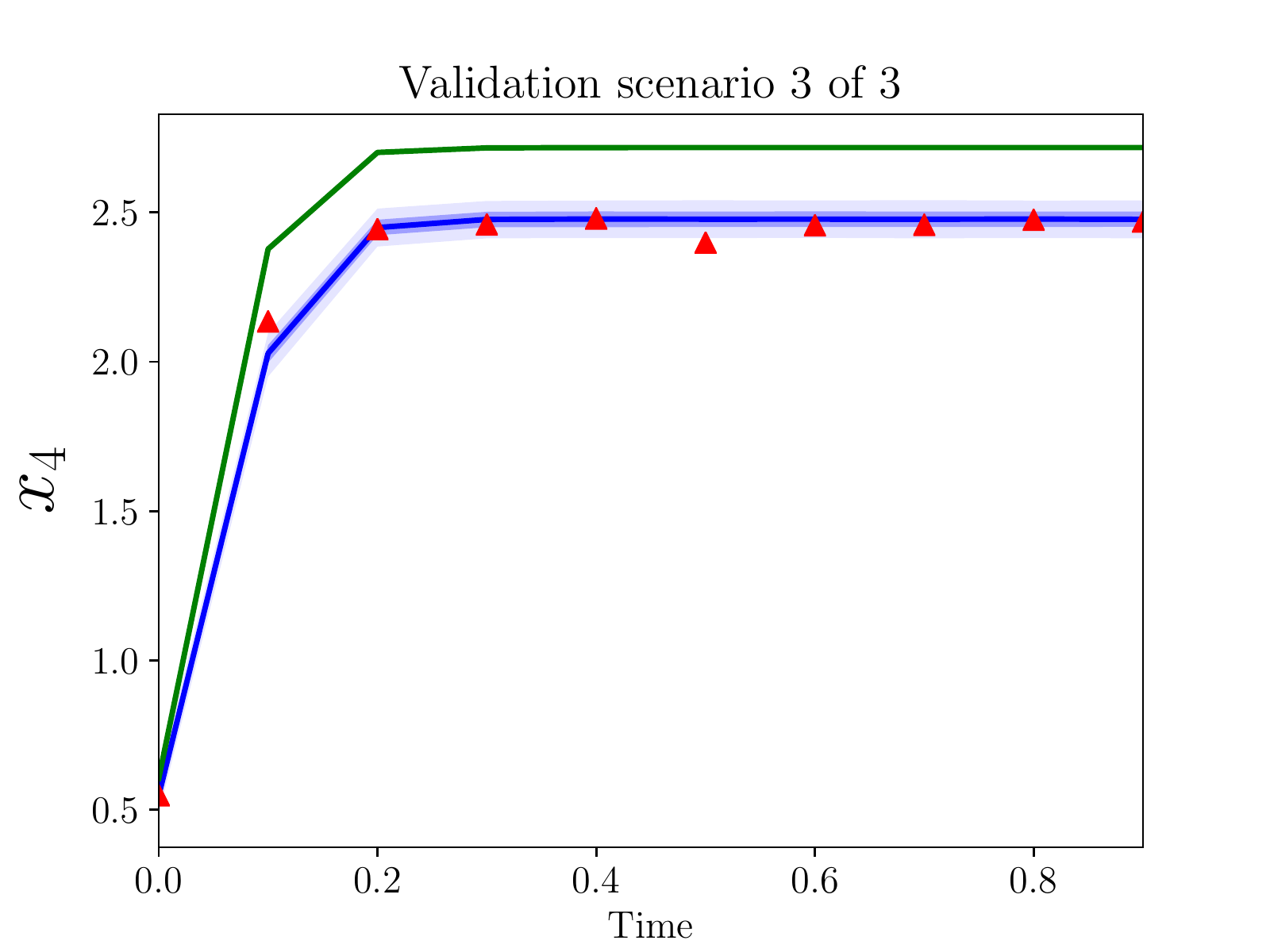}
    \end{subfigure}
    \caption{Reduced and enriched models, compared to observations, over three validation scenarios.
    $S=10, s =4$.  \label{fig:S10s4val3}}
\end{figure}

\begin{figure}[htb]
  \centering
    \begin{subfigure}{.24\textwidth}
  \centering
  \includegraphics[width=\textwidth]{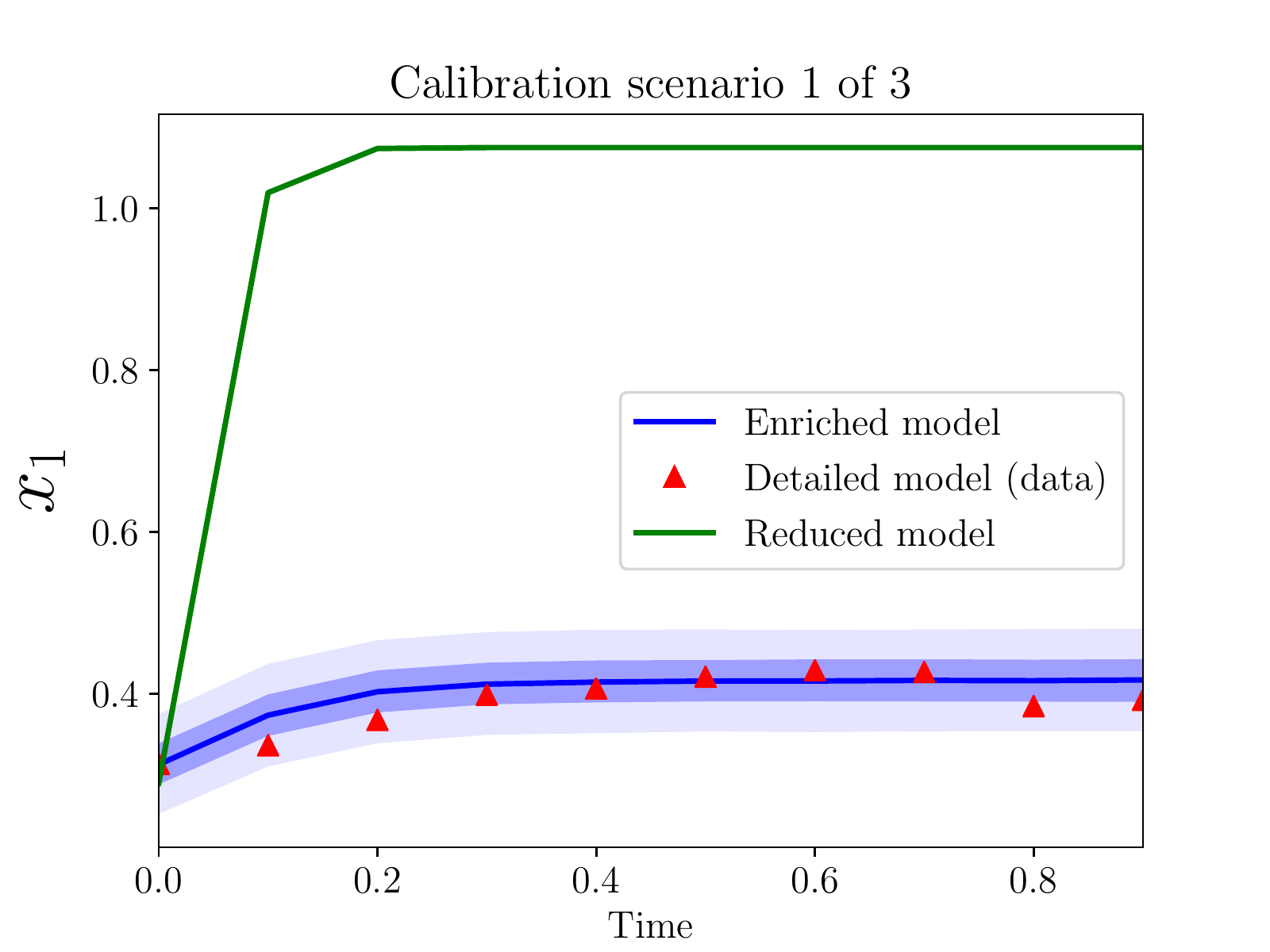}
    \end{subfigure}
    \begin{subfigure}{.24\textwidth}
  \centering
  \includegraphics[width=\textwidth]{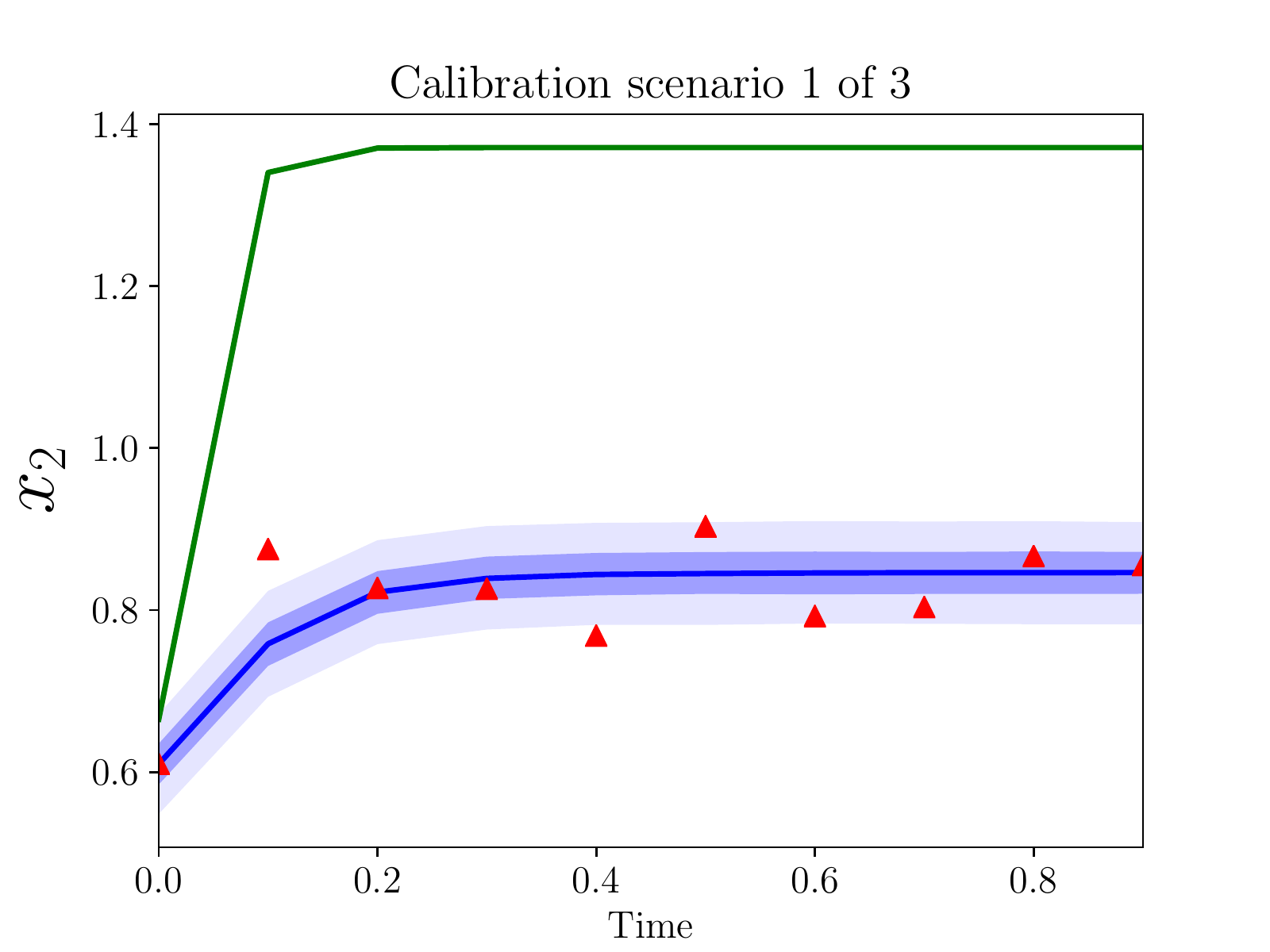}
    \end{subfigure}
    \begin{subfigure}{.24\textwidth}
  \centering
  \includegraphics[width=\textwidth]{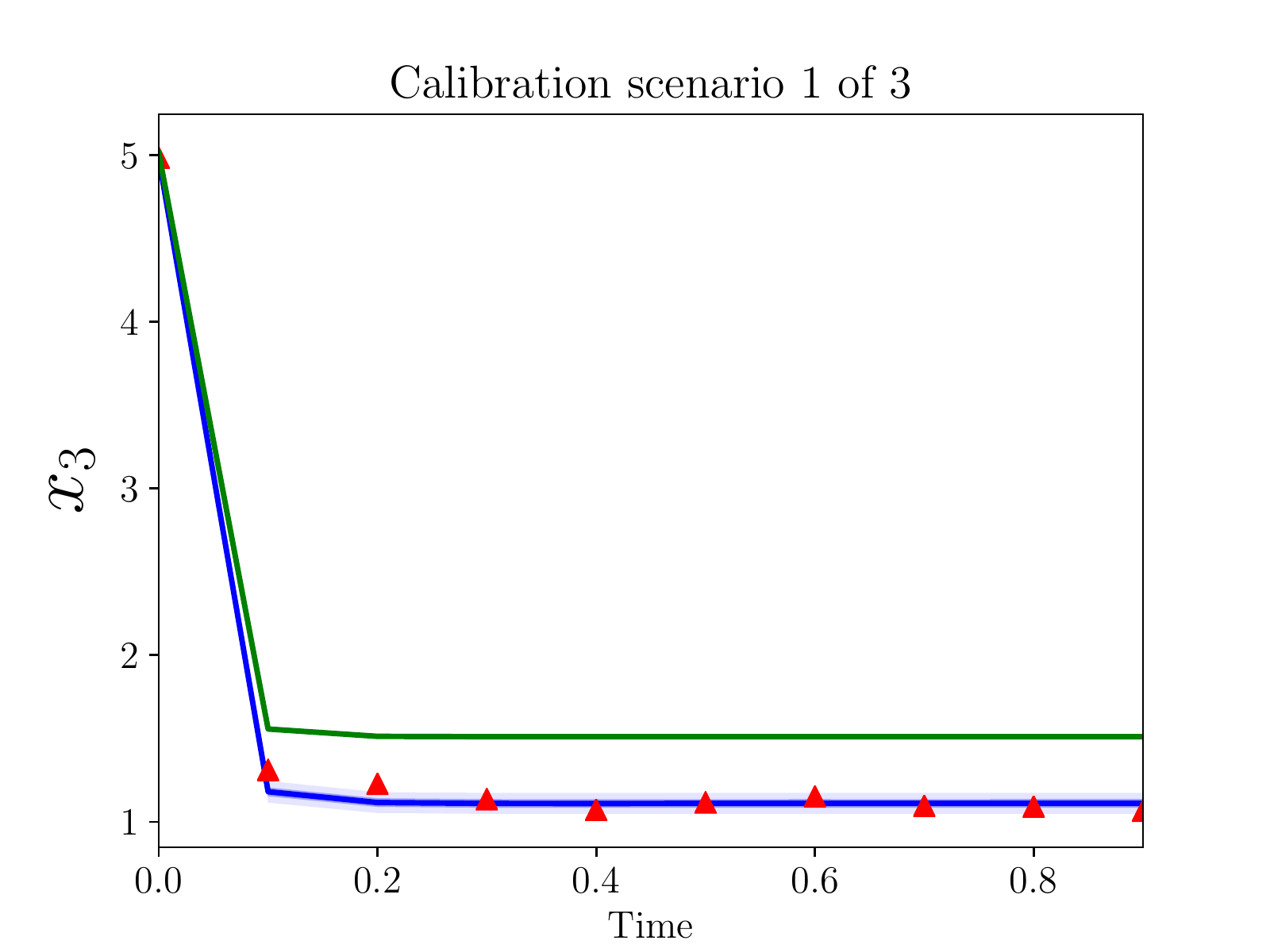}
    \end{subfigure}
    \begin{subfigure}{.24\textwidth}
  \centering
  \includegraphics[width=\textwidth]{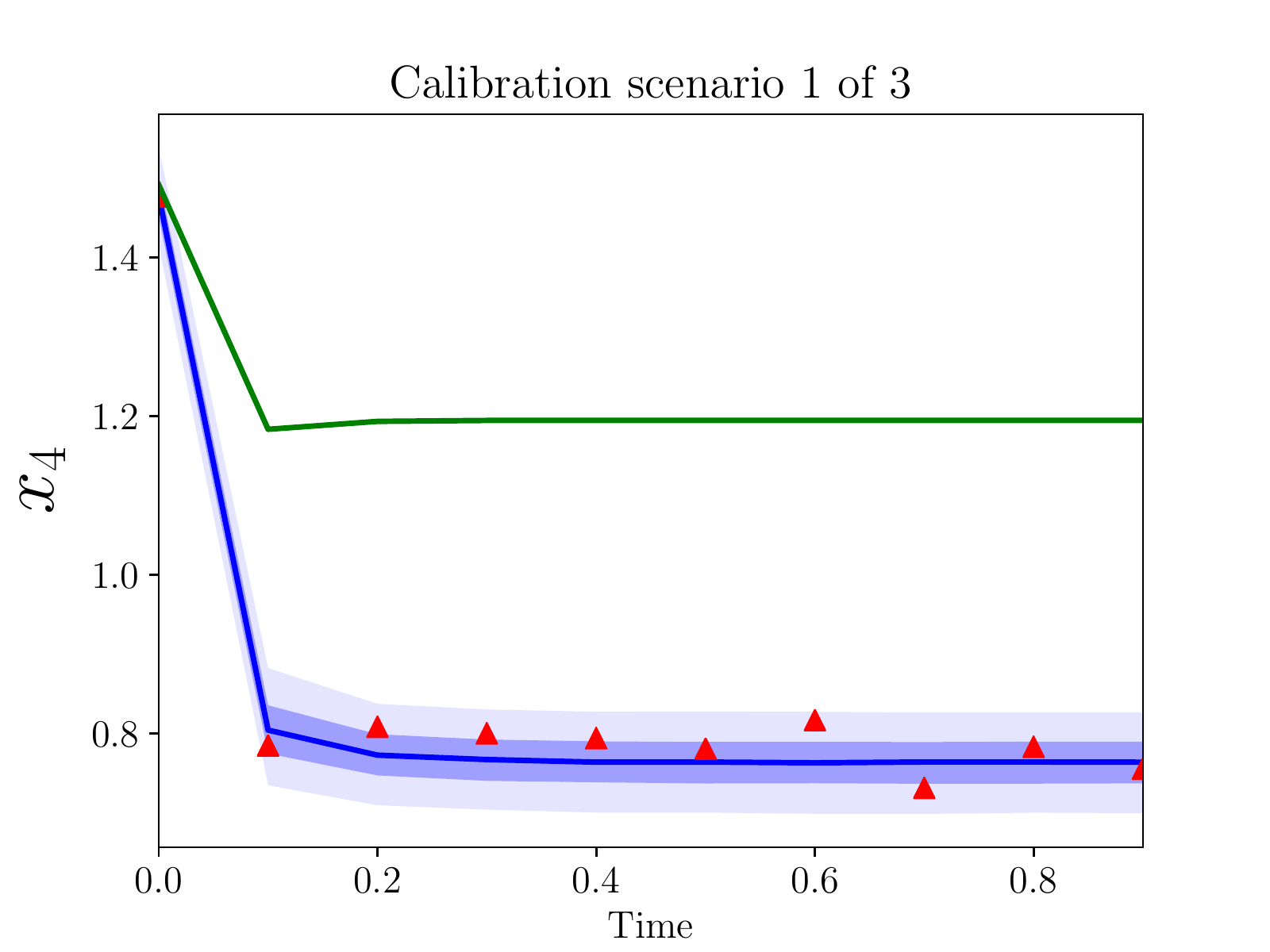}
    \end{subfigure}
    \begin{subfigure}{.24\textwidth}
  \centering
  \includegraphics[width=\textwidth]{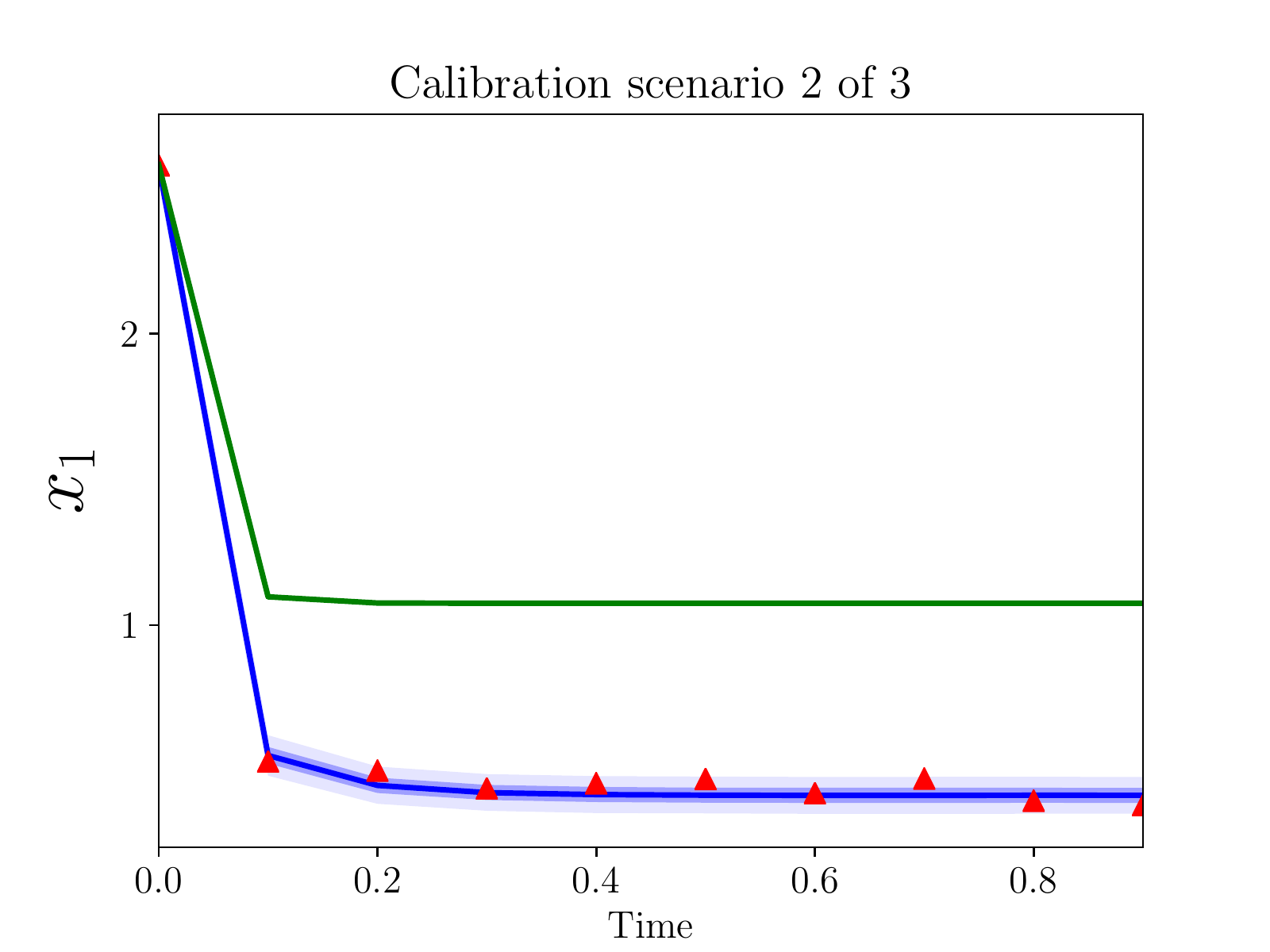}
    \end{subfigure}
    \begin{subfigure}{.24\textwidth}
  \centering
  \includegraphics[width=\textwidth]{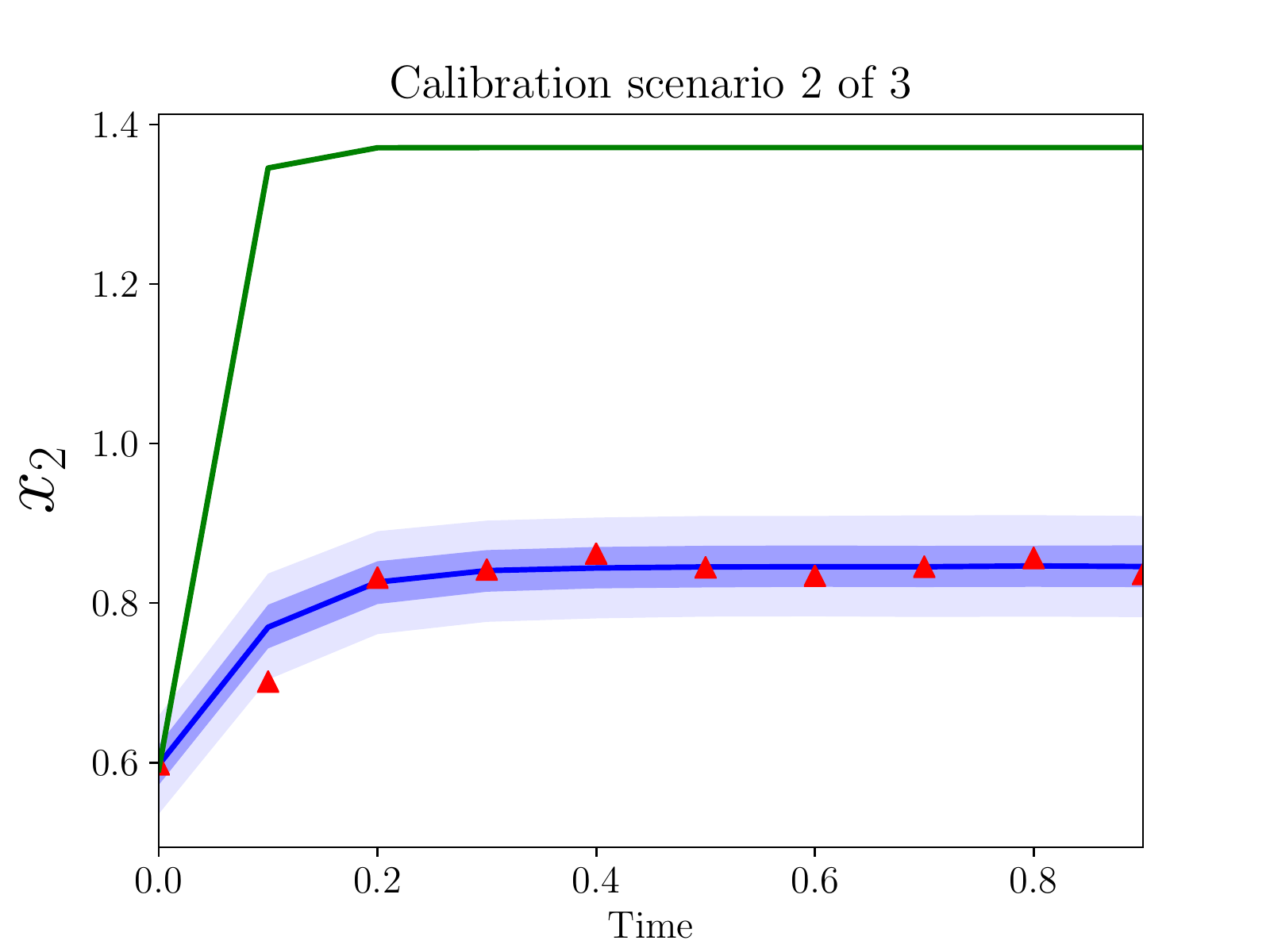}
    \end{subfigure}
    \begin{subfigure}{.24\textwidth}
  \centering
  \includegraphics[width=\textwidth]{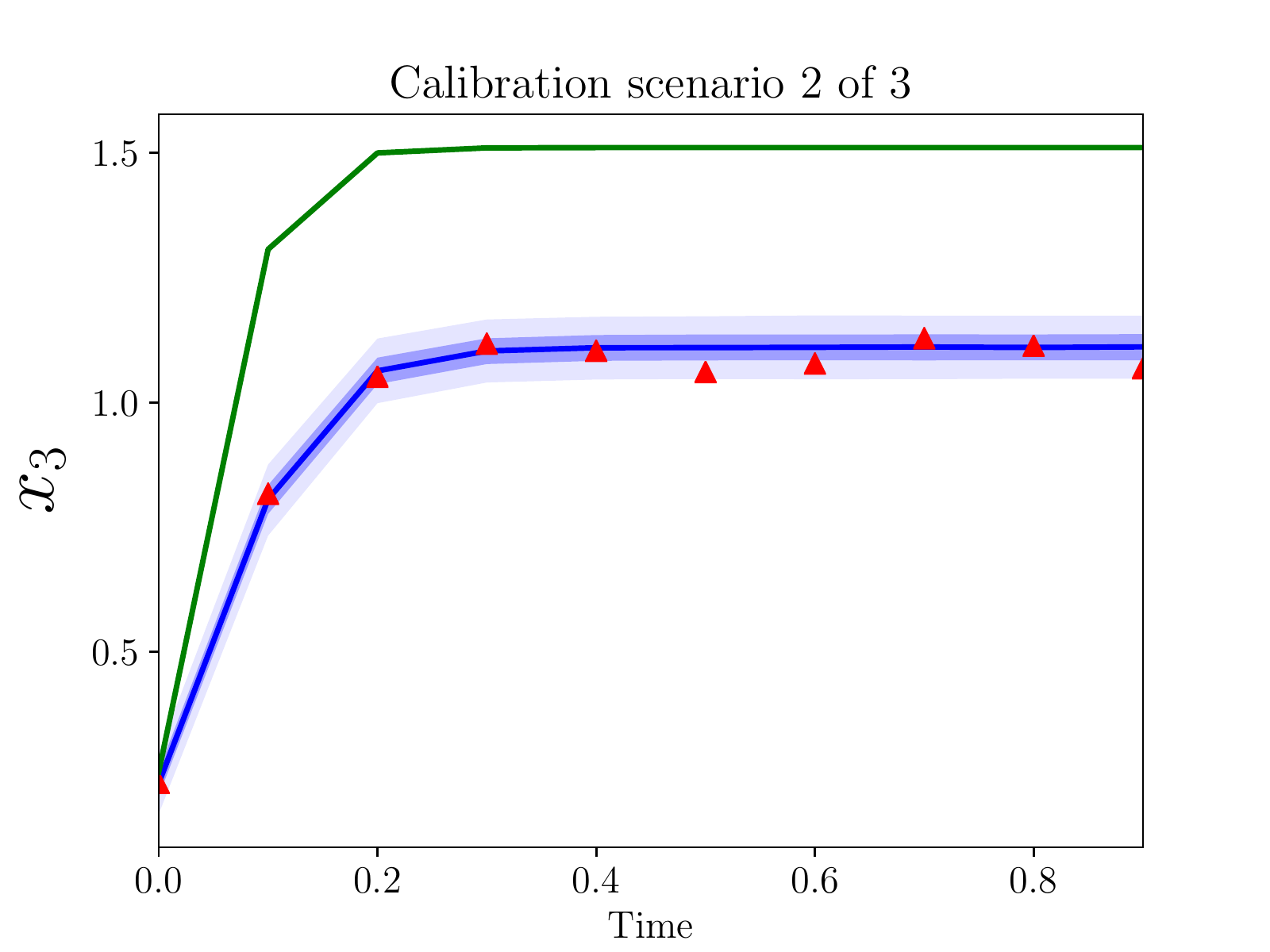}
    \end{subfigure}
    \begin{subfigure}{.24\textwidth}
  \centering
  \includegraphics[width=\textwidth]{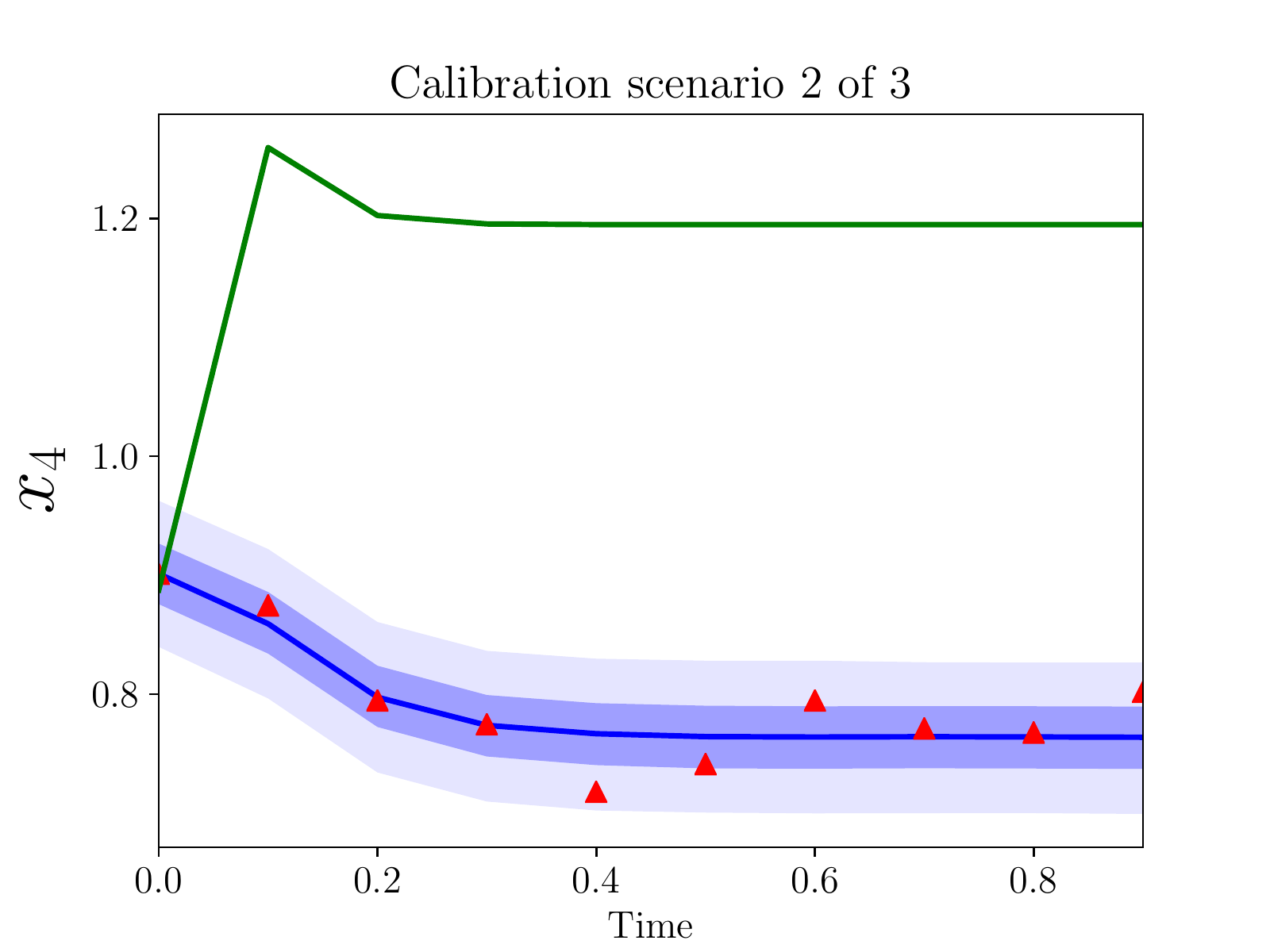}
    \end{subfigure}
    \begin{subfigure}{.24\textwidth}
  \centering
  \includegraphics[width=\textwidth]{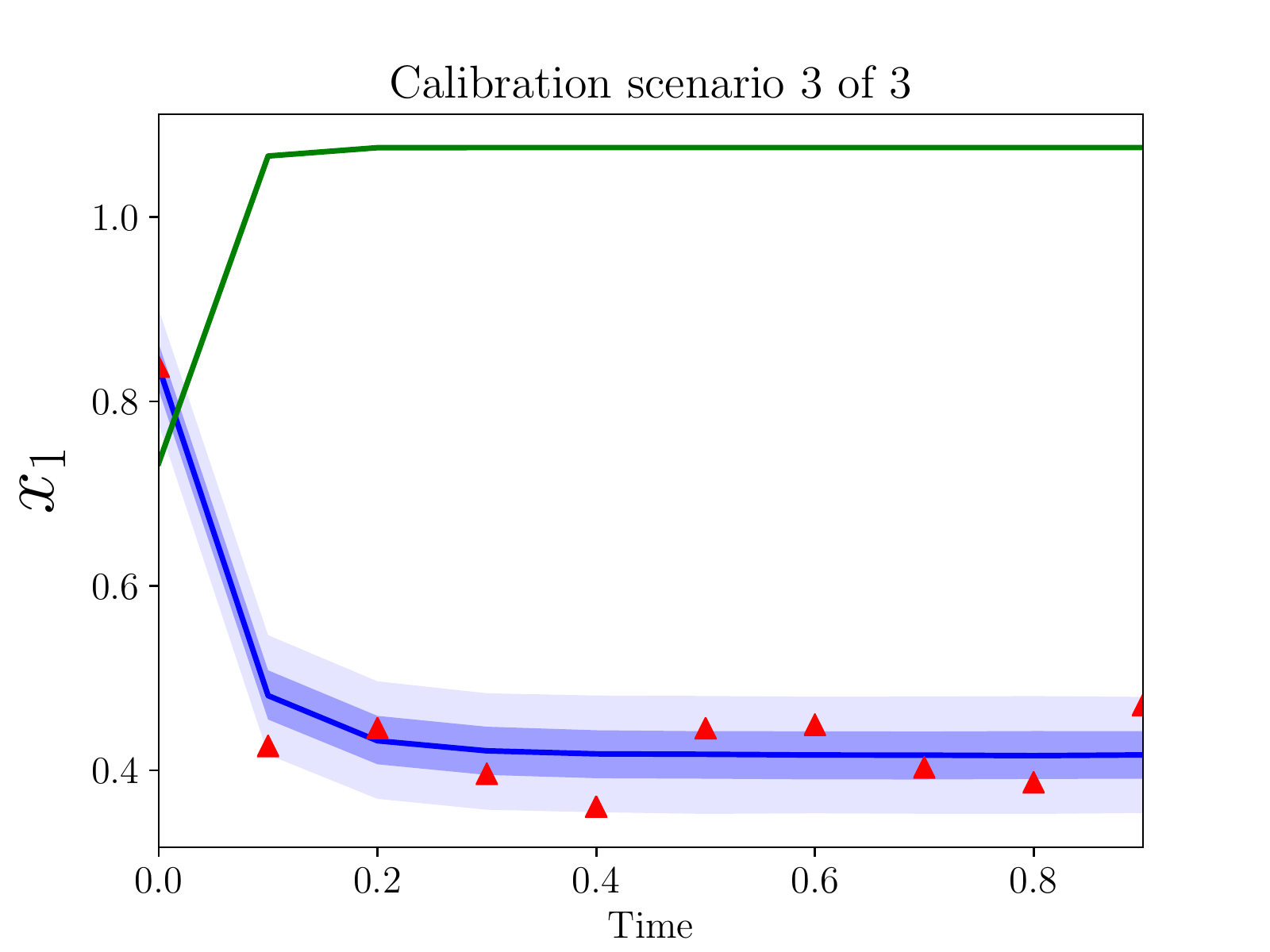}
    \end{subfigure}
    \begin{subfigure}{.24\textwidth}
  \centering
  \includegraphics[width=\textwidth]{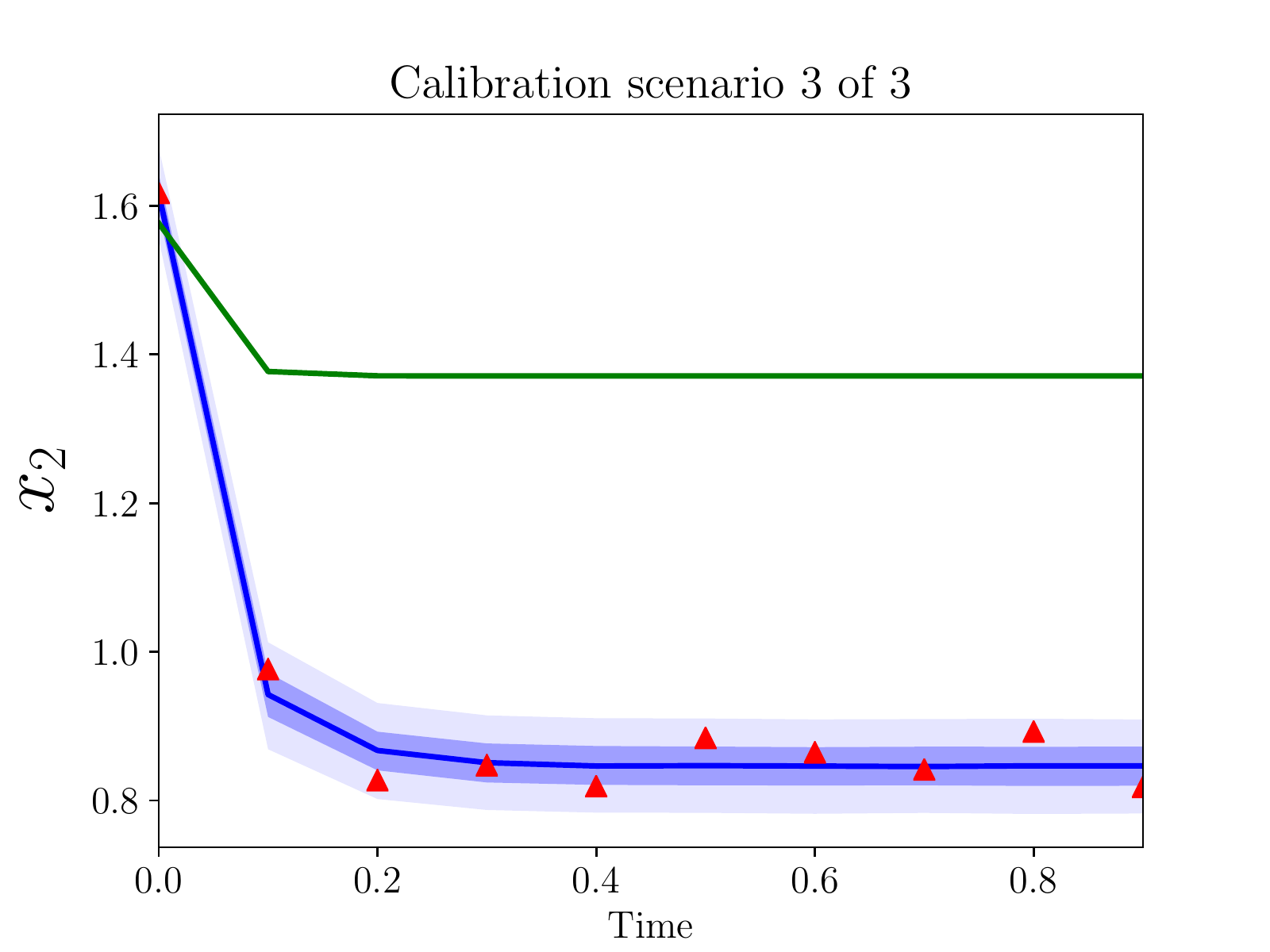}
    \end{subfigure}
    \begin{subfigure}{.24\textwidth}
  \centering
  \includegraphics[width=\textwidth]{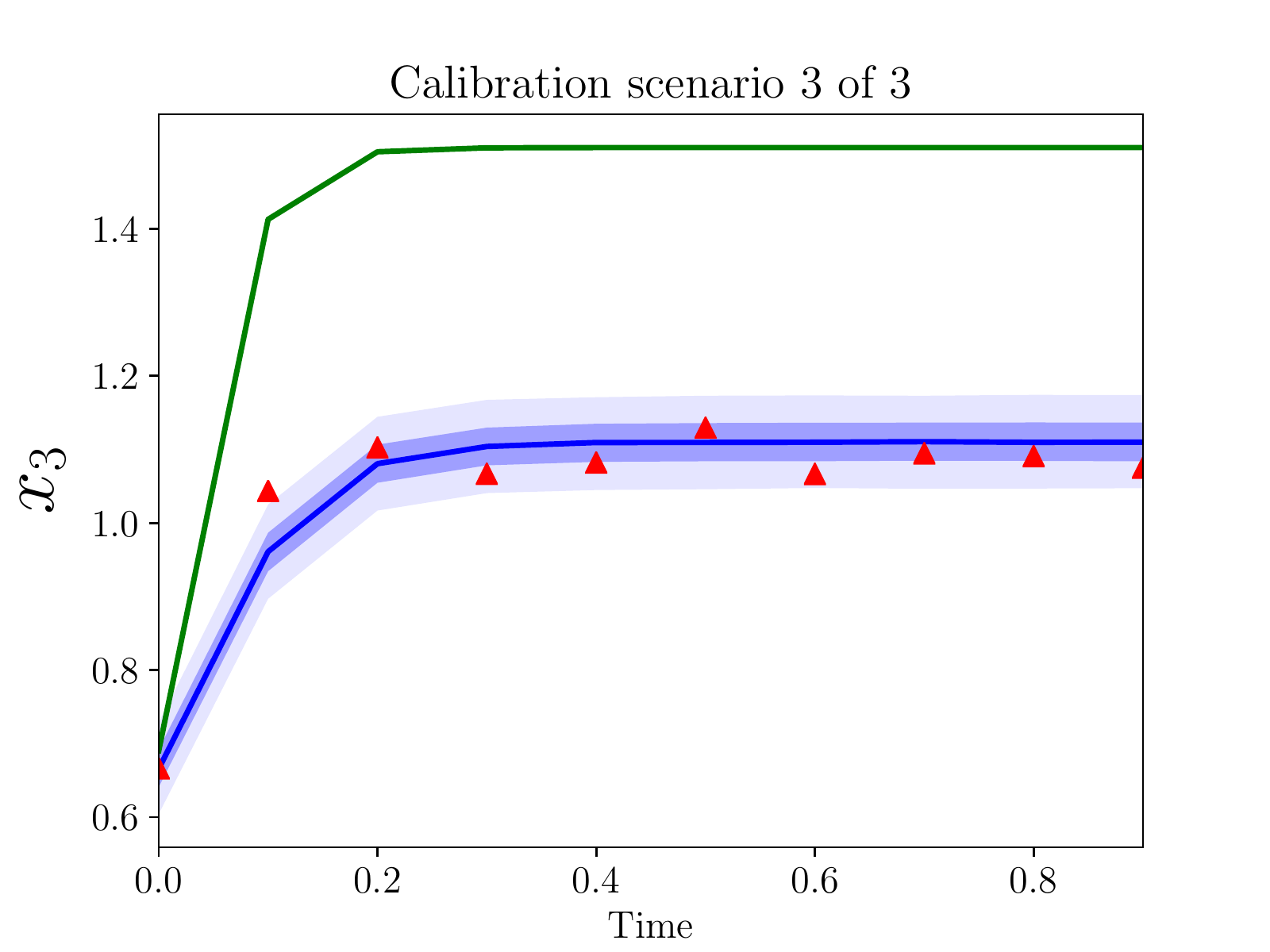}
    \end{subfigure}
    \begin{subfigure}{.24\textwidth}
  \centering
  \includegraphics[width=\textwidth]{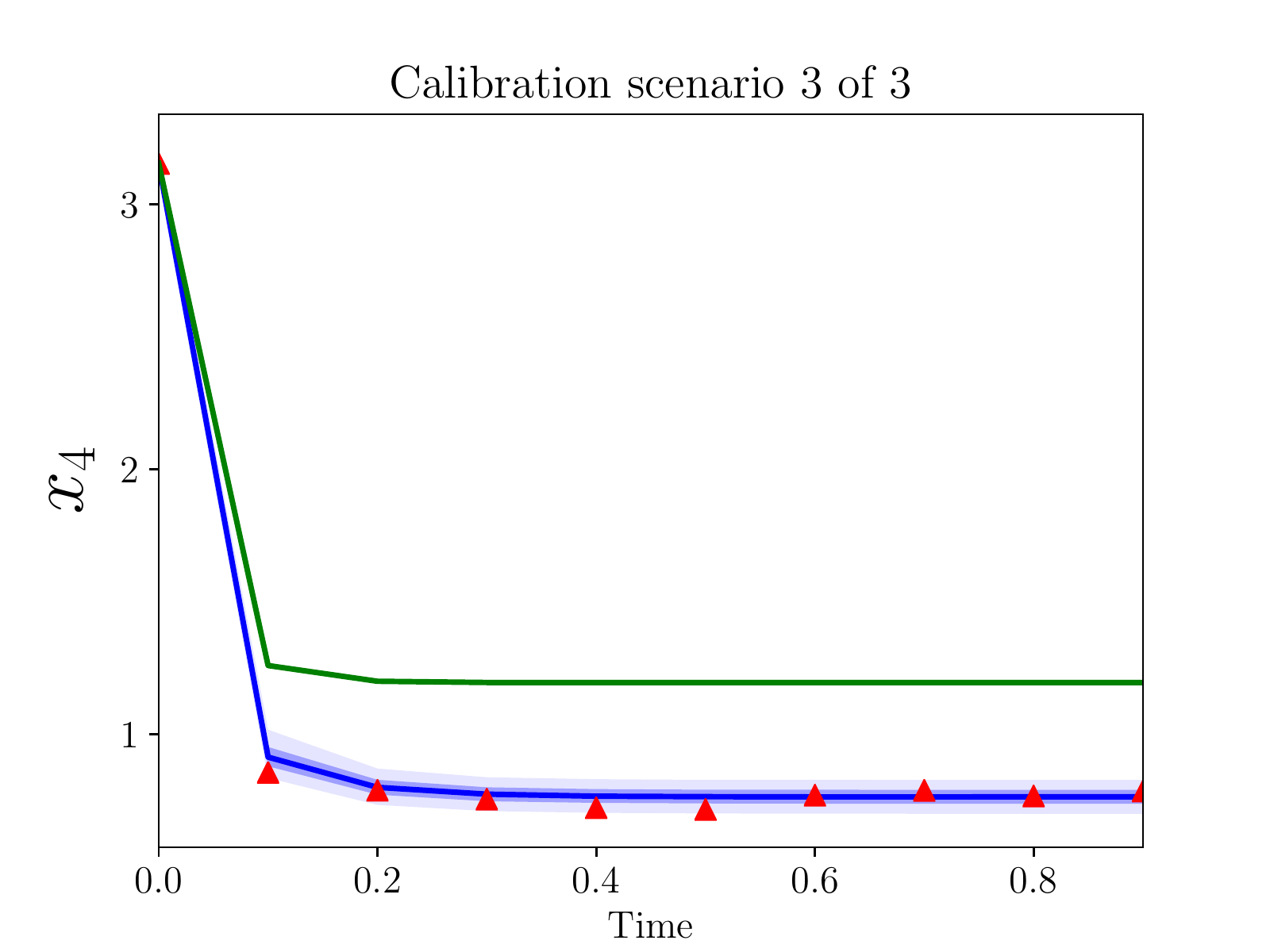}
    \end{subfigure}
    \caption{Reduced and enriched models, compared to observations, over three calibration
    scenarios. $S=20, s=4$.    \label{fig:S20s4cal3}}
\end{figure}

\begin{figure}[htb]
  \centering
    \begin{subfigure}{.24\textwidth}
  \centering
  \includegraphics[width=\textwidth]{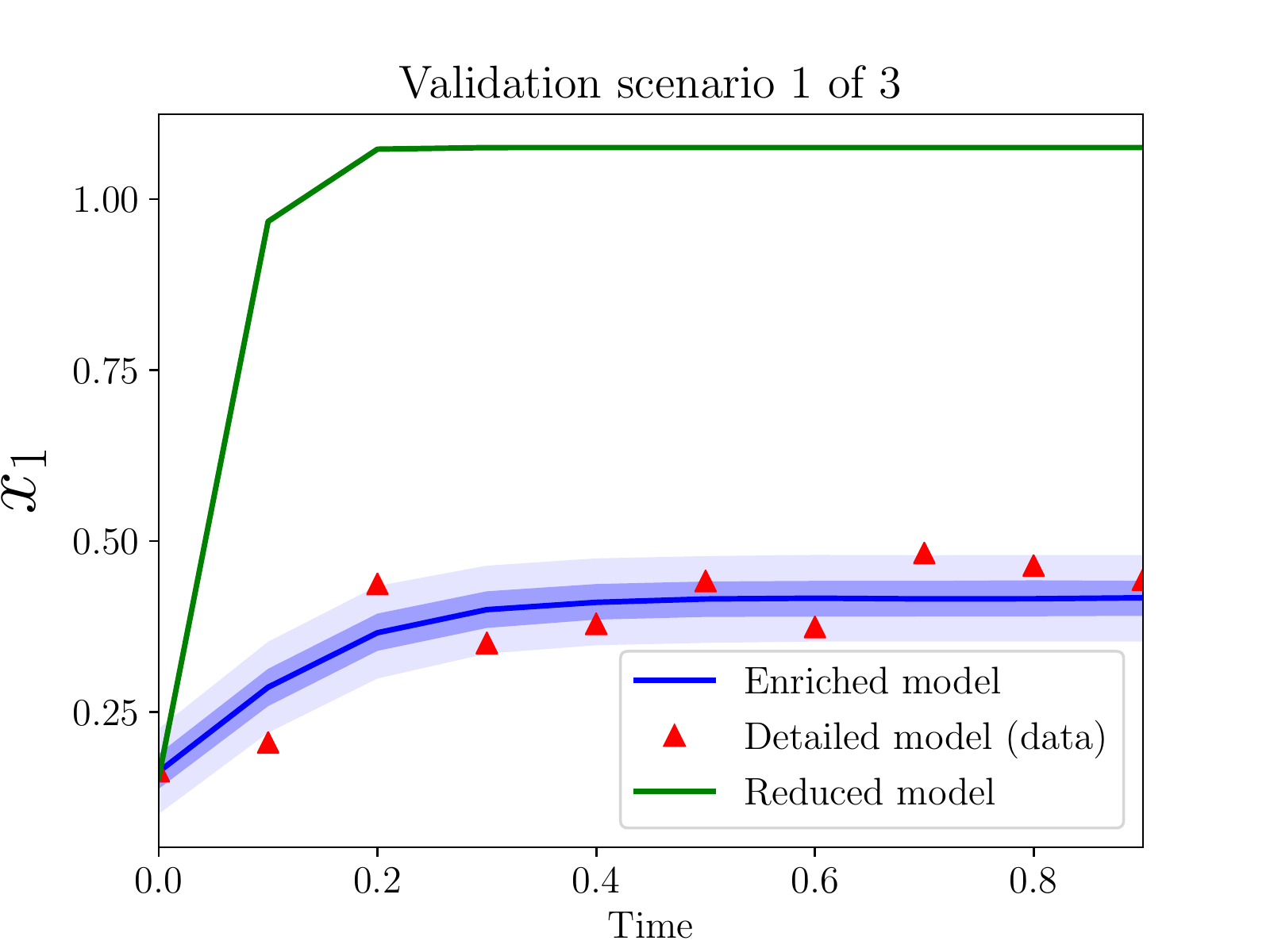}
    \end{subfigure}
    \begin{subfigure}{.24\textwidth}
  \centering
  \includegraphics[width=\textwidth]{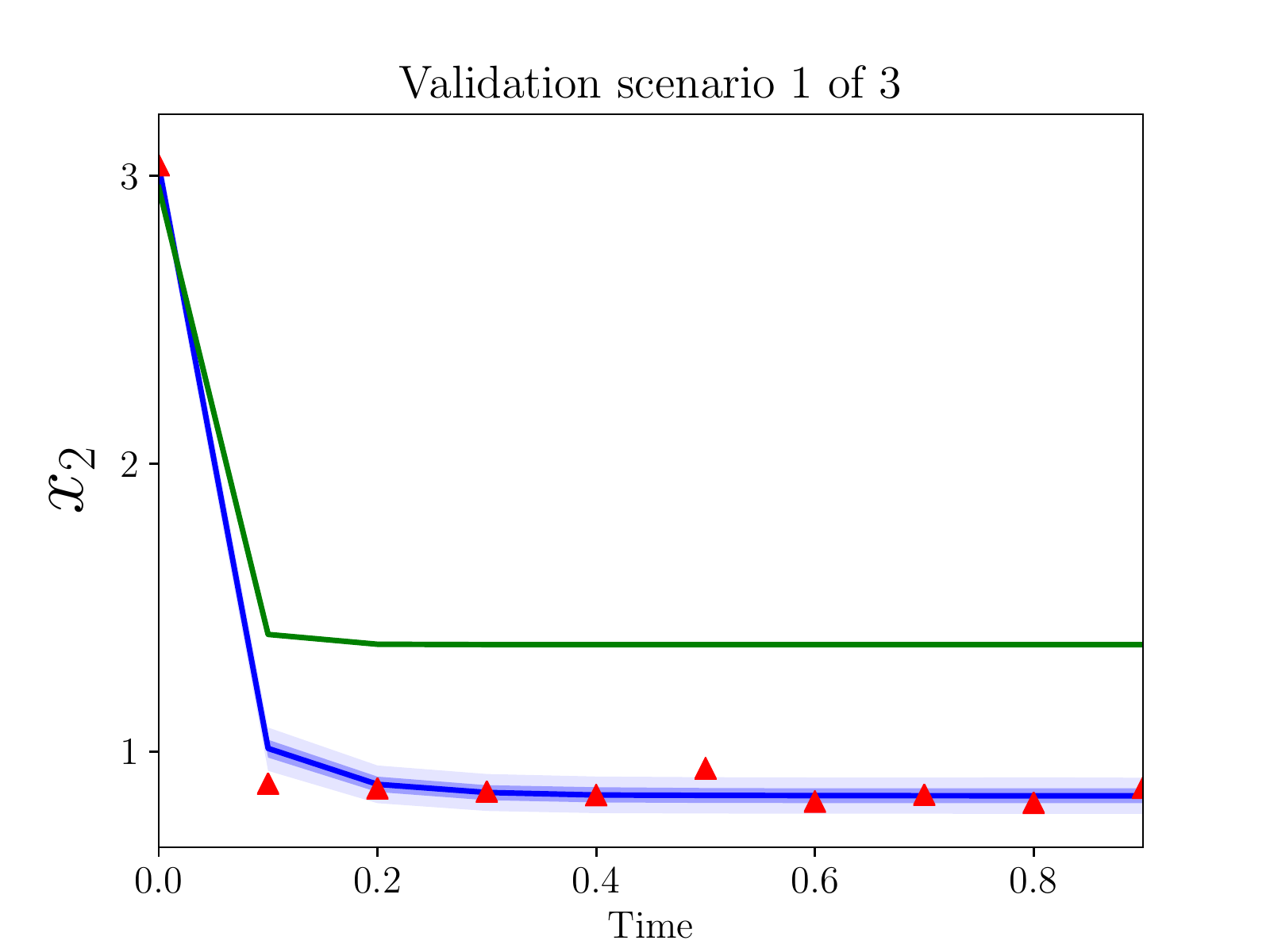}
    \end{subfigure}
    \begin{subfigure}{.24\textwidth}
  \centering
  \includegraphics[width=\textwidth]{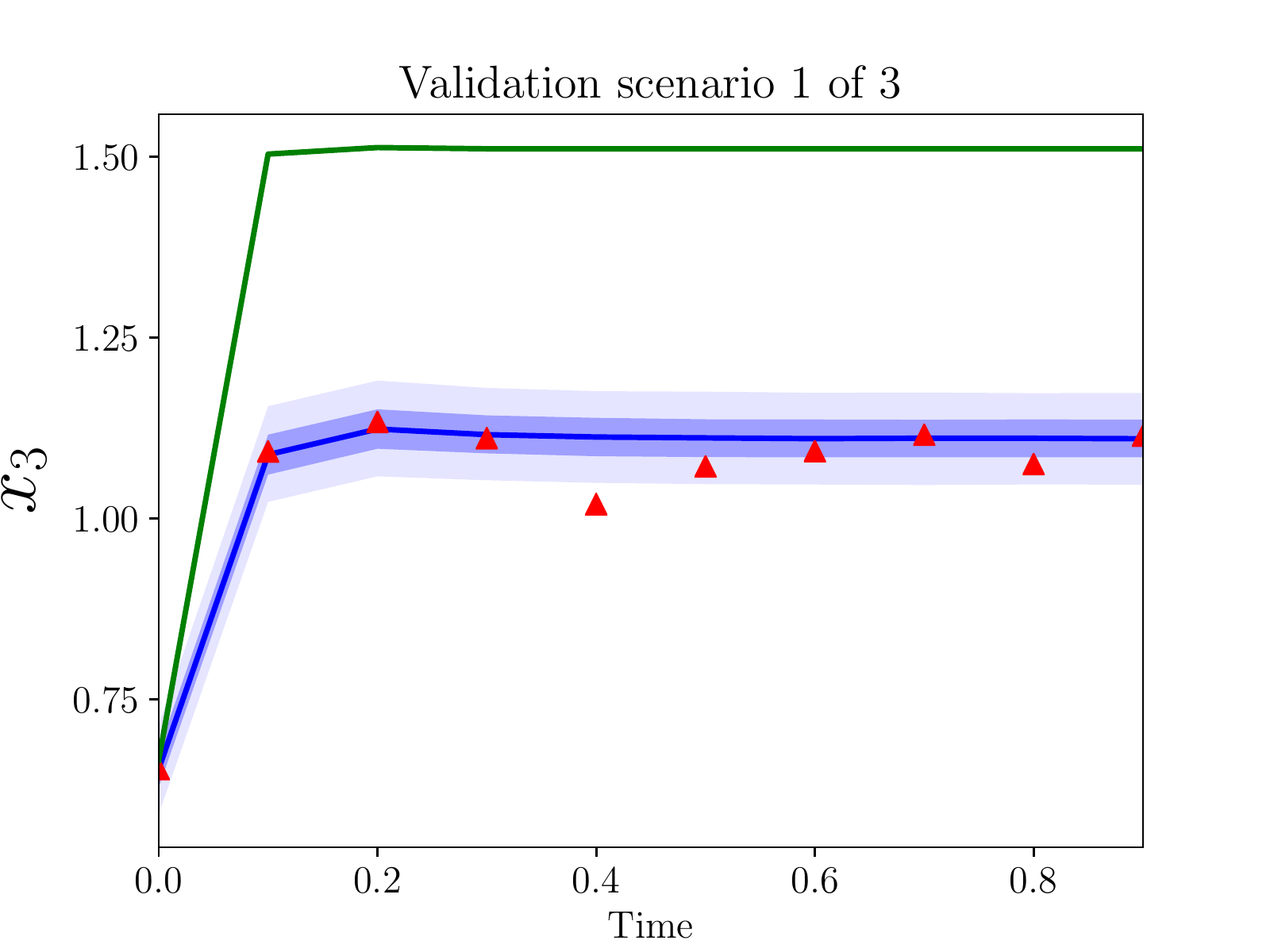}
    \end{subfigure}
    \begin{subfigure}{.24\textwidth}
  \centering
  \includegraphics[width=\textwidth]{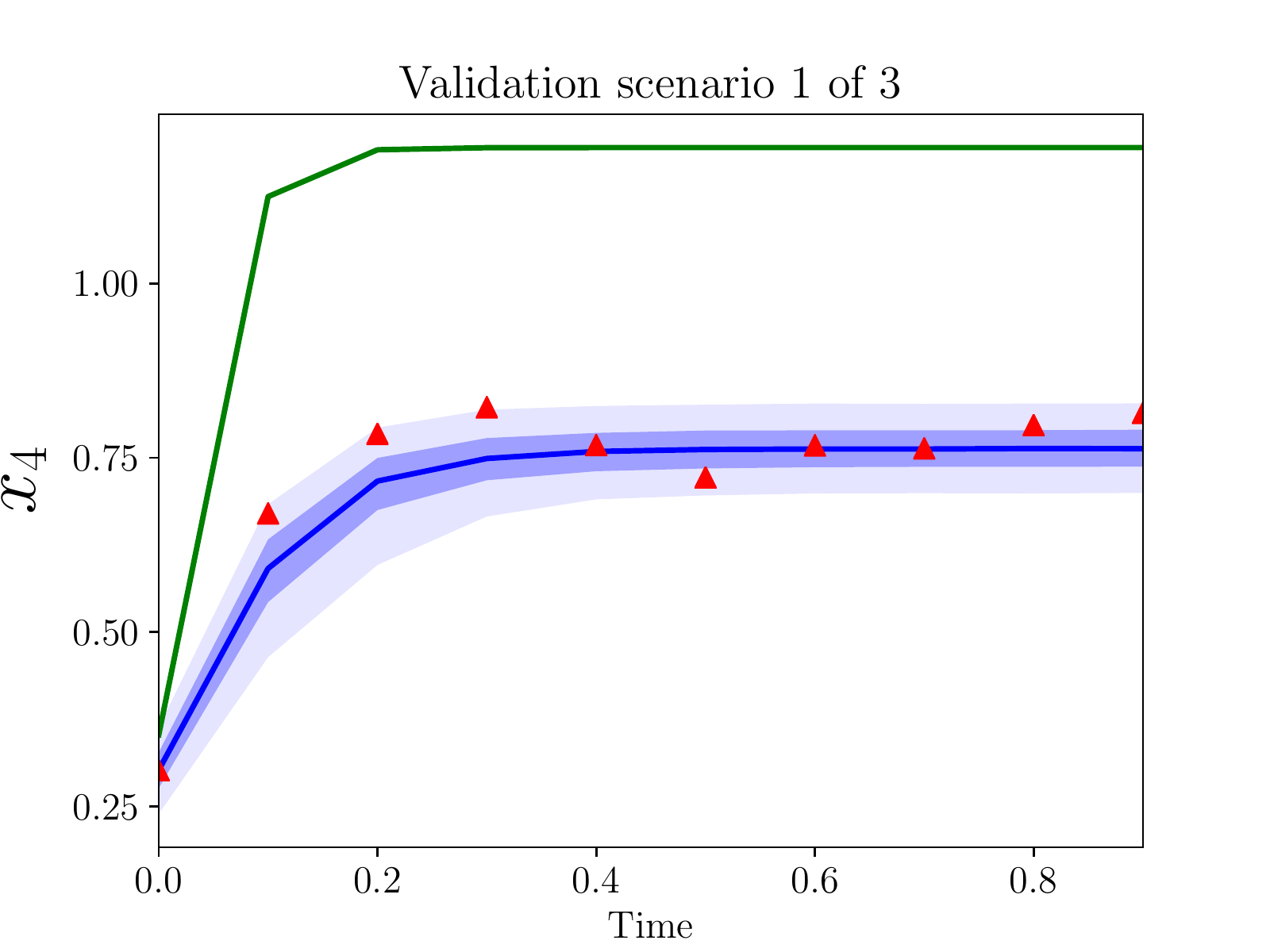}
    \end{subfigure}
    \begin{subfigure}{.24\textwidth}
  \centering
  \includegraphics[width=\textwidth]{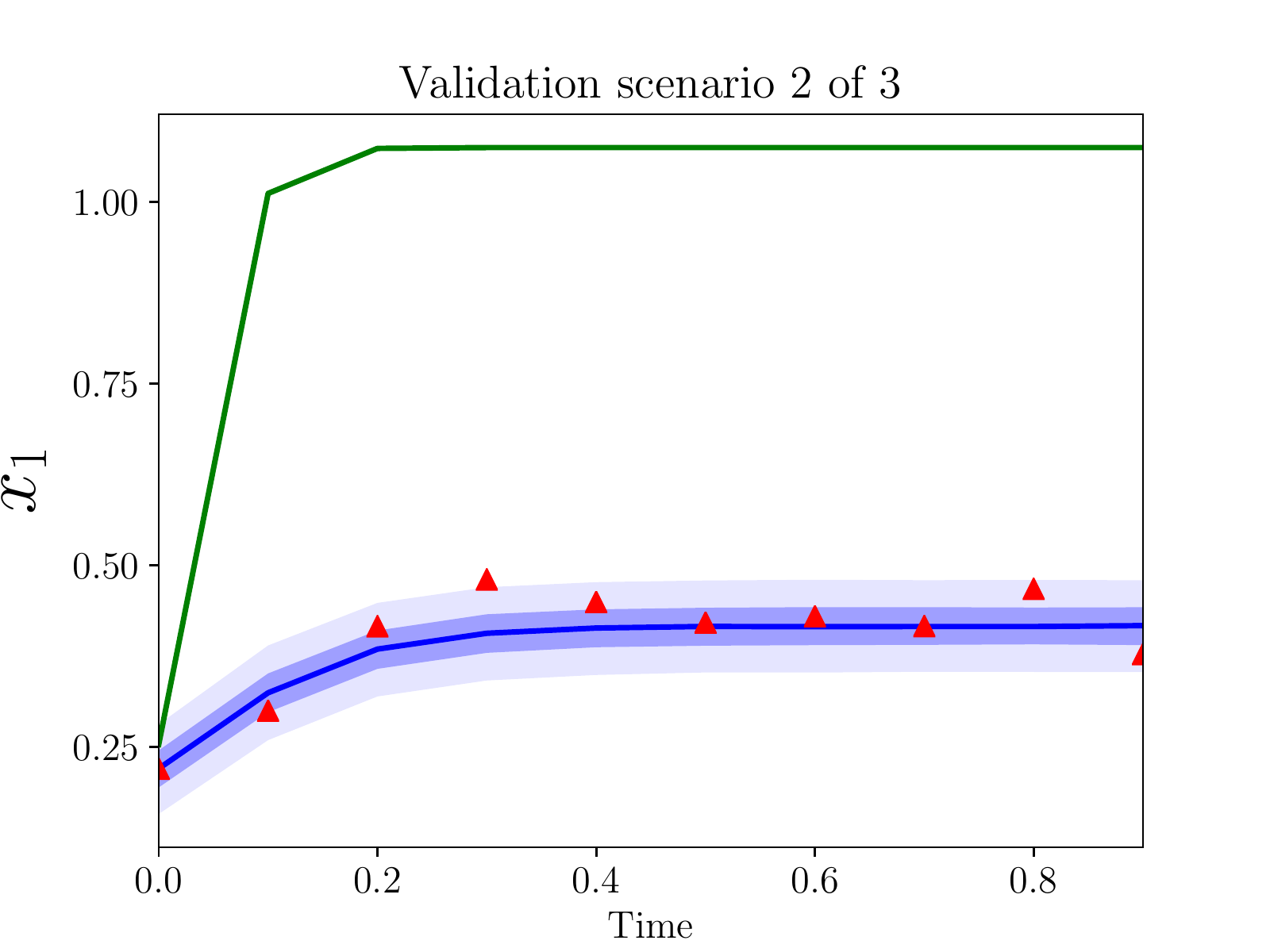}
    \end{subfigure}
    \begin{subfigure}{.24\textwidth}
  \centering
  \includegraphics[width=\textwidth]{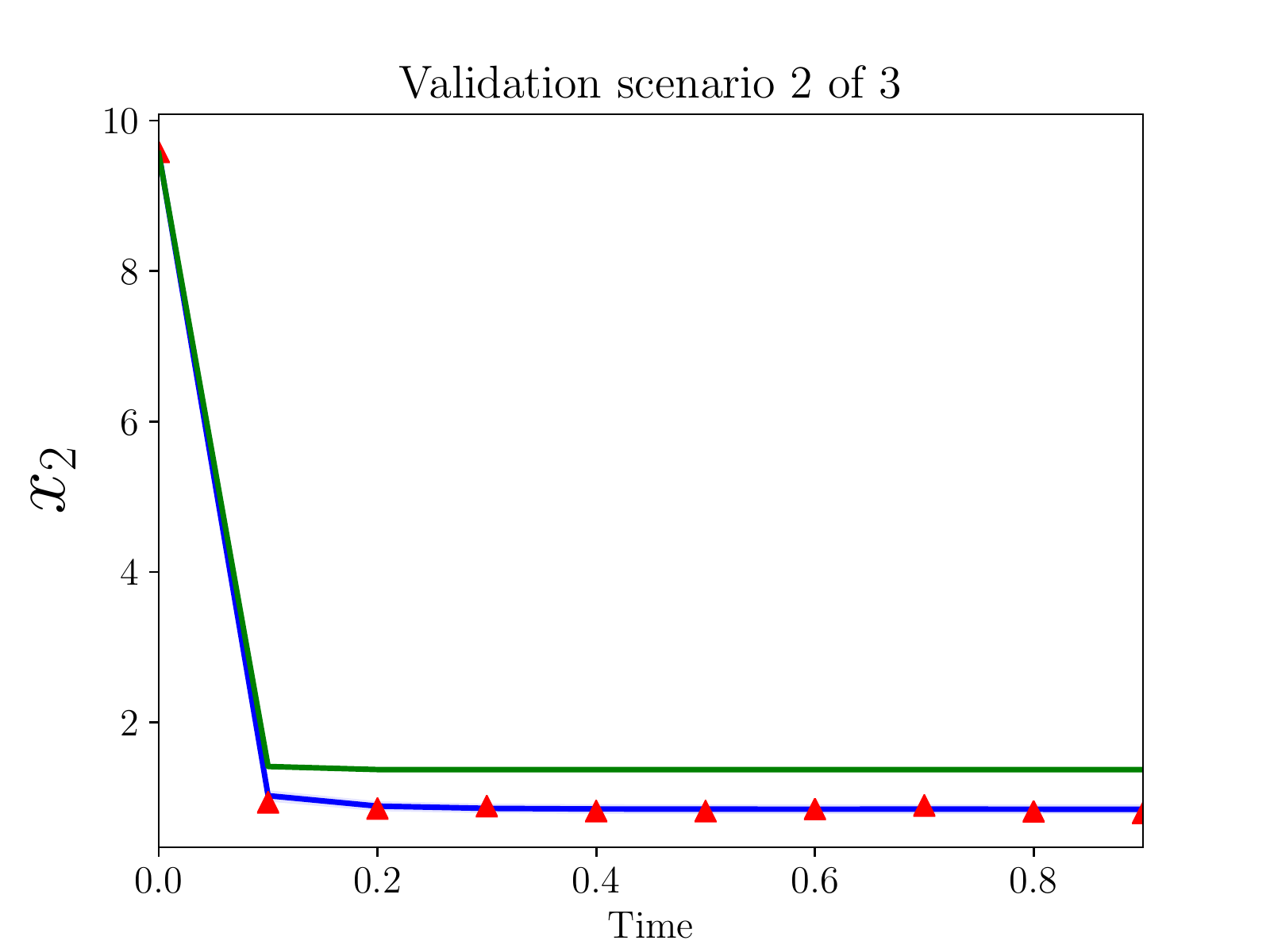}
    \end{subfigure}
    \begin{subfigure}{.24\textwidth}
  \centering
  \includegraphics[width=\textwidth]{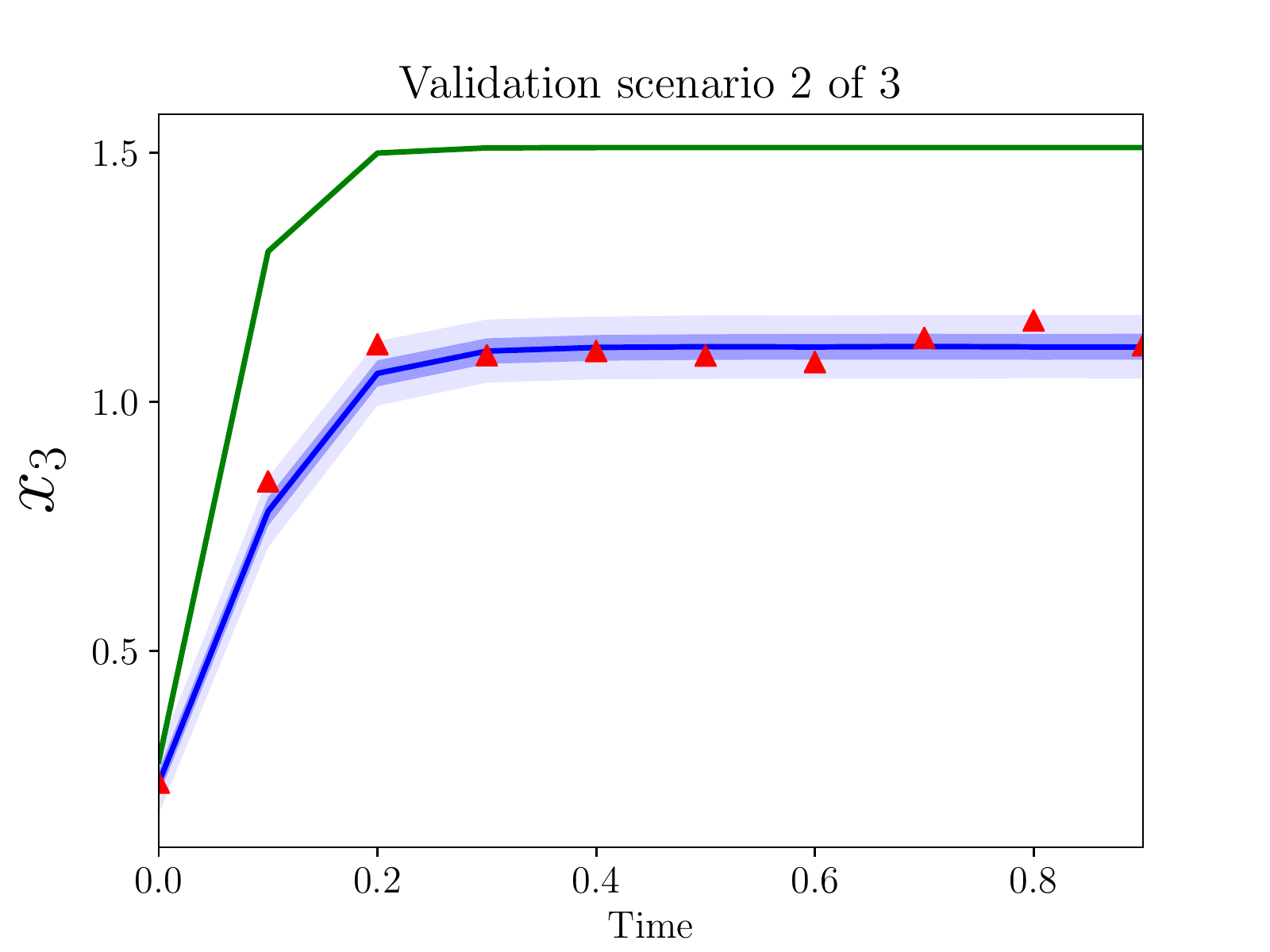}
    \end{subfigure}
    \begin{subfigure}{.24\textwidth}
  \centering
  \includegraphics[width=\textwidth]{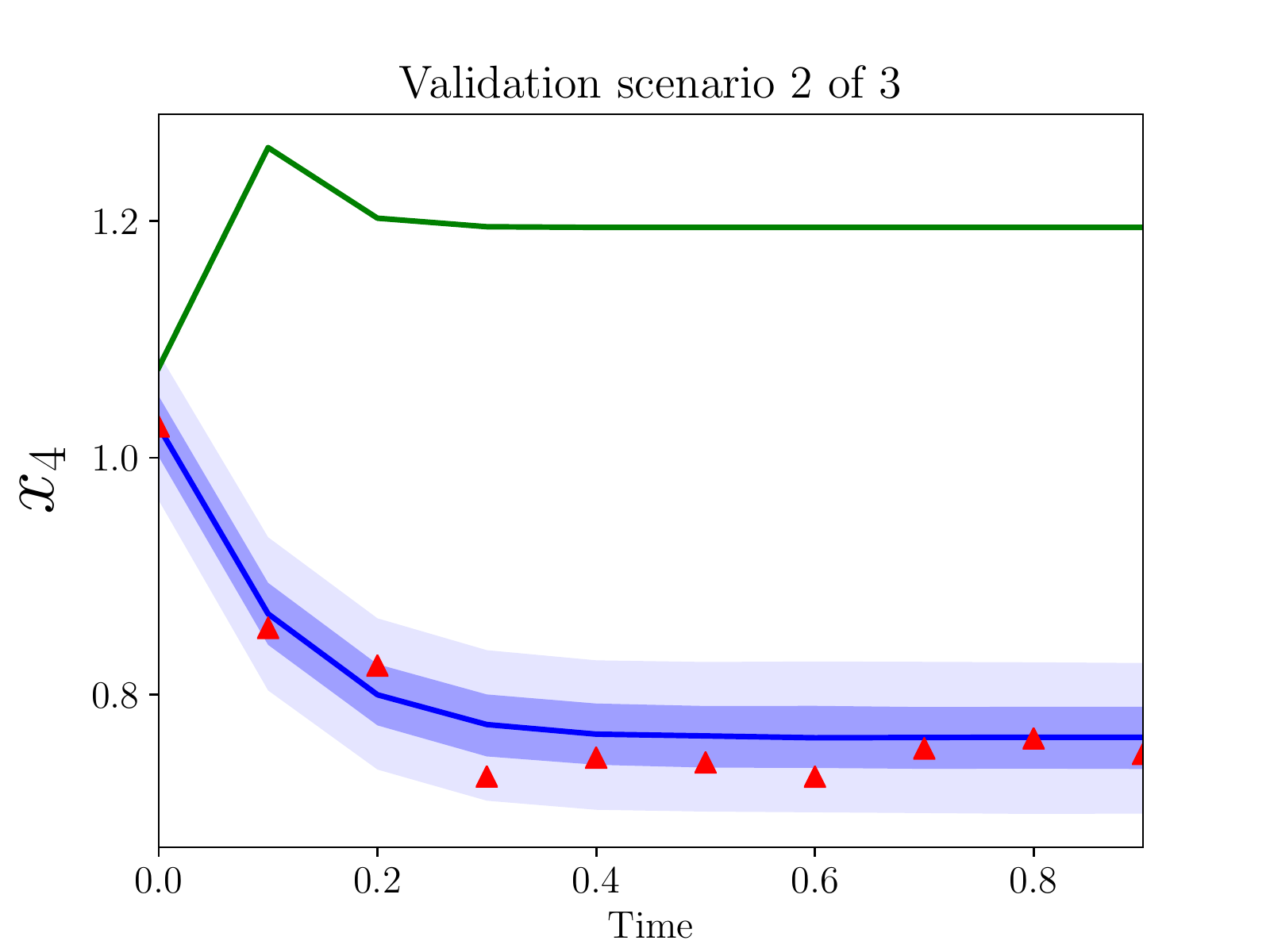}
    \end{subfigure}
    \begin{subfigure}{.24\textwidth}
  \centering
  \includegraphics[width=\textwidth]{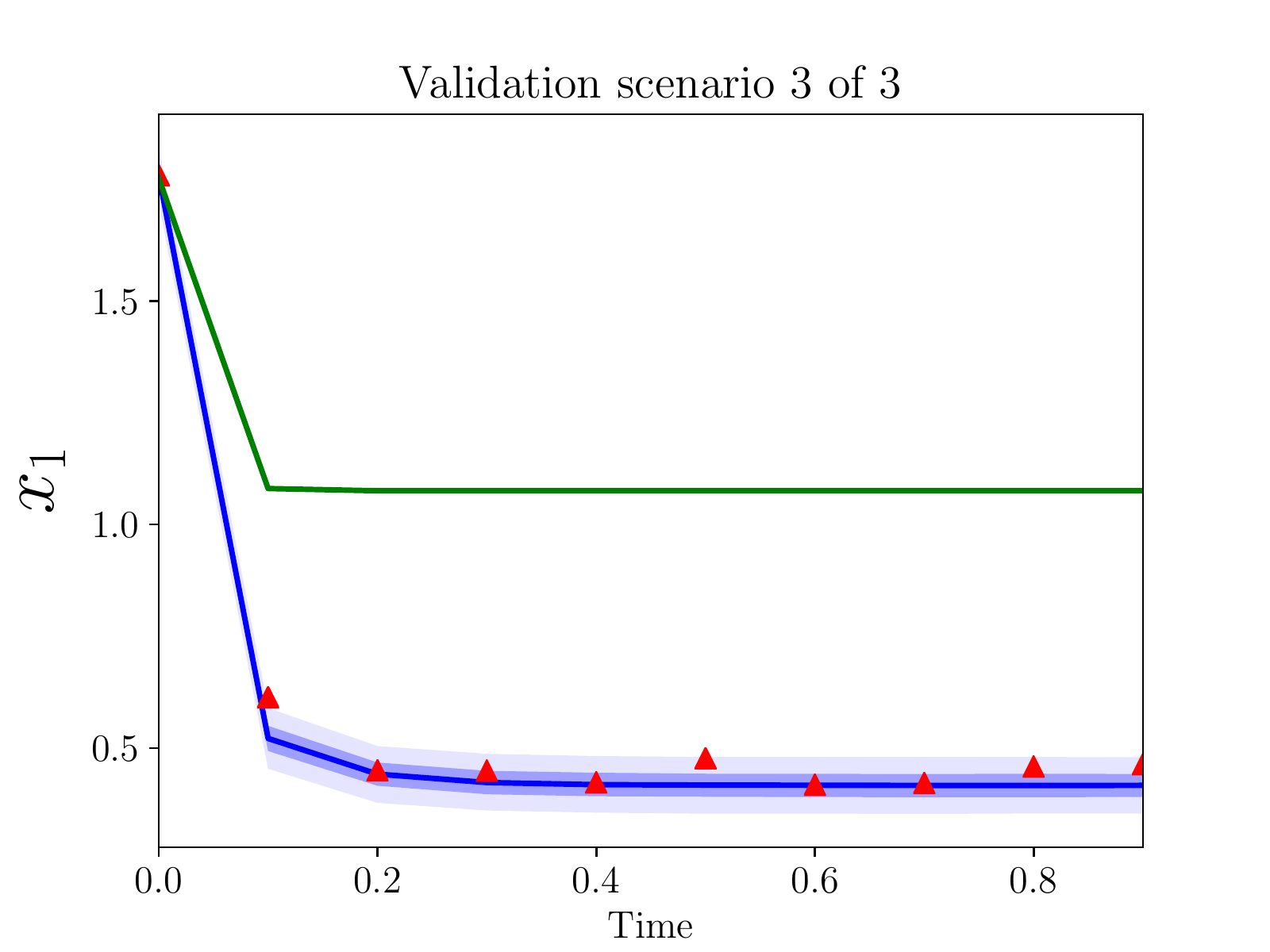}
    \end{subfigure}
    \begin{subfigure}{.24\textwidth}
  \centering
  \includegraphics[width=\textwidth]{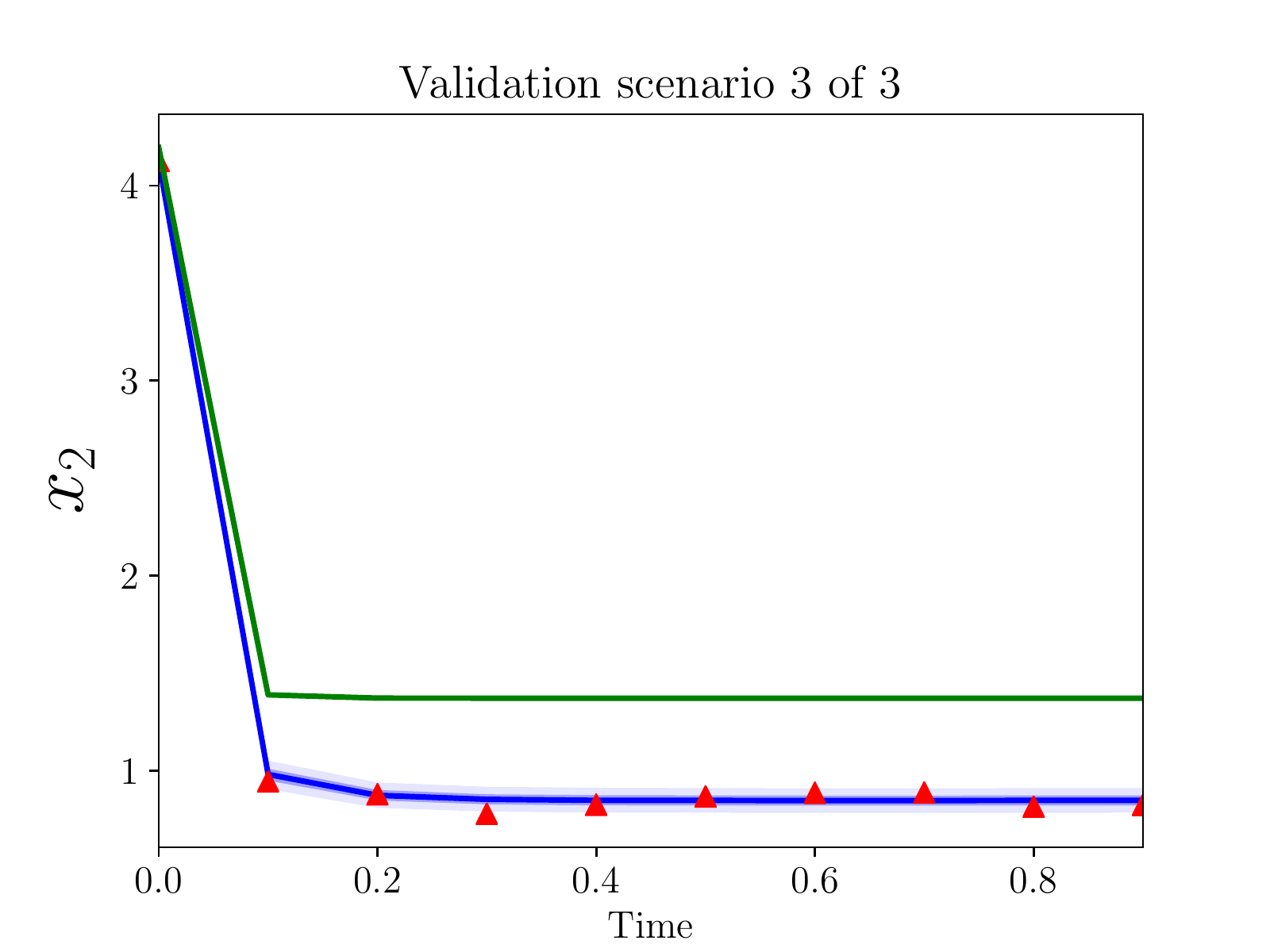}
    \end{subfigure}
    \begin{subfigure}{.24\textwidth}
  \centering
  \includegraphics[width=\textwidth]{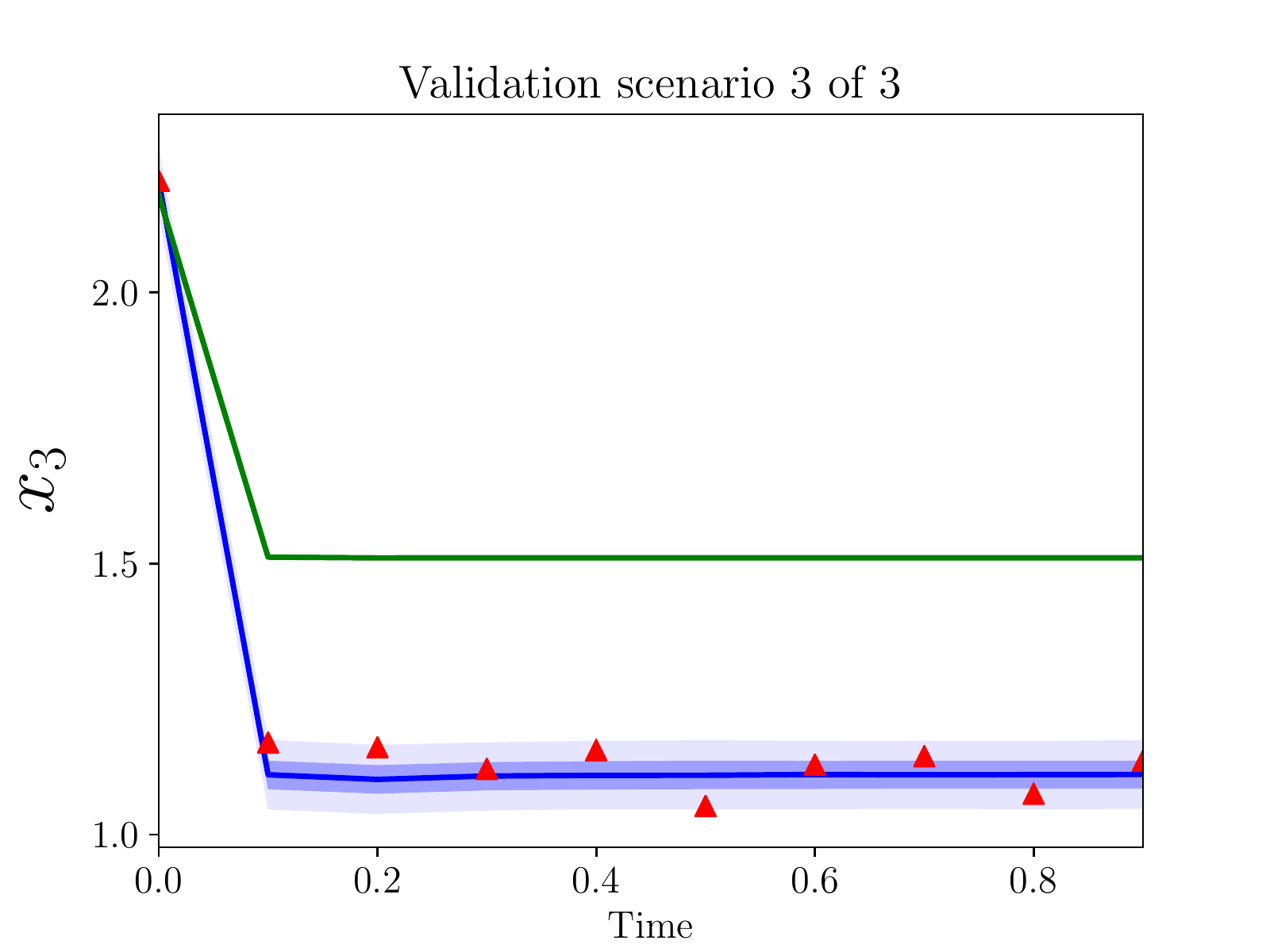}
    \end{subfigure}
    \begin{subfigure}{.24\textwidth}
  \centering
  \includegraphics[width=\textwidth]{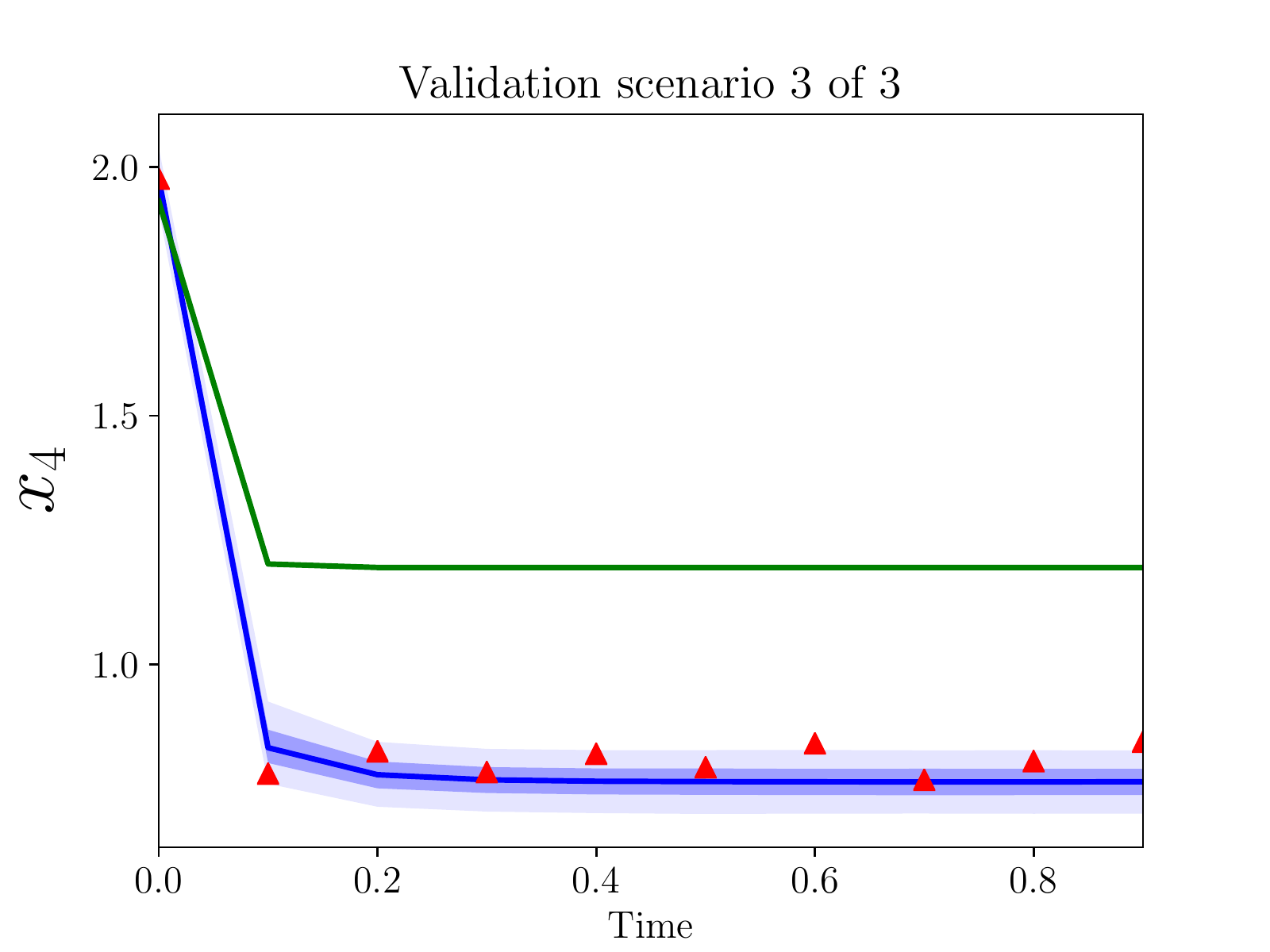}
    \end{subfigure}
    \caption{Reduced and enriched models, compared to observations, over three validation scenarios.
    $S=20, s = 4$. \label{fig:S20s4val3}}
\end{figure}

\subsection{Results for one realizaton of the detailed model}\label{ssec:num-one}
First, let us examine results for a single detailed and reduced model.  The detailed model is
generated according to Algorithm~\ref{alg:det}, with the following values:
\begin{equation}
    S = 10, \quad \sigma^2_B = 1,\quad \sigma^2_C = 1.
\end{equation}
Then the reduced model is generated according to Algorithm~\ref{alg:red} with $s=4$.

In this example, the observations from the detailed model are taken such that  $n_{\phi_c} = 3$,
$n_{\phi_v} = 3$, $T=10$, and $\sigma^2_\epsilon = 0.001$.

Figures~\ref{fig:S10s4cal3} and~\ref{fig:S20s4cal3} show trajectories for  $S=10$, $s=4$ and $S=20$, $s=4$,
 respectively. The reduced set variables are given for the three models: detailed, reduced,
and enriched. The 50\% and 95\% quantiles are plotted for the enriched model output. There is an
obvious discrepancy between the output from the detailed and reduced models, and the enriched model
is able to capture the bulk of this discrepancy. Nearly all of the observations from the detailed
model are contained within the model output bounds from the enriched model.

Figures~\ref{fig:S10s4val3} and~\ref{fig:S20s4val3} show the same results, but for validation scenarios. Recall that these
observations have not been used to calibrate the discprepancy operator.  The output of the enriched
model, at least to the eye, appears decent. The enriched model is greatly impoved in comparison to
the reduced model alone and, similarly to the calibration scenarios, captures the bulk
behavior of the detailed model in the validation scenarios.

In both above cases, there are a few observations which lie outside the predicted bounds of the
enriched model. This problem must be addressed more carefully with a quantitative validation process
as described in Section~\ref{sec:calval}. Additionally, these results only show the performance of
the discrepancy operator for a particular $S$ and $s$ and a single realization of $(D,R)$. The
agreement between trajectories from detailed and reduced models for different choices of $(D,R)$
are qualitatively similar, but some interesting differences appear by varying $s$ with respect to
$S$. In the next subsections, these statements are made much more precise.

\subsection{Results for many realizatons of the detailed model}\label{ssec:num-many}
\begin{figure}[htb]
  \centering
    \begin{subfigure}{.45\textwidth}
  \centering
  \includegraphics[width=\textwidth]{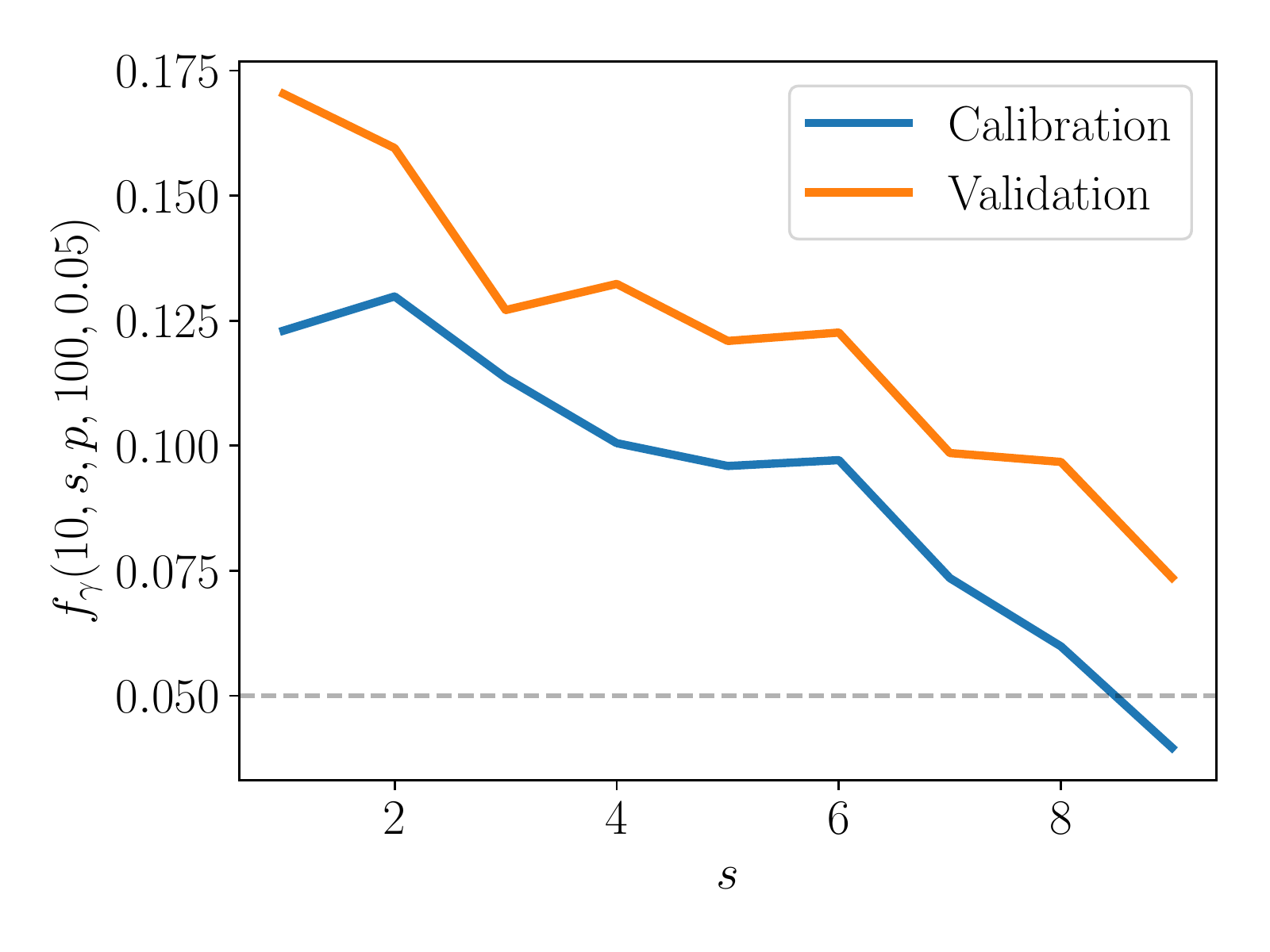}
        \caption{$\tau = 0.05$\label{fig:S10gamma05}}
    \end{subfigure}
    \begin{subfigure}{.45\textwidth}
  \centering
  \includegraphics[width=\textwidth]{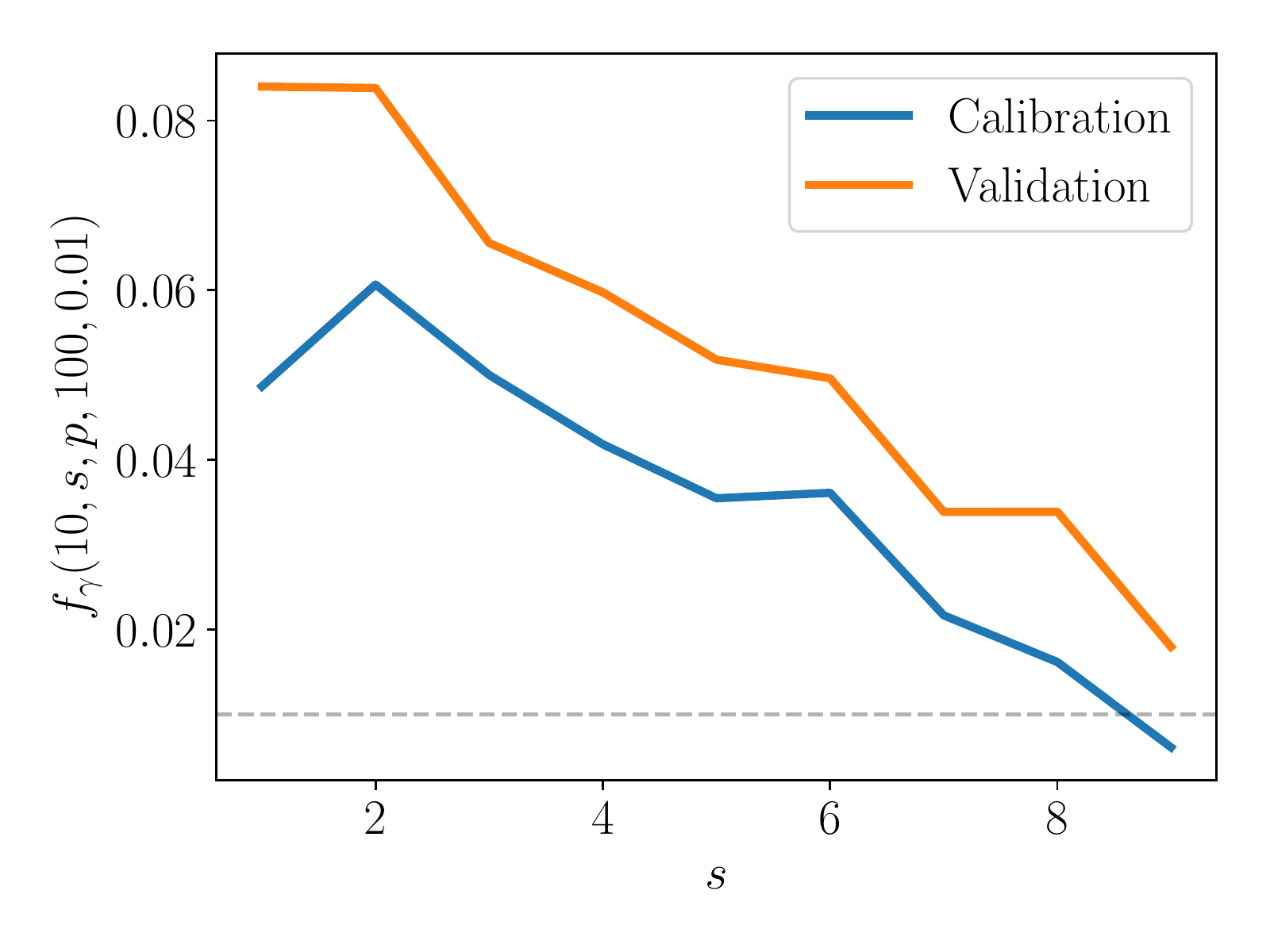}
    \caption{$\tau = 0.01$ \label{fig:S10gamma01}}
    \end{subfigure}
    \caption{Average fraction of $\gamma$-values below given threshold. $S=10$.\label{fig:S10gamma}}
\end{figure}

\begin{figure}[htb]
  \centering
    \begin{subfigure}{.48\textwidth}
  \centering
  \includegraphics[width=\textwidth]{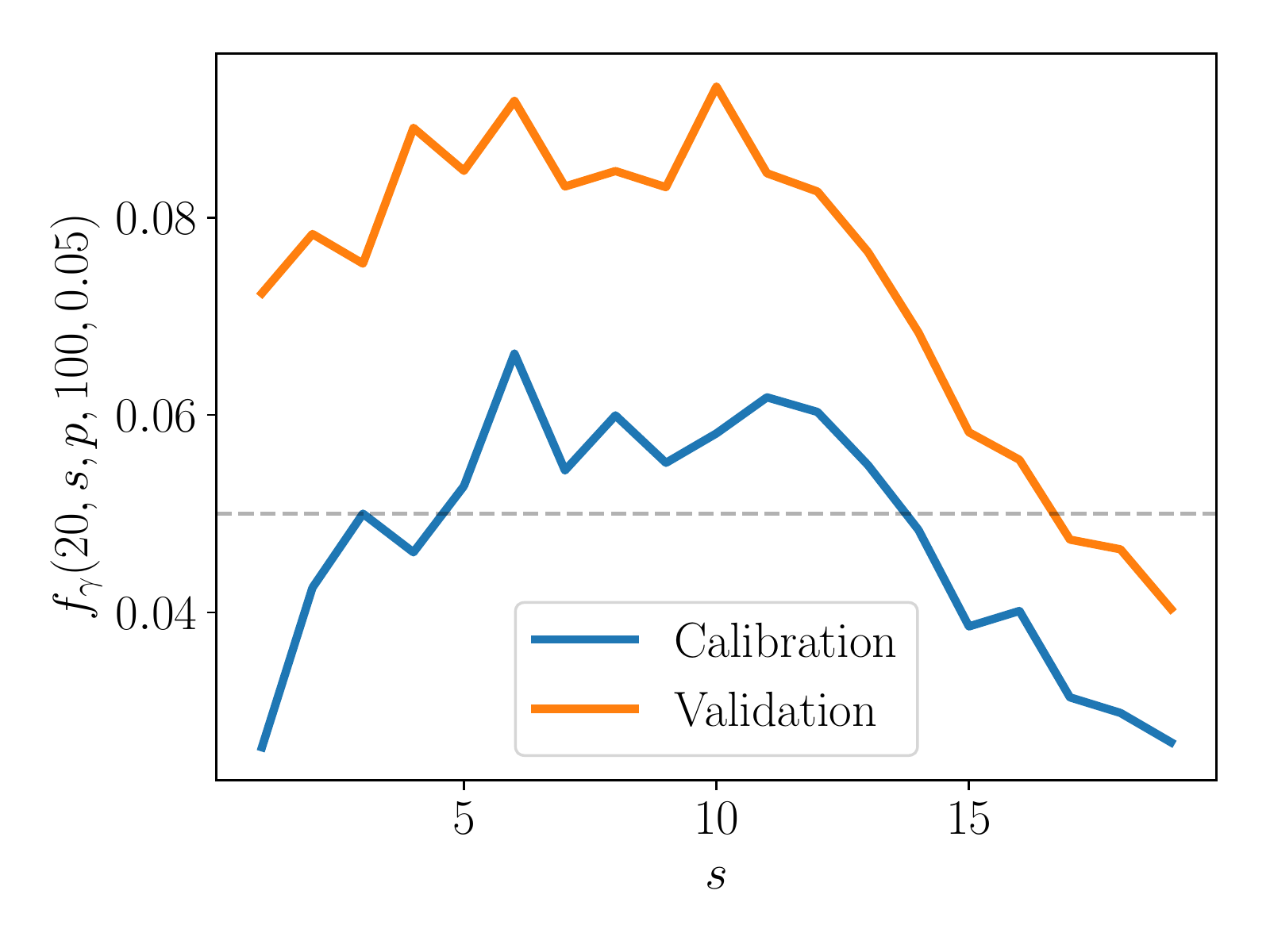}
        \caption{$\tau = 0.05$\label{fig:S20gamma05}}
    \end{subfigure}
    \begin{subfigure}{.48\textwidth}
  \centering
  \includegraphics[width=\textwidth]{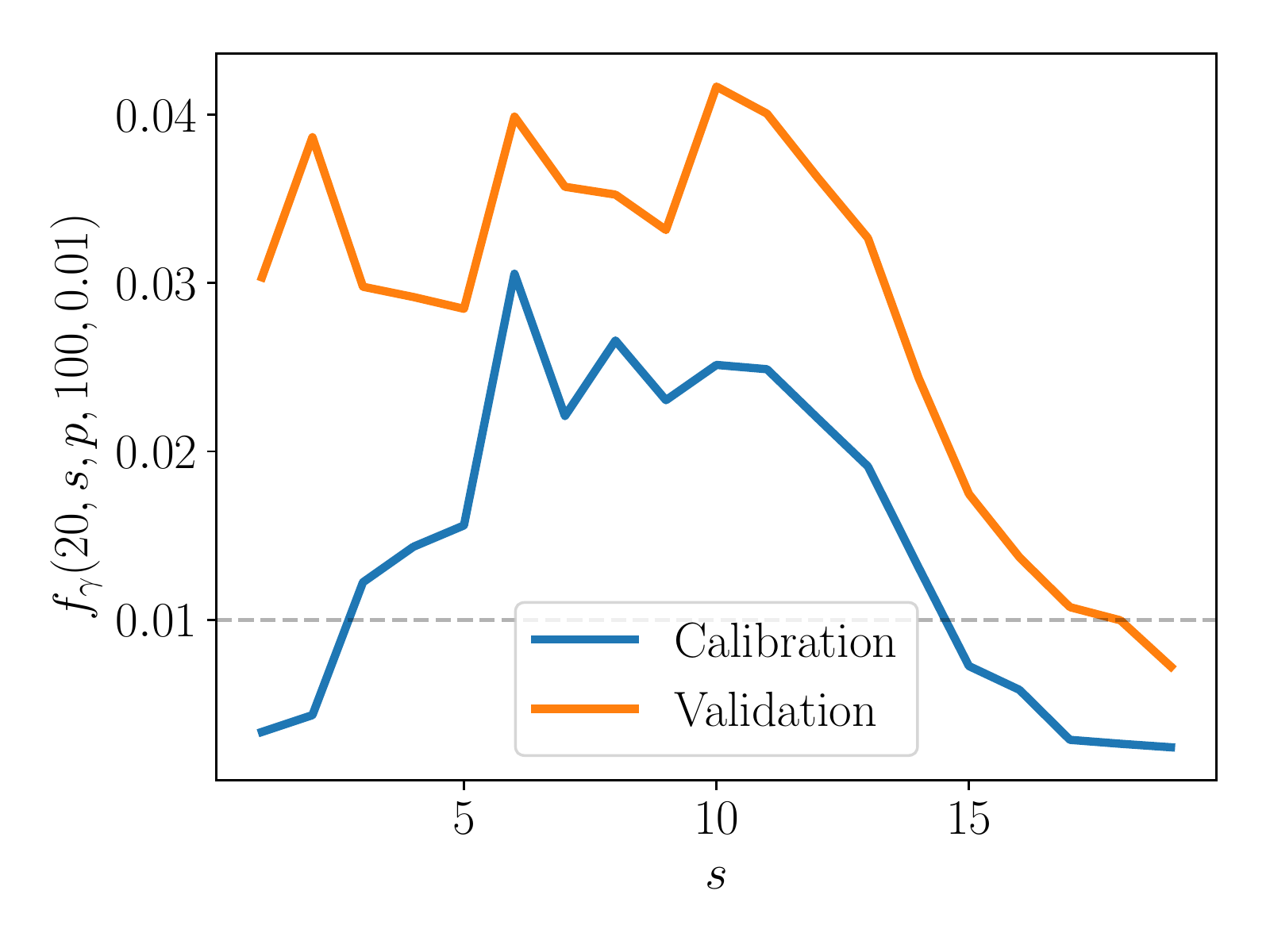}
    \caption{$\tau = 0.01$ \label{fig:S20gamma01}}
    \end{subfigure}
    \caption{Average fraction of $\gamma$-values below given threshold. $S=20$.\label{fig:S20gamma}}
\end{figure}

Now we examine the performance of the proposed discrepancy model in the context of
random forward models. To this end, three relevant concepts are detailed below.
\begin{enumerate}
    \item Let us quantify the average performance of the discrepany models. In this sense, we
        compute these $\gamma-$values for trajectories from $n_M$ realizations of detailed models,
        so that $n_M \gg 1$.

\item Note $\gamma$-values are computed with two types of data---calibration and validation data. To
    refer to these two types of data, we will use the variable $p = \{c,v\}$, so that $p=c$ denotes
        calibration data and $p=v$ denotes validation data.
    We must check how well the enriched model performs both in terms of the data that has been
        used to calibrate it, and also in terms of data that has not. Both types are shown in
        Figures~\ref{fig:S10gamma} and~\ref{fig:S20gamma}.

 \item Finally, let us examine how well the discrepancy operators perform for different pairs
     $(S,s)$.  We fix $S, s, p, n_M$ and then compute $\gamma-$values for all type $p$ observations
        over $n_M$ models, for a particular pair $(S,s)$. Call this set of $\gamma$-values
        $\Gamma(S,s,p,n_M)$.
        Now let $Q(S,s,p,n_M,\tau) = \{ \gamma_i : \gamma_i < \tau, \gamma_i \in \Gamma(S,s,p,n_M)
        \}$. Then the fraction of $\gamma$-values below the threshold $\tau$ is:
        \begin{equation} f_\gamma(S,s,p,n_M,\tau) = 
            \frac{|Q|}{|\Gamma|}.
        \end{equation}
        For example, if we want to compute $f_\gamma$ for all calibration data over $n_M$ model
        realizations, the denominator above is $|\Gamma| = sTn_{\phi_c}n_M$. The value $f_\gamma$ is
        plotted in Figures~\ref{fig:S10gamma} and~\ref{fig:S20gamma}, and $S$ is
        fixed at 10 and 20, respectively. Along the $x$-axis, $s$ ranges from 1 to $S-1$. Also shown
        are results for two values of $\tau$: 0.05 (shown in Figures~\ref{fig:S10gamma05}
        and~\ref{fig:S20gamma05}) and 0.01 (shown in Figures~\ref{fig:S10gamma01}
        and~\ref{fig:S20gamma01}).
\end{enumerate}

Let $\alpha = s/S$.  In the case that the model truly does represent the data-generating process and
in the limit of infinite observations, then this fraction of $\gamma$-values below the threshold is
equal to the threshold itself. That is
\begin{equation}
    \lim_{n_M \to \infty} f_\gamma(S, s, p, n_M,\tau) = \tau,
\end{equation}
when the model is a true match to the data-generating process.  Indeed $f_\gamma(10,s,c,100,\tau)$
approaches $\tau$ as $\alpha$ approaches $1$ (Figure~\ref{fig:S10gamma}). This suggests that the
enriched model is better able to capture the behavior of the detailed one as more species are
included in the reduced model, as one might expect.

Interestingly, in the $S=20$ case, $f_\tau$ peaks somewhere in the middle of the plot,
when $\alpha \approx 0.5$ (Figure~\ref{fig:S20gamma}). In other words, the enriched model is poorest for moderate $\alpha$, and
performs best as $\alpha$ approaches 1. Consider that when $\alpha$ is very low, only a few species
are included in the reduced model relative to the detailed one. But also, the discrepancy model has
only those few species to modify. On the other hand, when $\alpha$ is close to one, the reduced
model already includes much of the detailed model, and the discrepancy model must only fill a small
``gap'' between the two. For moderate $\alpha$, however, there are neither of these advantages---the
discrepancy model must account for the behavior of a large enough number of species, but the reduced
model is still significantly lacking compared to the detailed. A more rigorous study is clearly
needed to explain this behavior further.

\subsection{Relative model complexity}
\begin{figure}[htb]
  \centering
    \begin{subfigure}{.45\textwidth}
  \centering
  \includegraphics[width=\textwidth]{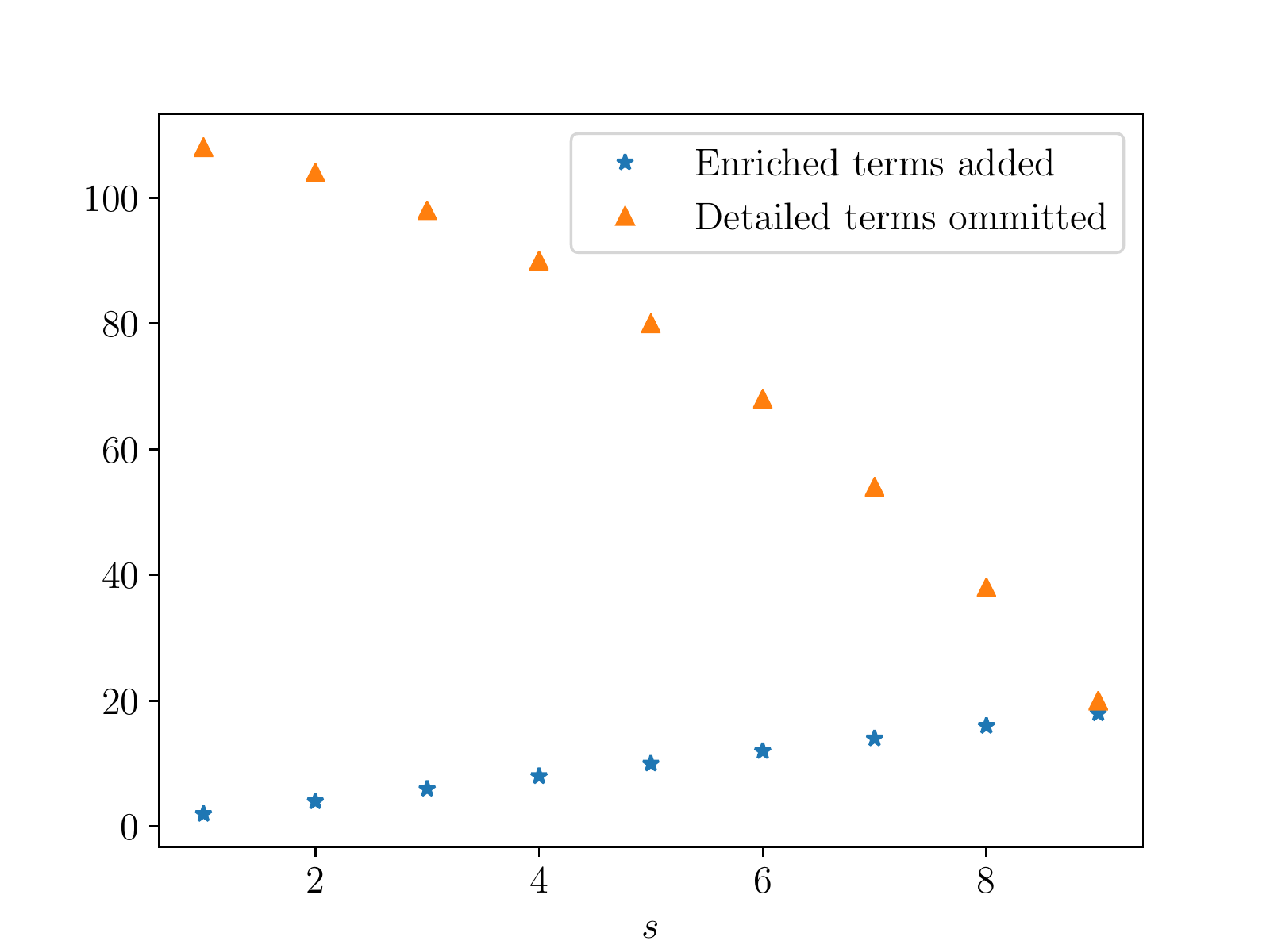}
        \caption{$S=10$}
    \end{subfigure}
    \begin{subfigure}{.45\textwidth}
  \centering
  \includegraphics[width=\textwidth]{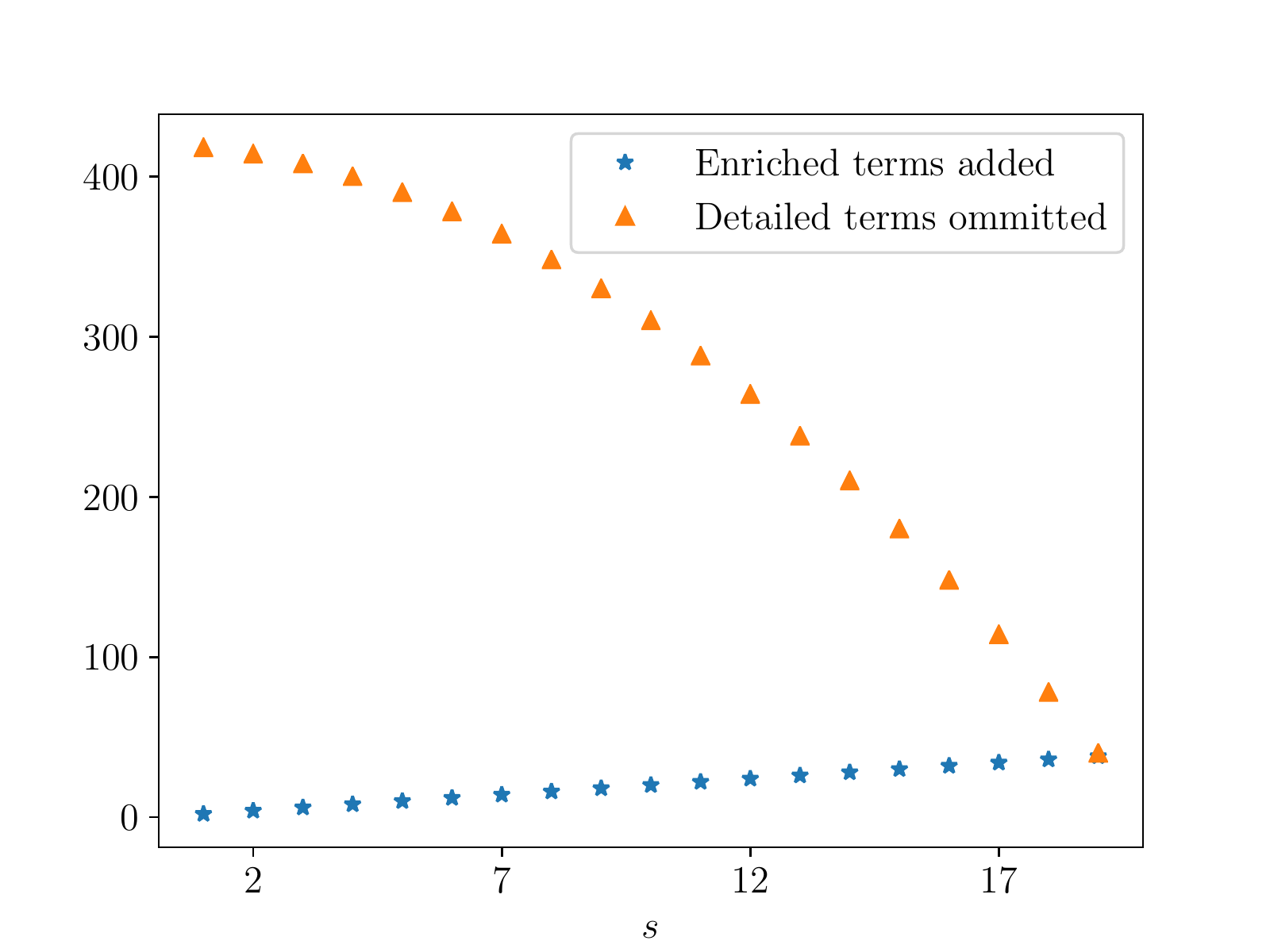}
    \caption{$S = 20$}
    \end{subfigure}
    \caption{Comparison of number of terms added by the enriched model and terms omitted from the
    detailed model.\label{fig:added}}
\end{figure}

A good discrepancy model should not overfit the data, and the best discrepancy model would be rich
enough to capture the relevant behavior of the detailed model without adding unnecessary complexity.
Although there are different ways one might measure complexity, here we measure the number of terms
introduced in the enriched model ($2s$) compared to those omitted from the detailed model. These
omitted terms include the $S^2 - s^2$ interspecific and intraspecific interaction terms, as well as
the $(S-s)$ growth rate terms. (Note the number of terms introduced is equal to the number of
enriched model parameters.)  For the cases $S=10, 20$, the absolute values are shown in
Figure~\ref{fig:added}.

\begin{figure}[htb]
  \centering
  \includegraphics[width=.5\textwidth]{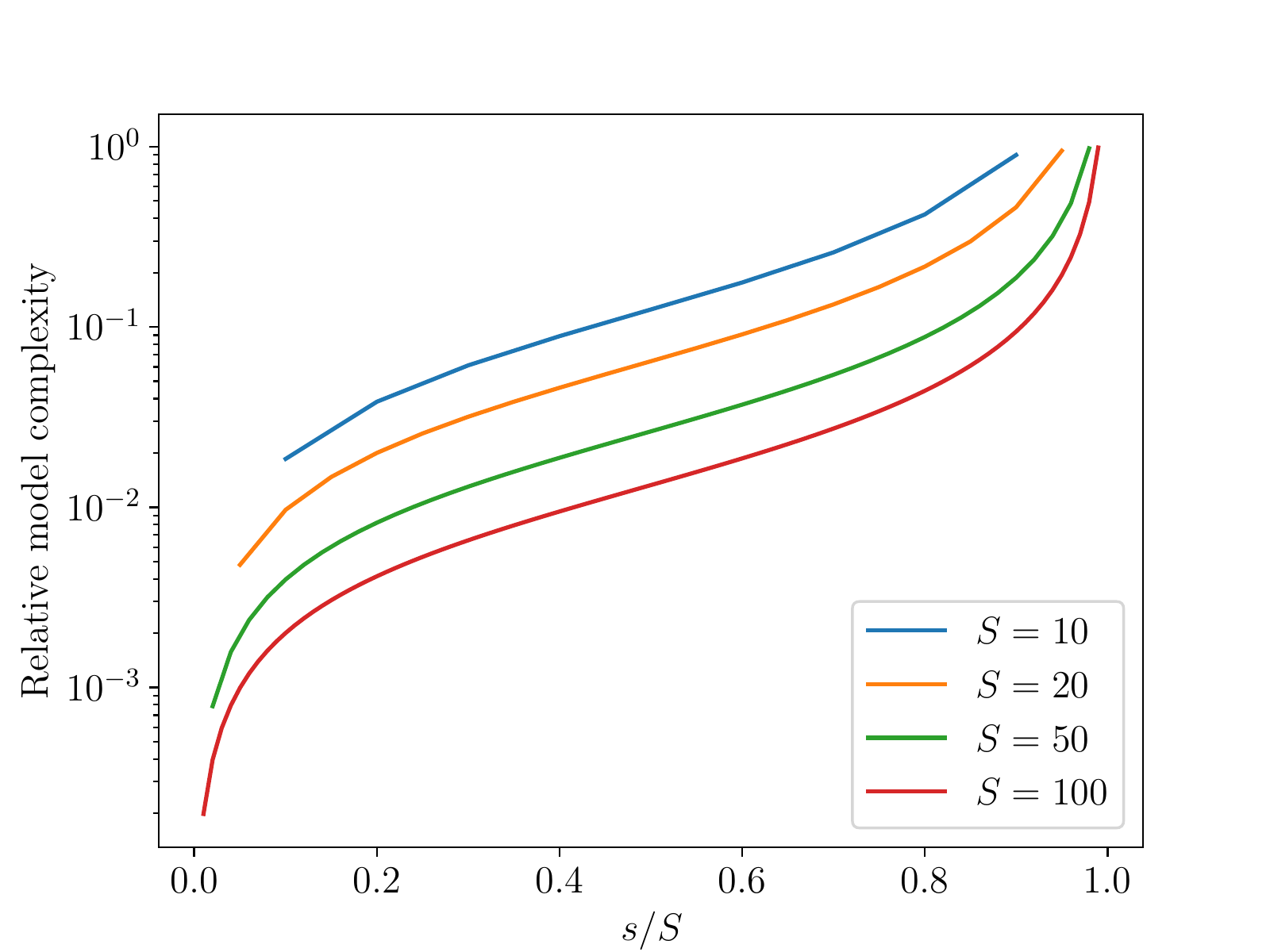}
  \caption{Enriched model complexity, measured as the ratio of enriched model terms added to detailed
    model terms omitted.\label{fig:rmc}}
 \end{figure}

Figure~\ref{fig:rmc} presents this information as a ratio of terms added relative to terms omitted
for various values of $S$. We call this ratio the \emph{relative model complexity}.
 The relative model complexity is plotted as $\alpha$ varies from $1/S$ to $(S-1)/S$ for
 a few different values of $S$. These include the two cases presented here ($S=10, 20$). We also
 show the relative model complexity for two higher values of $S$, namely $S=50$ and $S=100$. One
 might be interested in how this type of model complexity would scale for much larger systems.
 Moreover, if one knew a priori the true value of $S$ for some system, one could balance the
 effectiveness of the enriched model (as measured by $f_\tau$) against its relative model
 complexity.

Strikingly, the enriched models introduce many fewer terms than what the reduced models omit. For
example, in the two specific forward models shown in
Figures~\ref{fig:S10s4cal3}-\ref{fig:S20s4val3}, the relative model complexity is less than 0.1. And
yet the enriched model and observations do show surprisingly good agreement.

\section{Conclusion}
\label{sec:con}
This study takes an initial step towards representing model discrepancy in nonlinear dynamical
systems of interacting species. The proposed discrepancy model here is a linear operator, embedded
within the differential equations. The particular form is motivated by circumstances in which a set of
differential equations can be converted to a set of fewer equations; in this decoupling process, more
information must be introduced about the remaining set, such as memory or higher derivatives. In
this work, the discrepancy model is similarly constructed by introducing more information about the reduced set,
namely as a linear operator which acts on the remaining variables and (the absolute values of) their first derivatives.

As an initial study, the numerical results here do not show that the discrepancy models have
completely accounted for all of the discrepancy between the detailed and reduced models, for every
combination of $(S,s)$. However, the linear embedded discrepancy operators do show promise as
possible discrepancy models in the context of the generalized Lotka-Volterra equations. The results
also bring up as many new questions as they have answered.

For example, one question to pursue more rigorously is the following: What is the effective
dimension of the missing dynamics of the reduced model? That is, how many (and which) new random
variables need to be introduced to effectively (i.e., within some tolerance) capture the error of
the reduced model? The initial results here suggest that the discrepancy between the reduced and
detailed models can, under some conditions, be adequately described with a relatively small number
of discrepancy variables and parameters. But an outstanding question is whether or not some estimate
of this ``effective discrepancy dimension'' can be found a priori. Certainly, such an
estimate would heavily rely on given knowledge of the detailed and reduced models. 

Another avenue to explore is the design and analysis of more elaborate discrepancy representations
in the generalized Lotka-Volterra setting, including those with second (or higher) derivatives,
memory, nonlinear terms, or some combination of these.  Of course, a trade-off exists between the
richness of the discrepancy representation and the computational expense of both the forward and
inverse problems.

The detailed models (and thus also reduced models) investigated here are quite simple---the
interaction matrices are negative-definite, diagonally dominant, symmetric, with off-diagonal
entries sampled from identical distributions. An immediate next step in this research is to examine
the performance of linear embedded discrepancy operators after relaxing these restrictions on the
interaction matrix.

Finally, the two previous questions tackle problems in the same context (GLV equations) with more
complex discrepancy formulations or more complex interaction matrices. To understand the problem of
model discrepancy in a more general sense, another important and open direction to investigate is
the design and performance of embedded discrepancy operators in other physical settings. Doing so
would reveal what disciplines and classes of models could most benefit from embedded discrepancy
operators, and what we may hope to gain by their deployment.



\section*{Acknowledgments}
I would like to acknowledge Bob Moser and Youssef Marzouk for many helpful discussions about this
work.
\bibliographystyle{siamplain}
\bibliography{references}

\end{document}